\documentclass[aps, prd, amsmath, floats, floatfix, superscriptaddress,
nofootinbib,eqsecnum]{revtex4}
\usepackage{graphicx}
\usepackage{amsmath,amssymb}
\usepackage{amsfonts}
\usepackage{xspace} 
\usepackage{color}
\usepackage{dcolumn}
\usepackage{bm}

\def\laq{\raise 0.4ex\hbox{$<$}\kern -0.8em\lower 0.62ex\hbox{$\sim$}}
\def\gaq{\raise 0.4ex\hbox{$>$}\kern -0.7em\lower 0.62ex\hbox{$\sim$}}

\newcommand{\beq}{\begin{equation}}
\newcommand{\eeq}{\end{equation}}
\newcommand{\bea}{\begin{eqnarray}} 
\newcommand{\eea}{\end{eqnarray}}
\newcommand{\ba}{\begin{array}}
\newcommand{\ea}{\end{array}}

\newcommand{\mytextrm}[1]{{}}

\newlength{\sizeonefig}
\newlength{\sizetwofig}
\newlength{\sizeonefigb}
\newlength{\sizetwofigb}
\allowdisplaybreaks

\begin{document}

\title{Higher-order spin effects in the amplitude and phase of gravitational 
waveforms emitted by inspiraling compact binaries: Ready-to-use gravitational
waveforms}

\author{K. G. Arun}
\email{arun@iap.fr} 
\affiliation{Institut d'Astrophysique de Paris, UMR 7095-CNRS, Universit{\'e}
Pierre et Marie Curie, 98$^{\text{bis}}$ boulevard Arago, 75014 Paris, France}
\affiliation{LAL, Universit{\'e} Paris-Sud, IN2P3/CNRS, Orsay, France}

\author{Alessandra Buonanno}
\email{buonanno@umd.edu}
\affiliation{Maryland Center for Fundamental Physics, 
Department of Physics, University of Maryland, College Park, MD 20742, USA }

\author{Guillaume Faye}
\email{faye@iap.fr} 
\affiliation{Institut d'Astrophysique de Paris, UMR 7095-CNRS, Universit{\'e}
Pierre et Marie Curie, 98$^{\text{bis}}$ boulevard Arago, 75014 Paris,
France}

\author{Evan Ochsner}
\email{evano@umd.edu}
\affiliation{Maryland Center for Fundamental Physics, 
Department of Physics, University of Maryland, College Park, MD 20742, USA }

\begin{abstract}
We provide ready-to-use {\it time-domain} gravitational waveforms for spinning
compact binaries with precession effects through 1.5PN order in amplitude and
compute their mode decomposition using spin-weighted $-2$ spherical harmonics.
In the presence of precession, the gravitational-wave modes $(\ell,m)$ contain
harmonics originating from combinations of the orbital frequency and
precession frequencies. We find that the gravitational radiation from
binary systems with large mass asymmetry and large inclination angle can be
distributed among several modes. For example, during the last stages of
inspiral, for some maximally spinning configurations, the amplitude of the
$(2,0)$ and $(2,1)$ modes can be comparable to the amplitude of the $(2,2)$
mode. If the mass ratio is not too extreme, 
the $\ell=3$ and $\ell=4$ modes are
generally one or two orders of magnitude smaller than the $\ell = 2$ modes.
Restricting ourselves to spinning, non-precessing compact binaries, we apply
the stationary-phase approximation and derive the {\it frequency-domain}
gravitational waveforms including spin-orbit and spin(1)-spin(2) effects
through 1.5PN and 2PN order respectively in amplitude, and 2.5PN order in
phase. Since spin effects in the amplitude through 2PN order affect
only the first and second harmonics of the orbital phase, they do not extend
the mass reach of gravitational-wave detectors. However, they can interfere
with other harmonics and lower or raise the signal-to-noise ratio depending on
the spin orientation. These ready-to-use waveforms could be employed in the data-analysis of
the spinning, inspiraling binaries as well as in comparison studies at the
interface between analytical and numerical relativity.

\end{abstract}


\maketitle

\section{Introduction}
\label{intro}

Coalescing compact binaries made of neutron stars (NS) and/or black holes (BH)
can produce gravitational waves (GW) strong enough to be detected by
ground-based interferometers, such as LIGO~\cite{ligo}, Virgo~\cite{virgo} and
GEO~\cite{geo}, operating in the frequency range $10 \mbox{--}10^4$ Hz.
Moreover, supermassive BH binaries could be observed at lower frequencies
$10^{-5} \mbox{--} 10^{-1}$ Hz and up to cosmological distances by the
proposed laser space-based antenna LISA~\cite{lisa}. For detection purposes, 
matched filtering is applied to noisy data in order to
extract any signals that match members of the template
bank~\cite{LIGO40m99,LSCnsns,LSC05insp}.

Gravitational waves produced during the long inspiral phase can accurately be
modeled by the post-Newtonian (PN) approximation to general
relativity~\cite{Bliving}. As the BHs approach each other and their velocities
increase, the PN expansion is expected to become less and less reliable. 
Late in the evolution, non-perturbative information contained in
numerical-relativity (NR) simulations and PN-resummed
methods~\cite{BuonD00}, as well as perturbation theory 
need to be taken into account in building analytical
templates for inspiral, merger, and ringdown. In this paper, we shall limit the
discussion to the inspiral phase of coalescing BHs.

In constructing templates for detecting inspiraling signals, it is
recommended to account for all physical effects which contribute
significantly to the gravitational waveform. Those produced by the spins of
the binary constituents are among the most important ones, especially for
asymmetric compact binaries~\cite{BCPTV05}, such as NS-BH
binaries~\cite{BelczynskiSpinEvoln07}, and BH-BH binaries with component
masses $(m_1,m_2) \in [5,15]M_\odot \times [1,5]M_\odot$. For detecting such
systems, one may be able to employ phenomenological methods which capture the
essential features of spinning, precessing waveforms~\cite{BCV03b}. However,
parameter extraction~\cite{CF94,PW95,Sluys07MCMC} would warrant the inclusion
of as much information about the spins of the binaries as possible, so that
one should employ physical templates~\cite{PBCV04,BCPV04,Buonanno06} at the
highest PN order for this purpose.

For non-spinning compact binaries, the GW phase evolution has been computed
through 3.5PN order~\cite{BDI95,BDIWW95,B96, BFIJ02,BDEI04} and the $h_+$ and
$h_\times$ polarizations are available through 3PN order~\cite{BIWW96,ABIQ04,
  KBI07,K07,BFIS08}. For spinning, precessing binaries, the GW phase evolution
is known through 2.5PN order~\cite{FBBu06,BBuF06} for spin-orbit couplings,
and through 2PN order~\cite{KWWi93} for spin-spin couplings (spin(1)-spin(1)
and spin(2)-spin(2) contributions have been obtained in
Refs.~\cite{MVG05,RBK08}). Spin-orbit and spin(1)-spin(2) effects in the $h_+$
and $h_\times$ polarizations were computed through 1.5PN and 2PN order,
respectively, in Refs.~\cite{K95,WWi96}. \footnote{Note that spin-orbit
  effects through 2PN order in the $h_+$ and $h_\times$ polarizations were
  calculated in Ref.~\cite{OTO98}. However, Ref.~\cite{BBuF06} pointed out
  that a few multipole moments were computed incorrectly there.} More
recently, the spin(1)-spin(2) contributions at 3PN order in the conservative
two-body dynamics were found employing either effective-field theory
techniques~\cite{PR06,PR07,PR08b,ML08} or the Hamiltonian formalism of Arnowitt,
Deser and Misner~\cite{SHS07,SSH08,SHS08}. Now, spin(1)-spin(1) and
spin(2)-spin(2) effects at 3PN order in the conservative two-body dynamics are
also available~\cite{PR08a}. For including those higher-order spin effects in
the GW phase evolution and polarizations, the results
\cite{PR08a,PR08b,ML08,SHS07,SHS08} need to be extended to the non-conservative
dynamics, notably to the GW energy flux.

The importance of using templates that have amplitude corrections beyond the
leading PN order (henceforth referred to as Newtonian
approximation\footnote{Note that the leading PN order in the
    polarization amplitude is proportional to $1/c^4$ when one turns the
    fundamental constants on. However, being
    the leading term in a PN expansion, it is has become common to call it
    Newtonian.}) was emphasized by different authors in the context of
ground-based~\cite{SinVecc00a,Chris06,ChrisAnand06,ChrisAnand06b} and
space-based
detectors~\cite{SinVecc00b,MH02,HM03,AISS07,AISSV07,TriasSintes07,BHHS}, both
for detection and parameter estimation. So far, the effect of spins and
precession on parameter estimation was studied in
Refs.~\cite{Vecchio04,LangHughes06,LangHughes07,Sluys07MCMC}, but those
studies were limited to non-spinning and Newtonian GW
polarizations~\cite{BIWW96,ABIQ04,KBI07}.

In this paper we provide ready-to-use $h_+$ and $h_\times$ polarizations in
time domain for spinning, precessing binaries through 1.5PN order. The actual
computation of the gravitational waveform $h_{ij}$ through 1.5PN order was
done by Kidder~\cite{K95}, as well as Will and Wiseman~\cite{WWi96}, but the
ready-to-use $h_+$ and $h_\times$ polarizations at 1.5PN order were only
written explicitly for strictly circular orbits for which spins are aligned
with the orbital angular momentum. Recently, Ref.~\cite{MV08} has obtained 
the time-domain GW polarizations for generic orbits through 1.5PN order in the
  binary's comoving frame. The $h_+$ and $h_\times$ polarizations derived in
the present paper for spinning, precessing binaries through 1.5PN order
reduces to that of Refs.~\cite{WWi96,K95} in the aligned case except for a few
typographical errors which we correct.
 
In view of future studies at the interface between analytical and numerical
relativity~\cite{BCP07NR,Bertietal07,NRPNGoddard07,BuonEOB07,NRPNJena07,
NRPNCaltech07,HHBG07,Damour2007a,DN2007b,DN2008,UMD-CC08,BCGSB07} we decompose
the time-domain $h_+$ and
$h_\times$ polarizations in spin-weighted $-2$ spherical harmonics and compute
the modes, $h_{\ell m}$, to 1.5PN order. We then consider spinning,
non-precessing binaries for which we derive the Fourier domain representation
of the generated gravitational waveform within the stationary phase
approximation (SPA). We provide a very compact way of writing the Fourier
transforms of $h_+$ and $h_\times$ which can readily be used for data
analysis, for comparisons with numerical simulations, or for building
analytical frequency-domain templates including inspiral, merger and
ringdown~\cite{AjithNR07a,Pan07comparison}.
The impact of spinning, precessing
waveforms for parameter estimation will be investigated in a future paper.

The remainder of the paper is organized in the following way. In
Sec.~\ref{secI} we draw the source and detector frames, and introduce
conventions and notations. In Sec.~\ref{secII} we provide ready-to-use $h_+$
and $h_\times$ polarizations in time domain for nearly circular orbits. The
polarization modes $h_{\ell m}$ with respect to the spin-weighted $-2$ spherical
harmonics are derived in Sec.~\ref{secIII}. The features of the modes when
spins are present is then discussed in Sec.~\ref{secIIIbis}.
Section~\ref{secIV} focuses on spinning, non-precessing binaries. We compute
there the Fourier domain waveforms with spin effects through 2PN order in the
amplitude and 2.5PN order in the phase before discussing the main features
caused by higher harmonics. Finally, we summarize in Sec.~\ref{concl} our main
conclusions. Appendices~\ref{appB} and \ref{appD} present the GW polarizations
and modes for precessing binaries on nearly circular orbits through 1.5PN
order for generic inclination angles, whereas Appendix~\ref{appA} shows the PN
coefficients of the center-of-mass energy and radiative energy flux for
non-precessing, spinning binaries. Appendix~\ref{appC} gives explicitly
  the frequency domain amplitude coefficients with non-spin terms to 2.5PN
  and spin terms to 2PN order.

\section{Source frame, polarization and parameter conventions}
\label{secI}

To obtain the GW polarizations, it is useful to express the gravitational
strain tensor, $h_{ij}$, in an appropriate source frame. Next, one specifies
an orthonormal polarization triad composed of the direction of propagation
$\mathbf{\hat{N}}$ and two polarization vectors $\mathbf{\hat{P}}$ and
$\mathbf{\hat{Q}}$ which are used to construct the GW polarizations from the
strain tensor~\cite{FinnCh93}:
\begin{eqnarray}
\label{poldefs}
h_+ & = & \frac{1}{2}(\hat{P}^i\,\hat{P}^j - \hat{Q}^i\,\hat{Q}^j)\,h_{ij}\,, \\
h_{\times} & = & \frac{1}{2}(\hat{P}^i\,\hat{Q}^j + \hat{Q}^i\,\hat{P}^j)\,
h_{ij}\,. 
\end{eqnarray}
The gravitational strain measured by a detector is then given by
\begin{equation}
h_{\rm strain}(t) = F_+\,h_+(t) + F_\times\,h_\times(t)\,,\label{measuredstrain}
\end{equation}
where $F_+$ and $F_\times$ are the antenna response functions that
describe the detector's sensitivity to the two different
polarizations. For laser interferometers with arms at a right angle,
such as the LIGO and Virgo detectors, the antenna response functions
for a GW coming from the sky location $(\bar{\theta},\bar{\phi})$ in the
  spherical coordinate grid built from the arm basis, with 
polarization angle $\bar{\psi}$, are~\cite{FinnCh93} 
\bea
\label{antennaresponse1}
F_+ & = & \frac{1}{2}(1+\cos^2 \bar{\theta})\,\cos 2\bar{\phi}\,\cos
2\bar{\psi} - \cos \bar{\theta}\,\sin 2\bar{\phi}\,\sin 2\bar{\psi}\,,\\
\label{antennaresponse2}
F_\times & = & \frac{1}{2}(1+\cos^2 \bar{\theta})\,\cos
2\bar{\phi}\,\sin 2\bar{\psi} + \cos \bar{\theta}\,\sin
2\bar{\phi}\,\cos 2\bar{\psi}\,. 
\eea
Note that the strain measured in a given instrument, $h_{\rm strain}(t)$, is the same
regardless of convention, whereas the wave polarizations depend on the choice of
polarization vectors. Different choices of $\mathbf{\hat{P}}$ and
$\mathbf{\hat{Q}}$ give different polarizations, but there is a compensating
rotation of the polarization angle $\bar{\psi}$ so that $h_{\rm strain}(t)$ is unchanged~
\footnote{This can be seen explicitly from the relation linking
$h_\text{strain}(t)$ to the complex polarization $h(t)$ introduced in
Eq.~\eqref{hc}: $ h_\text{strain} = \Re \Big[h\, e^{2 i \bar{\Psi}}
(e^{2i \bar{\phi}}\, \cos^4 ({\bar{\theta}}/{2})+ e^{-2 i \bar{\phi}}\,
\sin^4 ({{\bar{\theta}}/{2}}) )\Big]$. The reader can easily check
the equivalence with Eqs.~\eqref{measuredstrain},
\eqref{antennaresponse1} and \eqref{antennaresponse2}.}. 
Here, we follow the convention of Refs.~\cite{WWi96,ABIQ04}, in which
\begin{equation}
\label{PandQ}
\mathbf{\hat{P}} = \frac{\mathbf{\hat{N}} \times
  \mathbf{J}_0}{|\mathbf{\hat{N}} \times \mathbf{J}_0|}\,,\quad \quad
\mathbf{\hat{Q}} = \mathbf{\hat{N}} \times \mathbf{\hat{P}}\,,
\end{equation}
where $\mathbf{J}_0$ is the unit vector along the initial total angular
momentum of the binary. In the absence of precession, the Newtonian orbital
angular momentum $\mathbf{L}_{N} = \mu \mathbf{r} \times \mathbf{v}$ (with
$\mathbf{r}$, $\mathbf{v}$, and $\mu$ being the binary separation vector,
velocity, and reduced mass, respectively) is parallel to $\mathbf{J}_0$. In
this case, $\mathbf{\hat{P}}$ coincides with the {\it ascending node} where
the orbital separation vector crosses the plane of the sky from below. In
the presence of precession, $\mathbf{\hat{P}}$ is still defined as
$\mathbf{\hat{N}} \times \mathbf{J}_0 / |\mathbf{\hat{N}} \times \mathbf{J}_0|$,
 but it is not in general the point
where the orbital separation vector ascends through the plane of the sky.~
\footnote{Note that Ref.~\cite{K95} chooses polarization vectors rotated by
${\pi}/{2}$ relative to ours. This results in an overall sign difference
from our polarizations, as can be seen by making the substitutions
$\mathbf{\hat{P}} \rightarrow \mathbf{\hat{Q}}$ and $\mathbf{\hat{Q}}
\rightarrow -\mathbf{\hat{P}}$, or by noting that GWs are spin-2 objects and
flip sign under a ${\pi}/{2}$ rotation. As mentioned, the polarization
angle of this convention is then rotated by ${\pi}/{2}$ relative to ours. This
flips the sign of the antenna response functions as well, and so the same strain
(\ref{measuredstrain}) is measured by either convention.}

\begin{figure}
\includegraphics[width=0.45\linewidth]{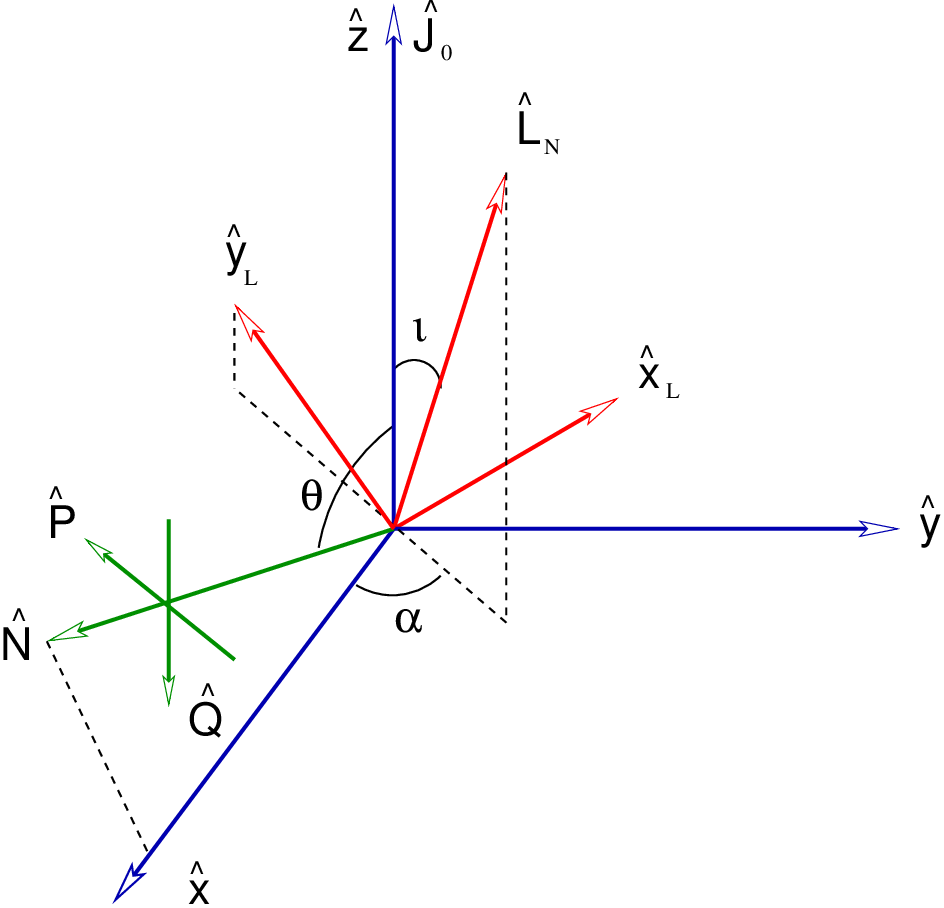}
\caption{We show (i) our source frame defined by the orthonormal basis 
$(\mathbf{\hat{x}},\mathbf{\hat{y}},\mathbf{\hat{z}})$, (ii) the instantaneous 
orbital plane which is described by the orthonormal basis 
$(\mathbf{\hat{x}}_{\rm L},\mathbf{\hat{y}}_{\rm L},\mathbf{\hat{L}}_{\rm N})$, 
(iii) the 
polarization triad $(\mathbf{\hat{N}},\mathbf{\hat{P}},\mathbf{\hat{Q}})$, 
and (iv) the direction 
of the total angular momentum at initial time $\mathbf{J}_0$. 
Dashed lines show projections into the $x\mbox{--}y$ plane.
\label{figure:SourceFrame}}
\end{figure}

For our source frame, we construct an adapted orthonormal basis
$(\mathbf{\hat{x}},\mathbf{\hat{y}},\mathbf{\hat{z}})$ (see
Fig.~\ref{figure:SourceFrame}). We take the $z$-axis to be along
$\mathbf{J}_0$ and the direction of GW propagation, $\mathbf{\hat{N}}$, to lie
in the $x\mbox{--}z$ plane, tilted by an angle $\theta$ from the
$z$-axis towards the $x$-axis. We describe the direction of the Newtonian
orbital angular momentum with the spherical coordinate angles
$(\iota,\alpha)$, where $\iota$ denotes the angle between the orbital angular
momentum and the $z$-axis while $\alpha$ is the angle between the $x$-axis and
the projection of the orbital angular momentum onto the $x\mbox{--}y$ plane.
For precessing binaries, as these angles vary in time, one must solve the
precession equations to find their evolution. Notice that this source frame is
the same as used in Ref.~\cite{K95}, and depicted in Fig.~2 of that paper.

We also find it useful to define basis vectors for the instantaneous orbital
plane. These vectors have an implicit time dependence through the angles
$(\iota,\alpha)$, and rotate about $\mathbf{\hat{L}}_{N}$ as it precesses.
Here are their components in the
$(\mathbf{\hat{x}},\mathbf{\hat{y}},\mathbf{\hat{z}})$ source basis:
\begin{eqnarray}
\label{orbitalplanebasis}
\mathbf{\hat{x}}_L &= & \frac{\mathbf{J}_0 \times \mathbf{\hat{L}}_{\rm
    N}}{|\mathbf{J}_0 \times \mathbf{\hat{L}}_{\rm N}|} = \left( - \sin \alpha
\ ,\ \cos \alpha \ ,\ 0\ \right)\,,\\ \mathbf{\hat{y}}_L &= &
\mathbf{\hat{L}}_{\rm N} \times \mathbf{\hat{x}}_L = \left( - \cos \iota
\, \cos \alpha \ , - \cos \iota\,\sin \alpha\ ,\ \sin \iota\ \right)\,.
\end{eqnarray}
As an initial condition, we take the orbital separation vector
$\mathbf{\hat{n}}$ to lie along $\mathbf{\hat{x}}_L$ at initial time, i.e.,
$\mathbf{\hat{n}}(t=0)\ = \mathbf{\hat{x}}_L(t=0)$. Then, we define the phase
$\Phi(t)$ to be the cumulative angle between
$\mathbf{\hat{x}}_L(t)$ and $\mathbf{\hat{n}}(t)$.
\begin{eqnarray}
\label{nandlambda}
\mathbf{\hat{n}}(t) & = & \mathbf{\hat{x}}_L(t) \,\cos \Phi(t) +
\mathbf{\hat{y}}_L(t)\,\sin \Phi(t)\,,\\ \boldsymbol{\hat{\lambda}}(t) & = & -
\mathbf{\hat{x}}_L(t)\,\sin \Phi(t) + \mathbf{\hat{y}}_L(t)\,\cos \Phi(t)\,.
\end{eqnarray}
We thus see that the phase $\Phi(t)$ measures how $\mathbf{\hat{n}}$ has
  rotated relative to the vector $\mathbf{\hat{x}}_L$. However, for a
  precessing binary, $\mathbf{\hat{x}}_L$ is itself rotating about
  $\mathbf{\hat{L}}_{\rm N}$. This means that the total rotation of
  $\mathbf{\hat{n}}$ about $\mathbf{\hat{L}}_{\rm N}$ can be decomposed as a
  rotation of $\mathbf{\hat{n}}$ in the comoving basis parametrized by
  $\Phi(t)$ times a rotation parametrized by a {\it precession phase} due to
  the movement of the orbital plane itself. In the non-precessing case we
have $\mathbf{J}_0\, ||\, \mathbf{\hat{L}}_{\rm N}$ and $\Phi(t)$ is expected
to become the standard orbital phase whose time derivative is the orbital
frequency. However, when $\mathbf{J}_0\, ||\, \mathbf{\hat{L}}_{\rm N}$ holds,
we cannot define $\Phi=0$ for $\mathbf{\hat{n}} = {\mathbf{J}_0 \times
  \mathbf{\hat{L}}_{\rm N}}/{|\mathbf{J}_0 \times \mathbf{\hat{L}}_{\rm N}|}$,
and we set $\Phi=0$ at the ascending node $\mathbf{\hat{N}} \times
\mathbf{\hat{L}}_{\rm N} = \mathbf{\hat{P}}$, where the orbital separation
crosses the plane of the sky from below. Now, $\Phi=0$ at the ascending
  node is achieved for $\mathbf{\hat{x}}_L = \mathbf{\hat{n}}$ and $
  \mathbf{\hat{P}}=\mathbf{\hat{n}}$, hence $\alpha=\pi$, so that the
  non-precessing regime is reached in the limit where $\iota=0$ and
  $\alpha=\pi$ for all time. This is applicable when the spins of the bodies
  are aligned or anti-aligned with the orbital angular momentum (or in the
  non-spinning limit). The waveforms are then greatly simplified.

We define the following mass parameters
\bea
M & = & m_1 + m_2\,, \\
\nu &=& \frac{m_1\,m_2}{(m_1+m_2)^2}\,,\\
\mathcal{M} &=& M\,\nu^{3/5}\,,\\
\delta & = & \frac{m_1-m_2}{m_1+m_2}\,,\\
\nu  &=&  \frac{1}{4}(1 - \delta^2).\label{massparams} 
\eea 
They are the total mass $M$, the symmetric mass ratio $\nu$, the fractional mass
difference $\delta$, the chirp mass ${\cal M}$. The
symmetric mass ratio is bounded according to $0 < \nu \leq 1/4$ and the
fractional mass difference satisfies $-1 < \delta < 1$.

The spin of a rotating compact body is of the order 
$S \sim m\,l\,v_{\rm spin}$ with $l \sim G\,m/c^2$. If the
compact body is {\it maximally }rotating, then $v_{\rm spin} \sim c$ and 
$S \sim \chi\,G\,m^2/c$. In words, from the PN point 
of view, the spin is formally of order 0.5PN. By contrast, if the compact 
body is {\it slowly} rotating, then $v_{\rm spin} \ll c$, and the 
spin is formally of higher PN order, $S  \sim \chi\,G\,m^2\,
v_{\rm spin}/c^2 \sim 1/c^2$. Throughout the paper, we use geometrical 
units where $G = c = 1$. Henceforth, we shall work with the 
spin vectors normalized by the component masses as
\beq \boldsymbol{\chi}_n =
\frac{\mathbf{S}_n}{m_n^2}\,, \quad n = 1,2\,,\label{spinvar} 
\eeq 
so that $|\boldsymbol{\chi}_n| \leq 1$ for objects that obey the Kerr
bound on rotational angular momentum. We also define symmetric and
anti-symmetric spin combinations as in Ref.~\cite{WWi96},
\bea \boldsymbol{\chi}_s & = & \frac{1}{2}\left(\boldsymbol{\chi}_1 +
\boldsymbol{\chi}_2 \right)\,,\\ \boldsymbol{\chi}_a & = &
\frac{1}{2}\left(\boldsymbol{\chi}_1 - \boldsymbol{\chi}_2
\right).\label{symspinvar} \eea

\section{Ready-to-use gravitational-wave polarizations for precessing binaries
on circular orbits through 1.5PN order: small inclination angles}
\label{secII}

The expression of the strain tensor $h_{ij}$ for generic orbits through 1.5PN
order was derived in Refs.~\cite{BDI95,WWi96} and is given by Eq.~(6.11) of
Ref.~\cite{WWi96}. In this section we compute ready-to-use polarizations in
time domain through 1.5PN order within the {\it adiabatic} regime where the
binary inspiral is modeled as a quasi-stationary sequence of orbits assumed to
be nearly circular. By nearly circular, we essentially mean an orbit that
would be exactly circular, with separation vector $ \mathbf{r}_0 $ of constant
radius $ r_0$, in the absence of spins and gravitational radiation. The
perturbation of the separation $ \delta r$ of such a motion is assumed to
remain small with respect to $ r_0 $ on timescales on which the
radiation-reaction effects can be neglected. This can only happen when the
precession angles are at most of the same order of magnitude as the relative
corrections induced by the spins in the dynamical quantities. Now, the
evolution of $ \delta r $ is governed by the radial part of the 1.5PN
perturbation of the force per mass unit given in Eq.~(2.1) of Ref.~\cite{K95}.
It turns out that this perturbation depends on the spin exclusively through
the two projections $(\mathbf{S}_n \cdot \mathbf{\hat{L}}_{\rm N})$ with $ n
=1,2 $ which are almost constant apart from remainders that will contribute at
higher orders in
our weak precession hypothesis. In order to write the equation for $ \delta r
$, we project the relative acceleration $\mathbf{a} $ in the
basis
$\{\mathbf{\hat{n}},\boldsymbol{\hat{\lambda}},\hat{\boldsymbol{L}}_{\mathrm{N}}\}$.
For the sake of convenience, we introduce an ``orbital''-like frequency
$\omega_{\rm orb} $, defined as $ \omega_{\rm orb}= (\mathbf{v} \cdot
\boldsymbol{\hat{\lambda}})/r$. The closure relation yields the following
decomposition for $\mathbf{v}$ and $\mathbf{a}$: 
\begin{align} \label{eq:kidder_basis}
\mathbf{v} &= \dot{r} \mathbf{\hat{n}} + \omega_{\rm orb} r
\boldsymbol{\hat{\lambda}} \, , \nonumber \\ 
\mathbf{a} &= (\ddot{r}-r \omega_{\rm orb}^2) \mathbf{\hat{n}} + (r
\dot{\omega}_{\rm orb} + 2 \dot{r} \omega_{\rm orb})
\boldsymbol{\hat{\lambda}} - r 
\omega_{\rm orb} \left(\boldsymbol{\hat{\lambda}} \cdot \frac{d\mathbf{\hat{L}}_{\rm
N}}{dt}\right) \mathbf{\hat{L}}_{\rm N} \, ,
\end{align}
with $\dot{r}\equiv dr/dt$. Splitting $ \mathbf{a} $ into an unperturbed part
$\mathbf{a}_0 $ plus a perturbation $\delta \mathbf{a} $ and using the
equations of motion, we find finally that $\delta r$ satisfies the equation
$\delta \ddot{r} + \omega_0^2 \delta r = \text{const.}$ where $ \omega_0
$ is the constant angular frequency of the background motion. A particular
solution is given by a constant perturbation, $ \delta r = \text{const}. $,
whereas the homogeneous solution satisfies an harmonic oscillator equation
independent of the spin.

By making the particular choice of a zero homogeneous solution, we can always
eliminate the oscillations of $r$ that are not directly linked to the non-zero
spins of the BHs. Based on these observations, we shall \emph{define}
precisely a nearly circular motion to be \emph{a perturbed circular motion
  whose homogeneous radial perturbation solution $(\delta r)_{\rm hom} $ is
  zero, as it would be for an exactly circular motion}.\footnote{Though
    this type of motion can exist and is more general than the
    spin-aligned or anti-aligned case, it does not necessarily represent yet the
    most likely evolution to be observed.} Assuming such a dynamics for
our binary system implies that both $ \delta r $ and $r=r_0 + \delta r$ must
be constant, provided we neglect higher order spin terms and radiative
effects. We can generalize nearly circular motions to the case where spin
precession angles are arbitrary in the absence of spin-spin interactions. This
is achieved by introducing the concept of spherical motion \emph{defined} as a
motion \emph{having a constant separation $r$}. It immediately follows from
Eq.~\eqref{eq:kidder_basis} that the full (conservative) acceleration is still
of the form $ - \omega_{\rm orb}^2 r \mathbf{n}$. Moreover, when
radiation-reaction effects are neglected, the orbital frequency computed from
the 1.5PN equations of motion keeps being almost constant \cite{KWWi93,K95},
even for precession angles that are no longer small. This can be seen
\cite{FBBu06} by noticing that the only possible non-constant terms in $
\omega_{\text{orb}} $ at the 1.5PN order come from the leading spin
contribution of the equations of motion, and thus, are of the form
$(\mathbf{S}_n \cdot \mathbf{\hat{L}}_{\rm N})$. Their time derivative reads
$(d\mathbf{\hat{L}}_\mathrm{N}/dt \cdot \mathbf{S}_n) +
(\mathbf{\hat{L}}_\mathrm{N} \cdot d\mathbf{S}_n/dt)$. The first term is zero
due to the precession equation $ d\mathbf{S}_n/dt = \boldsymbol{\Omega}_n
\times \mathbf{S}_n$, while the second term is a higher order correction
quadratic in spins because of the approximate conservation of
$\mathbf{\hat{L}}_N $. The treatment of the spin-spin dynamics is
more delicate. A possible way to proceed consists in averaging the time
dependent spin contributions in $\omega_{\rm{orb}}$ over one orbital
period~\cite{K95,Poisson98}.

Introducing the invariant velocity,
\beq
v \equiv (M \omega_{\rm{orb}})^{{1}/{3}}\,,
\eeq
we reduce Eq.~(6.11) of Ref.~\cite{WWi96} to nearly circular orbits and expand
it in powers of $v$ with the help of the relativistic extension of Kepler's
law linking $\omega_{\rm orb}$ and ${M}/{r}$ provided by Eq. (7.1) of
Ref.~\cite{WWi96}. Schematically, we obtain
\begin{equation}
\label{hij}
h_{ij} = \frac{2\,M\,\nu\,v^2}{D_L} \left [ \text{Q}_{ij} +
  \text{P}^{1/2}\text{Q}_{ij}\ v + \left(\text{P}^{1}\text{Q}_{ij} +
  \text{P}^{1}\text{Q}^{\rm SO}_{ij}\right)\ v^2 +
  \left(\text{P}^{3/2}\text{Q}_{ij} + \text{P}^{3/2}\text{Q}^{\rm tail}_{ij} +
  \text{P}^{3/2}\text{Q}^{\rm SO}_{ij}\right)\ v^{3} \right]_{\rm TT}\,,
\end{equation}
where SO indicates the spin-orbit terms; the tail integral
$\text{P}^{3/2}\text{Q}^{\rm tail}_{ij}$ given by Eq.~(6.11e) of
Ref.~\cite{WWi96} reads
\begin{equation}
\text{P}^{3/2} \text{Q}_{ij}^{\rm tail} = 4 \left[\pi (\hat{\lambda}^{i}
  \hat{\lambda}^j - 
  \hat{n}^i \hat{n}^j) + 12 \ln \left(\frac{v}{v_0}\right) \hat{\lambda}^{(i}
  \hat{n}^{j)}\right]_\text{TT} \, ,
\end{equation} 
$v_0$ being an arbitrary numerical constant reflecting the freedom in the
choice of the radiative time origin. The symbol TT on the square bracket
indicates the transverse trace-free projection in the plane orthogonal to the
direction $\mathbf{\hat{N}}$ of the observer. We remind the reader that the
non-spinning contributions to Eq.~(\ref{hij}) are known through 3PN
order~\cite{K07,BFIS08}.

Apart from the spins, there are four vectors that appear in the
expressions for the $\text{P}^n\text{Q}_{ij}$'s in Eq.~(\ref{hij}). In the
source frame constructed in Sec.~\ref{secII}, they have the following
$(x,y,z)$ components
\begin{equation}
\label{nvec}
\mathbf{\hat{n}} = \left(-\sin \alpha\,\cos \Phi-\cos \iota\,\cos \alpha\,\sin
\Phi\, ,\, \cos \alpha\,\cos \Phi-\cos \iota\,\sin \alpha\,\sin \Phi\, ,\,\sin
\iota\,\sin \Phi \right)\,,
\end{equation}
\begin{equation}
\label{lvec}
  \boldsymbol{\hat{\lambda}} = \left(\sin \alpha\,\sin \Phi - \cos \iota\,\cos
  \alpha\,\cos \Phi\, ,\,-\cos \alpha\,\sin \Phi-\cos \iota\,\sin \alpha\,\cos
  \Phi\, ,\,\sin \iota\,\cos \Phi\right)\,,
\end{equation}
\begin{equation}
\label{Nvec}
\mathbf{\hat{N}} =  \left(\sin \theta\, ,\, 0\, ,\,\cos \theta\right)\,,
\end{equation}
\begin{equation}
\label{Lvec}
\mathbf{\hat{L}}_{\rm N} = \left(\sin \iota\,\cos \alpha\, ,\,\sin \iota\,\sin
\alpha\, ,\,\cos \iota\right)\,,
\end{equation}
where $\Phi $ is the phase defined in Eq.~(\ref{nandlambda}) that
  measures how $\mathbf{\hat{n}}$ has rotated relative to the vector
  $\mathbf{\hat{x}}_L$. As $\mathbf{\hat{x}}_L$ is itself rotating about
  $\mathbf{\hat{L}}_{\rm N}$ for a precessing binary, the orbital frequency,
  or the total angular velocity of $\mathbf{\hat{n}}$ about
  $\mathbf{\hat{L}}_{\rm N}$, is the angular velocity of the motion of the
  binary within its instantaneous orbital plane, plus a precession velocity
  due to the movement of the orbital plane itself. To derive the relationship
  between the phase $\Phi(t)$ and the orbital phase (or {\it carrier phase}),
  we compute the derivative of $\mathbf{\hat{n}}(t)$ by means of
  Eqs.~(\ref{orbitalplanebasis}), (\ref{nandlambda}), obtaining
\begin{equation}
\label{dndt}
\frac{d \mathbf{\hat{n}}}{dt} = \left(\frac{d\Phi}{dt} + 
\cos \iota \,\frac{d\alpha}{dt}\right)\,\boldsymbol{\hat{\lambda}} 
+\left(\frac{d\iota}{dt}\,\sin \Phi - \sin \iota\,\cos \Phi\,\frac{d\alpha}{dt}
\right)\,\mathbf{\hat{L}}_{\rm N}\,.
\end{equation}
By imposing $\mathbf{\hat{L}}_{\rm N} = \mathbf{\hat{n}} \times
  \mathbf{v}/| \mathbf{\hat{n}} \times \mathbf{v}|= \mathbf{\hat{n}} \times
  d\mathbf{\hat{n}}/dt/| \mathbf{\hat{n}} \times d\mathbf{\hat{n}}/dt|$, and
  using Eq.~(\ref{nvec}) as well as Eq.~(\ref{Lvec}), we find that the term
  proportional to $\mathbf{\hat{L}}_{\rm N}$ in Eq.~(\ref{dndt}) must be zero.
  Thus, we have $d\mathbf{\hat{n}}/dt = \dot{r} \mathbf{r}/r^2 +
  \dot{\mathbf{r}}/r \equiv
  (\mathbf{v}\cdot\boldsymbol{\hat{\lambda}})\boldsymbol{\hat{\lambda}} $,
  where $\mathbf{v}\cdot\boldsymbol{\hat{\lambda}}$ is the orbital frequency
  $\omega_{\rm orb}$ defined before Eq.~\eqref{eq:kidder_basis}, which may be
  now interpreted as the angular velocity with which $\mathbf{\hat{n}}$
  rotates about $\mathbf{\hat{L}}_{\rm N}$. Identification with
  Eq.~\eqref{dndt} leads to the relation
\begin{equation}
\label{omegaorb}
\omega_{\rm{orb}} = \dot{\Phi} + \cos \iota\,\dot{\alpha}\,;
\end{equation}
the phase $\Phi(t)$ being simply the integral 
\begin{equation}
\label{phiorb}
\Phi(t) = \int^t_0 \left [\omega_{\rm orb}(t') -
\cos \iota(t') \,\dot{\alpha}(t')\right ]\,dt'\,.
\end{equation}
Due to the freedom in the choice of the time origin by the radiative observer,
$h_{ij}$ depends on an undetermined time scale or, equivalently, on an
arbitrary reference orbital frequency $\omega_0$. The constant $\omega_0$ is
actually associated to the presence of gravitational-wave tails and appears
solely in logarithms of the form $\ln (\omega_{\rm orb}/\omega_0)$. Such
contributions may be absorbed in the orbital phase by a redefinition of $\Phi$
into a shifted phase $\Psi$~\cite{BIWW96}. Through 1.5PN order in the
shift, we can pose
$\Psi = \Phi - 2v^3\,\ln (\omega_{\rm orb}/\omega_0)$.
By plugging Eqs.~(\ref{nvec})--(\ref{Lvec}) into Eq. (\ref{hij}), taking the 
combinations given in Eq. (\ref{poldefs}), and collecting terms by powers in
$v$, we obtain the waveform polarizations 
\begin{equation}
h_{+,\times}\ =\
\frac{2\,M\,\nu\,v^2}{D_L}\,\left[H_{+,\times}^{(0)}+ H_{+,\times}^{(1/2)} +
H_{+,\times}^{(1/2,{\rm SO})} +  H_{+,\times}^{(1)} + H_{+,\times}^{(1,{\rm SO})} +
H_{+,\times}^{(3/2)} + H_{+,\times}^{(3/2,{\rm SO})}  \right]\,.
\label{h}
\end{equation}
The Newtonian, 0.5PN and 1PN order terms were already computed explicitly
  in Refs.~\cite{K95,WWi96} [see in particular Eqs. (B2), (B3) of
  Ref.~\cite{K95}], but as a series expansion of ${M}/{r}$ rather than $v$. Let
  us list for the reader convenience a few typographical errors we found
  there. In Eq. (4.9d) of Ref.~\cite{K95}, the factor of $(1/6)(149 - 6\nu)$
has to be replaced with $(1/6)(149 - 36\nu)$; in Eq. (B2c) $Q_+$ must be
changed to $-Q_+$; in Eq. (B3c) the right parenthesis is missing in the
expression $(\cos^2 i\,\sin^2\alpha + \cos^2\alpha)$; at last, in Eq. (B3j)
$c\,d$ should be read as $-c\,d$. In Ref.~\cite{WWi96}, Eq. (F14b) must be
multiplied by $3 \nu$; in Eq. (F20) $-3 \nu$ has to be replaced with $+3 \nu$;
there should be an overall minus sign in front of Eq. (F25c). If we re-expand
Kidder's polarizations in $v$, and correct all the previous typos, we obtain
complete agreement with both results through 1PN order.\footnote{It is also
  worth noting that Ref.~\cite{K95} sets the origin of phase to be at a point
  referred to as the ascending node and defined to be the point where the
  orbital separation crosses the $x\mbox{--}y$ plane. This is in fact the same
  as our phase origin, $\mathbf{\hat{x}}_L = \mathbf{J}_0 \times
  \mathbf{\hat{L}}_{\rm N} / |\mathbf{J}_0 \times \mathbf{\hat{L}}_{\rm N}|$,
  but to reduce the possibility of confusion, we do not call this point the
  ascending node. We reserve this term to mean the point where the separation
  vector crosses \emph{the plane of the sky} from below.}

\begin{figure}
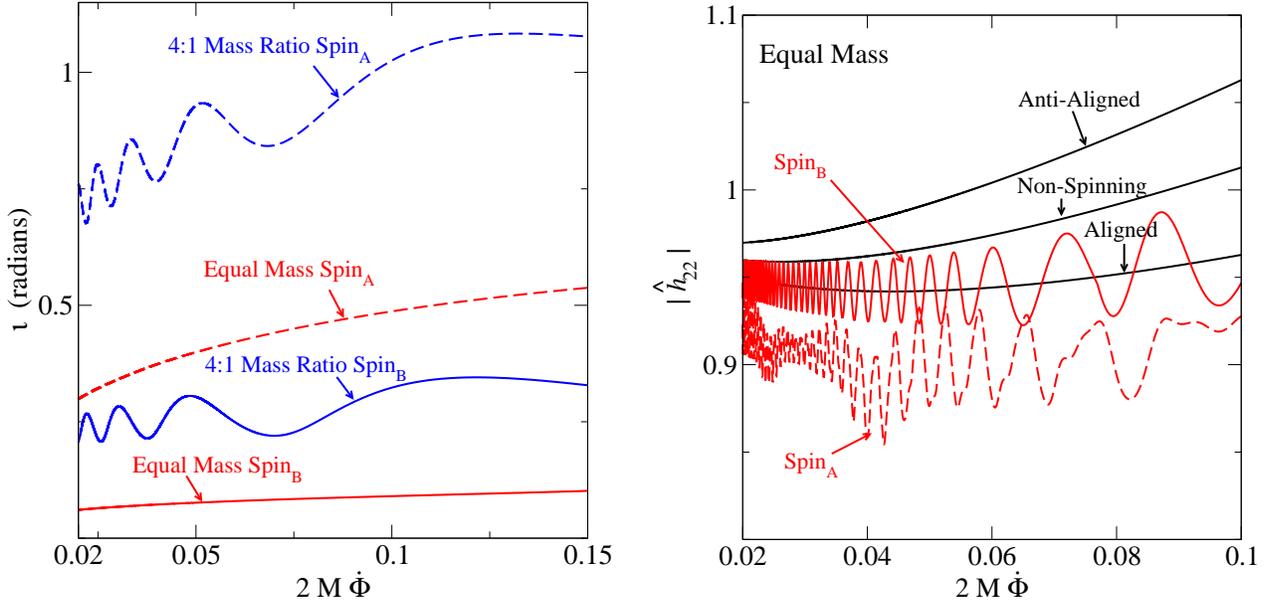

\includegraphics[width=0.45\linewidth]{IotaPlot.eps} \hspace{0.3cm}
\includegraphics[width=0.45\linewidth]{h22ComparisonPlot.eps}
\caption{The left panel shows the inclination angle of the orbital angular 
momentum relative to the total angular momentum, $\iota$, 
as a function of $2 M\dot{\Phi}$ for binaries with mass ratios 1:1 
and  4:1, having initial spin orientations relative to the 
orbital angular momentum: 
${\rm Spin_{\rm A}}=\{\theta_1=\pi/2,\phi_1=0,\theta_2=\pi/2,\phi_2=\pi/2\}$ and 
${\rm Spin_{\rm B}}= \{\theta_1=\pi/6,\phi_1=\pi/4,\theta_2=\pi/6,\phi_2=\pi\}$. 
The right panel compares the modulus of $h_{22}$ for the two 
precessing spin configurations with the non-spinning, aligned and 
anti-aligned cases for equal masses. The computations use waveforms accurate to 
1.5PN in amplitude and phase evolved with the precession equations at 1.5PN 
order [see Eqs.~(\ref{phiorb}), (\ref{preceq}), (\ref{angmomdot}), 
and (\ref{omegadot})]. Note that these plots (and those of Figs. 3 and 4) 
begin at $2\,M\, \dot{\Phi} = 0.02$ which is approximately where the 
dominant second harmonic from a binary of total mass $16\,M_\odot$ 
enters the LIGO band at 40 Hz and where the second harmonic from a binary of 
total mass $6.5 \times 10^6\,M_\odot$ enters the LISA band at $10^{-4}$ Hz.
\label{figure:h22}}
\end{figure}

The lengthy expression for the GW polarizations can be reduced to a much more
compact form by noticing (see also Sec. IVD in Ref.~\cite{K95}) 
that in the limit $S \ll L$ the angle $\iota$ can be considered a 0.5PN
order correction. This can be seen from 
\beq \label{eq:sin_iota}
\sin \iota = \frac{| \mathbf{J}_0 \times \mathbf{L}|}{ J_0\,L}\,,
\eeq
if we neglect radiation reaction effects, i.e., we assume $\mathbf{J}_0 =
\mathbf{J}$, and use $\mathbf{J} = \mathbf{L} + \mathbf{S}_1 + \mathbf{S}_2$
and $\mathbf{S}_n= {\cal O}(1/c)$. We may then replace $\sin \iota $ and $\cos
\iota$ in $h_{+,\times}$ with their Taylor series expansions in $\iota$,
\begin{eqnarray}
\label{iotaexpansion}
\sin \iota & = & \iota - \frac{\iota^3}{6} + {\cal O} (\iota^5)\,,\\
\cos \iota & = & 1 - \frac{\iota^2}{2} + {\cal O} (\iota^4)\,.
\end{eqnarray}
However, the assumption $S \ll L$ becomes less and less reliable for smaller
mass ratio binaries. In fact, as a first approximation, we have $S_n/L =
(m_n/M)^2\,\chi_n\,v/\nu$ with $v = (GM\omega_{\rm orb}/c^3)^{1/3}$. Thus,
even if $S_n \sim {\cal O}(1/c)$, $L$ can become comparable to $S_n$ when
$\nu$ is sufficiently small. Moreover, we have assumed $\mathbf{J}_0 =
\mathbf{J}$ in Eq.~\eqref{eq:sin_iota}, but the latter is not exact
when radiation reaction is included, and it can be strongly violated in
presence of transitional precession~\cite{ACST94}. For these reasons, though
we have decided to list in this section the GW polarizations expanded in
$\iota$, we display in Appendix~\ref{appB} the full expressions for generic
inclination angles. For the $\iota$-expanded polarizations, we find
\begin{subequations} 
\label{eq:Hplus_prec_iotaexpanded}
\begin{align}
  H_{+}^{(0)} & = - \bigg(c_{\theta }{}^2+1\bigg) \cos 2 (\alpha +\Psi )\,, \\
  H_{+}^{(1/2)} & = v \, \delta\, s_{\theta } \, \Bigg[\bigg(\frac{c_{\theta
  }{}^2}{8}+\frac{5}{8} \bigg) \cos (\alpha +\Psi ) -\frac{9}{8}
  \bigg(c_{\theta}{}^2 + 1\bigg) \cos 3(\alpha +\Psi) \Bigg]\,, \\ H_{+}^{(1)}
  & = v^2 \Bigg[ \bigg(- \frac{c_{\theta }{}^4}{3} + \frac{3
  c_{\theta}{}^2}{2} + \frac{19}{6} + \bigg(c_{\theta }{}^4 + \frac{11
  c_{\theta}{}^2}{6} - \frac{19}{6}\bigg) \nu \bigg) \cos 2 (\alpha +\Psi )
  \nonumber\\ & + \frac{4}{3}\bigg(1-c_{\theta }{}^4\bigg)
  \bigg(3\nu-1\bigg)\, \cos 4 (\alpha +\Psi )\Bigg]\,, \\ H_{+}^{(3/2)} & =
  v^3 \Bigg[\delta \, s_{\theta } \bigg(\frac{c_{\theta }{}^4}{192}-\frac{5
  c_{\theta }{}^2}{16}-\frac{19}{64}+\bigg(-\frac{c_{\theta
  }{}^4}{96}-\frac{c_{\theta }{}^2}{8}+\frac{49}{96}\bigg) \nu \bigg) \cos
  (\alpha +\Psi )-2 \pi ( c_{\theta }{}^2 + 1) \cos 2 (\alpha +\Psi )
  \nonumber \\ & +\delta \, s_{\theta}\, \bigg(\bigg(-\frac{81 c_{\theta
  }{}^4}{128}+\frac{45 c_{\theta }{}^2}{16}+\frac{657}{128}\bigg) +
  \bigg(\frac{81 c_{\theta }{}^4}{64}+\frac{9 c_{\theta
  }{}^2}{8}-\frac{225}{64}\bigg) \nu \bigg) \cos 3 (\alpha +\Psi ) \nonumber
  \\ & +\delta \, s_{\theta}\, \frac{625}{384}\bigg(1 - c_{\theta
  }{}^4\bigg)\bigg(2 \nu -1\bigg)\, \cos 5 (\alpha +\Psi )\Bigg]\,, \\
  H_{+}^{(1/2,{\rm SO})} & = - 2 \iota \, c_{\theta } s_{\theta } \cos (\alpha
  + 2 \Psi )\,, \\ H_{+}^{(1,{\rm SO})} & = v^2 \Bigg[\bigg(c_{\theta
  }\bigg(\chi _a^x + \delta \chi_s^x\bigg) - s_{\theta}\bigg(\chi _a^z +
  \delta \chi_s^z\bigg)\bigg) \cos (\alpha +\Psi ) - c_{\theta } \bigg(
  \chi_a^y + \delta \chi_s^y\bigg) \sin (\alpha + \Psi )\Bigg] \nonumber \\ &
  + v \, \iota \, \delta \, c_{\theta } \, \Bigg[ \frac{1}{4} s_{\theta}{}^2
  \cos \Psi - \bigg(\frac{c_{\theta }{}^2}{8}+\frac{5}{8}\bigg) \cos (2 \alpha
  + \Psi ) + \bigg(-\frac{9}{8} + \frac{27 c_{\theta }{}^2}{8} \bigg) \cos (2
  \alpha +3 \Psi ) \Bigg] \nonumber \\ & + \iota^2 \Bigg[-\frac{3}{2}s_{\theta
  }{}^2 \cos 2 \Psi + \frac{1}{2} (c_{\theta}{}^2 + 1) \cos 2 (\alpha
  +\Psi)\Bigg]\,, \\ H_{+}^{(3/2,{\rm SO})} & = v^3 \Bigg[
  s_{\theta}\,c_{\theta}\,\bigg(2 \delta\,\chi_a^x + (2 - \nu)\,\chi_s^x\bigg)
  + \bigg(\frac{4}{3}\,(1+c_{\theta}{}^2)\,\delta\,\chi_a^z +
  \frac{4}{3}\,\bigg( (1+c_{\theta}{}^2) + \nu\,(1-5 c_{\theta}{}^2)\bigg)
  \,\chi_s^z \nonumber\\ & - s_{\theta}\,c_{\theta}\,\bigg(2 \delta\,\chi_a^x
  + (2 + 7 \nu)\, \chi_s^x\bigg)\bigg)\,\cos 2(\alpha + \Psi) -
  s_{\theta}\,c_{\theta}\, \bigg(2 \delta\,\chi_a^y + (2 -
  \nu)\,\chi_s^y\bigg)\, \sin 2(\alpha + \Psi)\Bigg]\nonumber\\ & + v^2 \iota
  \, s_{\theta } \, \Bigg[c_{\theta } \, \bigg(-c_{\theta }{}^2 + 4 + \bigg(3
  c_{\theta }{}^2+\frac{2}{3}\bigg) \nu \bigg) \cos (\alpha +2 \Psi ) +
  c_{\theta } \, \bigg(-\frac{c_{\theta }{}^2}{3} - 1 + \bigg(c_{\theta }{}^2
  + 3 \bigg) \nu \bigg) \cos (3 \alpha +2 \Psi ) \nonumber \\ & + c_{\theta
  }^3 \bigg(\frac{16}{3}- 16 \nu \bigg) \cos (3 \alpha + 4 \Psi) - (\chi _a^y
  + \delta \chi _s^y) \sin \Psi \Bigg] + v\, \iota ^2 s_{\theta }\, \delta\,
  \Bigg[\bigg(-\frac{3 c_{\theta }{}^2}{16} + \frac{9}{16}\bigg) \cos (\alpha
  -\Psi ) \nonumber\\ & - \bigg(\frac{11 c_{\theta}{}^2}{32} +
  \frac{23}{32}\bigg) \cos (\alpha +\Psi) + \frac{27}{32} \bigg(c_{\theta
  }{}^2 + 1 \bigg) \cos 3 (\alpha +\Psi ) -\frac{1}{32} (c_{\theta}{}^2 + 1)
  \cos (3 \alpha + \Psi) \nonumber \\ & + \bigg(-\frac{45}{32} + \frac{135
  c_{\theta }{}^2}{32}\bigg) \cos (\alpha + 3 \Psi )\Bigg] + \iota ^3
  c_{\theta } s_{\theta } \, \Bigg[\frac{1}{2} \cos (\alpha -2 \Psi ) +
  \frac{5}{6} \cos (\alpha +2 \Psi )\Bigg]
\end{align}
\end{subequations}
\begin{subequations}
\label{eq:Hcross_prec_iotaexpanded}
\begin{align}
H_{\times}^{(0)} & = - 2 c_{\theta } \sin 2 (\alpha +\Psi ) \,, \\
H_{\times}^{(1/2)} & = v \,  \delta \, c_{\theta } s_{\theta } \,
\Bigg[- \frac{9}{4} \sin 3 (\alpha +\Psi)+\frac{3}{4} \sin (\alpha
+\Psi )\Bigg]\,, \\
H_{\times}^{(1)} & =  v^2 c_{\theta } \, \Bigg[\bigg(-\frac{4 c_{\theta
  }{}^2}{3}+\frac{17}{3}+\bigg(-\frac{13}{3}+4 c_{\theta }{}^2\bigg) \nu
\bigg) \sin 2 (\alpha +\Psi )+ s_{\theta }{}^2 \bigg(-\frac{8}{3} + 8 \nu
\bigg) \sin 4 (\alpha +\Psi )\Bigg]\,, \\ 
H_{\times}^{(3/2)} & =  v^3 c_{\theta} \, \Bigg[\delta \, s_{\theta } \,
\bigg(\bigg(-\frac{21}{32}+\frac{5 c_{\theta }{}^2}{96}\bigg)
+\bigg(-\frac{5 c_{\theta }{}^2}{48}+\frac{23}{48}\bigg) \nu
\bigg) \sin (\alpha +\Psi ) - 4 \pi \sin 2 (\alpha +\Psi 
) \nonumber \\ & + \delta  \, s_{\theta }  \, \bigg(\bigg(-\frac{135 c_{\theta
  }{}^2}{64}+\frac{603}{64}\bigg) +\bigg(-\frac{171}{32}+\frac{135 c_{\theta
  }{}^2}{32}\bigg) \nu  \bigg) \sin 3 (\alpha +\Psi 
) \nonumber\\ & +\delta  \, s_{\theta }  \,
\bigg(\frac{625}{192}\,(2 \nu -1)\, s_{\theta}{}^2\bigg) \sin 5
(\alpha +\Psi )\Bigg]\,, \\ 
H_{\times}^{(1/2,{\rm SO})} & =  - 2  \iota \, s_{\theta } \sin (\alpha +2 \Psi )\,,
\\
H_{\times}^{(1,{\rm SO})} & = v^2\,\Bigg[(\chi_a^y+\delta \chi_s^y) \cos
(\alpha +\Psi ) + c_{\theta } \bigg(c_{\theta } (\chi_a^x +\delta
\chi_s^x)-s_{\theta } \, (\chi_a^z + \delta \chi_s^z) \bigg) \sin (\alpha
+\Psi )\Bigg] \nonumber \\ & + \iota \,v\, \delta \, \Bigg[s_{\theta}{}^2 \sin
\Psi -\bigg(\frac{c_{\theta }{}^2}{2}+\frac{1}{4}\bigg) \sin (2 \alpha +\Psi
)+\bigg(-\frac{9}{4} + \frac{9 c_{\theta }{}^2}{2}\bigg) \sin (2 \alpha +3
\Psi )\Bigg] + \iota ^2 \, c_{\theta } \sin 2 (\alpha +\Psi )\,, \\
H_{\times}^{(3/2,{\rm SO})} & = v^3\, \Bigg[s_{\theta } \, \bigg(2 \delta
\chi_a^y + (2 - \nu)\,\chi_s^y\bigg) \, \bigg(1 + \cos 2 (\alpha +\Psi )\,\bigg)
\nonumber \\ & + \bigg(\frac{8}{3}\,c_{\theta}\, \delta\, \chi_a^z +
c_{\theta}\,\bigg(\frac{8}{3} - (\frac{4}{3} +
4c_{\theta}^{2})\nu\,\bigg)\,\chi_s^z - s_{\theta }\,\bigg(2\delta\,\chi_a^x +
\bigg(2 + (3 + 4c_{\theta}{}^2)\nu\,\bigg)\,\chi_s^x \bigg)\bigg) \sin 2
(\alpha +\Psi )\Bigg] \nonumber \\ & +\iota \, v^2\,s_{\theta }\,\Bigg[
\bigg(c_{\theta } \, (\chi_a^x + \delta \chi_s^x) - s_{\theta } (\chi_a^z+
\delta \chi_s^z)\bigg) \sin \Psi + \bigg(\bigg(-3 c_{\theta
}{}^2 + 6\bigg) +\bigg(-\frac{16}{3} + 9 c_{\theta }{}^2\bigg) \nu \bigg) \sin
(\alpha +2 \Psi ) \nonumber \\ & + \bigg(-\bigg(c_{\theta
}{}^2+\frac{1}{3}\bigg) + \bigg(3 c_{\theta }{}^2+1\bigg) \nu \bigg) \sin (3
\alpha +2 \Psi )+ \bigg(\bigg(-\frac{8}{3} + 8 c_{\theta }{}^2\bigg) +\bigg(-
24 c_{\theta }{}^2 + 8\bigg) \nu \bigg) \sin (3 \alpha +4 \Psi
)\Bigg] \nonumber \\ & +\iota ^2 v \, \delta \, c_{\theta } s_{\theta }\,
\Bigg[\frac{3}{8} \sin (\alpha -\Psi ) - \frac{17}{16} \sin (\alpha +\Psi ) +
\frac{27}{16} \sin 3 (\alpha +\Psi ) - \frac{1}{16} \sin (3 \alpha +\Psi ) +
\frac{45}{16} \sin (\alpha +3 \Psi)\Bigg] \nonumber \\ & +\iota ^3
\Bigg[\frac{1}{2} s_{\theta } \sin (\alpha -2 \Psi ) + \frac{5}{6} s_{\theta }
\sin (\alpha +2 \Psi )\Bigg] \,,
\end{align}
\end{subequations}
where $s_\theta$ and $c_\theta$ are shorthand notations for $\sin \theta$ 
and $\cos \theta$ respectively. In Sec.~\ref{secIII} (see
Fig.~\ref{figure:h22}), we shall discuss typical variations of the inclination
angle $\iota$ depending on spin orientations and binary mass ratios. Note that
whereas the terms of $H_{+,\times}^{(1,{\rm SO})}$ linear in
$\boldsymbol{\chi}_n$ depend on the first harmonic of the orbital frequency,
those of $H_{+,\times}^{(3/2,{\rm SO})}$ depend on its zeroth and second
harmonic, and so do the terms of $H_{+,\times}^{(2,{\rm SS})}$ quadratic in
the spin components, although we do not use them here. We include these 2PN SS
polarization corrections when constructing frequency-domain waveforms for
binaries having their spins aligned or anti-aligned with the orbital angular
momentum in Sec.~\ref{secIV}. The harmonic dependence of the polarization
corrections produced by the spins can be understood from the explicit
expression for $h_{ij}$ shown in Eqs.~(4.9c) and (4.9d) of Ref.~\cite{K95} or
Eqs. (F15a)-(F15c) of Ref.~\cite{WWi96}. The 1PN SO contributions are
proportional to the components of the orbital separation vector,
$\mathbf{\hat{n}}$, which are themselves proportional to $\sin \Phi$ and $\cos
\Phi$, so that $H_{+,\times}^{(1,{\rm SO})}$ depend on the first harmonic of
the orbital phase. Next, the 1.5PN SO and 2PN SS contributions are
proportional to products of the orbital separation or instantaneous velocity
unit vectors, $\mathbf{\hat{n}}$ or $\boldsymbol{\hat{\lambda}}$, and to
products of $\sin \Phi$ or $\cos \Phi$. These can be re-expressed in terms of
$\sin 2\Phi$, $\cos 2\Phi$ or constant quantities independent of $\Phi$, so
that $H_{+,\times}^{(3/2,{\rm SO})}$ and $H_{+,\times}^{(2,{\rm SS})}$ depend on the 
zeroth and second harmonics of the orbital phase. Because the expressions for
$h_{ij}$ in Refs.~\cite{K95,WWi96} are expanded in $(M/r)$, while we use an
expansion in $v=(M\omega_{\rm orb})^{1/3}$, one has to convert from one
expansion to the other by using Eqs. (7.1) and (F20) of Ref.~\cite{WWi96}. In
doing so, the $v$-expansion gains additional 1.5PN SO and 2PN SS corrections
proportional to the Newtonian order term depending on the second harmonic of
the orbital phase. The 1PN SO term is left unchanged.

Ready-to-use time-domain GW polarizations for spinning, precessing binaries
through 1.5PN order in phase and amplitude can be obtained by solving
numerically the following equations: (i) The spin precession
equations~\cite{FBBu06,BBuF06}
\begin{subequations}
\label{preceq}
\bea
\frac{d \mathbf{S}_1}{d t} &=& \mathbf{\Omega}_1\times \mathbf{S}_1\,,\\ 
\frac{d \mathbf{S}_2}{d t} &=& \mathbf{\Omega}_2\times \mathbf{S}_2\,,
\eea
\end{subequations}
where at 1.5PN order 
\beq
\mathbf{\Omega}_{1,2} = \omega_{\rm orb}^{5/3}\,\left (\frac{3}{4} +
  \frac{\nu}{2} \mp \frac{3}{4} \delta \right )\,\mathbf{\hat{L}}_{\rm N}\,; 
\eeq
(ii) The evolution equation for the Newtonian angular momentum
\beq
\dot{\mathbf{\hat{L}}}_{\rm N} = -\frac{v}{\nu}\,(\dot{\mathbf{S}}_1 +
\dot{\mathbf{S}}_2)\,;
\label{angmomdot} 
\eeq
(iii) The equation for the orbital frequency 
\begin{equation} 
\frac{\dot{\omega}_{\rm orb}}{\omega_{\rm orb}^2}=\frac{96}{5}\,\nu\,v^{5}\Bigg
\{1-\left (\frac{743}{336}+ \frac{11}{4} \nu \right ) \,v^{2} + 
\left[\left(\frac{19}{3}\,\nu - \frac{113}{12}\right)
\boldsymbol{\chi}_s\cdot\mathbf{\hat{L}}_{\rm N} - \frac{113}{12}\,\delta\,
\boldsymbol{\chi}_a\cdot\mathbf{\hat{L}}_{\rm N}\right]\,v^3 
+ 4\pi v^3 \Bigg \}\, .
\label{omegadot}
\end{equation}
Integrating $\omega_{\rm orb}$ yields the orbital phase $\Phi_{\rm
  orb}(t)\equiv \int^t_0 \omega_{\rm orb}(t') dt'$. The GW polarizations~(\ref{h})
through 1.5PN order in phase and amplitude are computed by solving numerically
Eqs.~(\ref{phiorb}), (\ref{preceq}), (\ref{angmomdot}), and (\ref{omegadot}).
In order to compute the GW polarizations~(\ref{h}) through 1.5PN order in
amplitude, but at the highest available PN order in phase, one should replace
Eqs.~(\ref{preceq}), (\ref{angmomdot}) with Eq.~(7.5) in Ref.~\cite{BBuF06}
and Eq.~(32) in Ref.~\cite{BCP07NR}, respectively.

\section{Gravitational-wave modes for precessing binaries on nearly circular 
orbits through 1.5PN order: small inclination angles}
\label{secIII}

Due to the spin-2 nature of GWs, it is convenient to decompose the waveform
components in the dyad $\{(\mathbf{P}+ i \mathbf{Q})/\sqrt{2},(\mathbf{P}-i
\mathbf{Q})/\sqrt{2} \}$ with respect to an orthonormal basis of spin $ \pm 2
$ functions that are defined on the 2-sphere and belong to an irreducible
representation of SO(3). Most commonly, the complex polarization 
\beq 
h = h_{jk} \frac{P^j- i Q^j}{\sqrt{2}} \frac{P^k-i Q^k}{\sqrt{2}} = h_+
- i h_\times\,,
\label{hc}
\eeq 
is expanded into the set of spin-weighted $-2$ spherical harmonics. Like
the standard spherical harmonics, these functions of the two angles of
spherical coordinates are labeled by a pair of integers, say $ (\ell, m) $,
with $ \ell \ge 2$ and $ m \le |\ell| $. The spin-weighted $-s$ spherical
harmonics associated to any such pair are given by \footnote{Our definition of
  ${}_{-s} Y^{\ell m}(\theta, \phi) $ differs from that of Ref.~\cite{GMNRS67}
  by a factor 
  $(-1)^m$ so that ${}_{0} Y^{\ell m}(\theta,\phi) $ coincides with the most
  broadly used definition of $ Y^{\ell m}(\theta,\phi)$;
  for the d-matrix $d^\ell_{m'm}(\theta)$, we adopt the same convention as
  Landau-Lifchitz~\cite{LLvol3}.} \cite{GMNRS67}
\beq 
{}_{-s} Y^{\ell m}(\theta,\phi) = (-1)^s \sqrt{\frac{2\ell + 1}{4\pi}}\,
d_{sm}^\ell(\theta)\, e^{i m \phi}\,, 
\eeq
with
\beq
d_{sm}^\ell(\theta) = \sum_{k=\max(0,m-s)}^{\min(\ell+m,\ell-s)}
\frac{(-1)^k}{k!} 
  \frac{\sqrt{(\ell + m)! (\ell - m)! (\ell + s)! (\ell - s)!}}{(k - m + s)!
    (\ell + m - k)!  (\ell - k - s)!} \left(\cos \frac{\theta}{2}
  \right)^{2\ell+m-2k-s} \left( \sin \frac{\theta}{2} \right)^{2k-m+s} \,,
\eeq
and the orthogonality relation holds
\begin{equation} 
\label{eq:surface_integral}
\int d\Omega  \, {}_{-s}Y^{\ell m}(\theta,\phi) \, {}_{-s}Y^{\ell' m' *}
(\theta,\phi) = \delta^{\ell \ell'} \delta^{m m'}\,,
\end{equation}
where $d\Omega = \sin \theta\, d\theta\,d\phi $ denotes the element of solid
angle and $ \delta^{\ell \ell'}$ is the Kronecker symbol. The integration is
performed over the unit sphere, so that $ 0 \le \theta \le \pi $ and $ 0 \le
\phi \le 2\pi$. The mode expansion of the complex polarization (\ref{hc}) is
then of the form
\begin{equation}\label{eq:mode_definition}
h(\theta,\phi) = \sum_{\ell = 2}^{+\infty} \sum_{m=-\ell}^{\ell} h_{\ell m}\, 
{}_{-2}Y^{\ell m}(\theta,\phi) \,.
\end{equation}
The GW modes $h_{\ell m}$ are extracted using the orthogonality property
(\ref{eq:surface_integral}) by means of the surface integral
\begin{equation}
\label{eq:hlm}
h_{\ell m} = \int d \Omega \, h(\theta, \phi) \, {}_{-2}{Y}^{\ell
  m*}(\theta,\phi) \,, 
\end{equation}
where the star on the spin-weighted $-2$ harmonic indicates the complex
conjugation. Therefore, the calculation of $ h_{\ell m} $ requires the
knowledge of the polarizations $ h_+ $ and $ h_\times $ for an arbitrary value
of the azimuthal angle $ \phi $ of the direction $ \mathbf{\hat{N}} $. In
Sec.~\ref{secII} we have computed $ h_+ $ and $ h_\times $ only for $ \phi = 0
$; however, a specific choice of the $x$-axis orientation \emph{cannot} be
responsible for any information loss. Thus, we must be able to recover $
h(\theta,\phi) $ from the expression of $ h(\theta,0) $ alone.
 
The quantity $h$ at a given point depends on a number of parameters, such as $
\iota $ or the spin variables, and can actually be regarded as a function of
$\theta $, $ \phi $ as well as a function of the whole set of parameters that
possess a geometrical character. More precisely, we may write $h(\theta,\phi)
= \overline{h}(\theta,\phi,\iota, \alpha, \Phi, \chi_n^x, \chi_n^y,
\chi_n^z)$. Let us now introduce the projection basis $ (\mathbf{\hat{x}}',
\mathbf{\hat{y}}', \mathbf{\hat{z}}'=\mathbf{\hat{z}}) $ obtained by applying
a rotation of angle $ \phi $ about the $z$-axis on the vectors of the original
basis $ (\mathbf{\hat{x}}, \mathbf{\hat{y}}, \mathbf{\hat{z}})$. Let us also
associate to each variable of $ \overline{h} $ a primed counterpart, which is
defined in the same way as the unprimed variable but refers to the new basis
rather than the original one. For instance, $ \alpha' $ denotes the azimuthal
angle of the orbital angular momentum measured from the fixed vector $
\mathbf{\hat{x}}' $ instead of $ \mathbf{\hat{x}} $. In particular, we have
$\theta' = \theta $, $ \phi' = 0 $, $ \iota' = \iota $, $ \alpha'=\alpha -
\phi $. The phase $\Phi$, defined as the angle $ (\mathbf{\hat{L}}_{\rm N}
\times \mathbf{\hat{z}}, \mathbf{\hat{n}})= ( \mathbf{\hat{L}}_{\rm N} \times
\mathbf{\hat{z}}', \mathbf{\hat{n}})$, is not affected by the transformation:
$\Phi' = \Phi$. The $x'$ and $y'$ spin components can be obtained from
the 2-dimensional formula for a passive rotation of angle $ \phi$, that is
\bea
{\chi'}_n^x &=& \chi_n^x\,\cos \phi + \chi_n^y\,\sin \phi \, , \\
{\chi'}_n^y &=& - \chi_n^x\, \sin \phi + \chi_n^y\,\cos \phi \, ,
\eea
while the third component is left unchanged. With our conventions, the
  polarization vectors in the new basis remain equal to $\mathbf{P}$ and
  $\mathbf{Q}$ respectively. Therefore, the complex polarization is identical
  to that of the old frame. Moreover, by construction of the primed variables,
  the functional dependence of $h$ is the same as before, meaning
  that $ h = \overline{h}(\theta, \phi, \iota, \alpha,\Phi, \chi_n^x,
  \chi_n^y, \chi_n^z) = \overline{h}(\theta', \phi', \iota', \alpha', \Phi',
  {\chi'}_n^x, {\chi'}_n^y, {\chi'}_n^z) $. Hence the important relation
\begin{equation} \label{eq:h_phi}
   h \equiv \overline{h}(\theta, \phi, \iota, \alpha,\Phi, \chi_n^x, \chi_n^y,
   \chi_n^z) = \overline{h}(\theta, 0, \iota, \alpha-\phi, \Phi, \cos \phi
   \chi_n^x + \sin
   \phi \chi_n^y, - \sin \phi \chi_n^x + \cos \phi \chi_n^y, \chi_n^z)\,, 
\end{equation}
where the function $ \overline{h}(\theta, 0, \iota, \alpha, \chi_n^x,
\chi_n^y, \chi_n^z) $ is given by Eqs.~\eqref{eq:Hplus_prec_iotaexpanded},
\eqref{eq:Hcross_prec_iotaexpanded} for the $\iota$-expanded expressions or by
Eqs.~\eqref{eq:Hplus_prec}, \eqref{eq:Hcross_prec} for the full ones.

  At 1.5PN order, the GW polarizations decompose into a sum of 3 terms, $
  h_0(\theta, \iota,\alpha, \Phi) + \sum_{n=1,2}
  \boldsymbol{\chi}_n.\mathbf{h}_n(\theta, \iota, \alpha, \Phi) $, which shows
  that $ h$ may be written as 
  \beq 
  h_0(\theta, \iota, \alpha-\phi, \Phi) + \sum_{n=1,2} \left[e^{i \phi} 
  {\zeta}^*_n k_n(\theta, \iota, \alpha-\phi, \Phi) + e^{-i \phi}
  \zeta_n {k}^*_n(\theta, \iota, \alpha-\phi, \Phi) + \chi^z_n h^z_n(\theta,
  \iota, \alpha-\phi, \Phi)\right]\,, 
  \eeq with $ \zeta_n = (\chi^x_n + i \chi^y_n)/\sqrt{2} $
  and $ k_n = (h^x_n + i h^y_n)/\sqrt{2} $. Each mode $ h_{\ell m} $ splits
  accordingly into 7 contributions: the spin-free term and 6 terms
  proportional to each of the spin variable components. These contributions are
  parametrized by a vector weight $ m'=-1,0,1 $, as well as the body label $
  n=1,2 $ of the spins; $ n = 0$ refers to quantities entering the spin-free
  part of $ h $ for which we also set $ m'= 0 $. As a result, for precessing
  binaries, the integral to compute $h_{\ell m}$ takes the form:
\begin{equation}
h_{\ell m} = \sum_{m'=-1}^{1} \sum_{n=1}^{2} X_{m',n} \int d\Omega  \,
K_{m',n}(\theta, \iota, \alpha - \phi, \Phi) e^{i (-m'-m)\phi} \,
{}_{-2}{Y}^{\ell m*}(\theta,0) \, , 
\end{equation}
where $ X_{0,0} = 1 $, $ X_{0,{n'}} = \chi^z_{n'}$ (for $ n'=1,2 $),
$X_{-1,n'} = \zeta^*_{n'} $, $ X_{1,n'} = \zeta_{n'} $, $ K_{0,0} = h_0$, $
K_{0,n'} = h^z_{n'} $, $ K_{-1,n'} = k_{n'}$ and $ K_{1,n'} ={k}^*_{n'} $. By
means of the change of variable $ \phi \rightarrow \phi+\alpha $, we are able
to factor out a complex exponential $ e^{-i (m+m')\alpha} $ which contain all
the dependence in $ \alpha $. Let us now focus henceforth on the case where
the waveform has been expanded in powers of $\iota$. As we shall explicitly
see below, [see Eq.~(\ref{eq:modes})], the $ h_{\ell m} $'s are then made of:
(i) a spin-free piece proportional to $ e^{-i m \alpha} $, (ii) two spin
pieces proportional to $ e^{-i (m-1) \alpha} $ and to $ \zeta_1^* $ or $
\zeta_2^*$ respectively, (iii) two spin pieces proportional to $ e^{-i (m+1)
  \alpha} $ and to $\zeta_1 $ or $\zeta_2 $ respectively, (iv) two spin pieces
proportional to $ e^{-i m \alpha} $ and to $ \chi^z_1$ or $ \chi^z_2$
respectively. In contrast to what happens in the non-spinning case, $ h_{\ell
  m} $ is not in general proportional to $ e^{-i m \Phi} $ except for the
terms that are free of $ \iota $, since both $ \hat{\mathbf{n}} $ and $
\hat{\boldsymbol{\lambda}} $ reduce to trigonometric functions of $ \Phi +
\alpha $ as $ \iota \to 0 $. The contributions to the polarization modes that
are linear in $ \iota $ involve couplings of the type $e^{-i (m+m') (\Phi +
  \alpha)} \cos \Phi $ or $e^{-i (m+m') (\Phi + \alpha)} \sin \Phi $ because
the terms of first order in $ \iota $ entering $\hat{\mathbf{n}}$ and
$\hat{\boldsymbol{\lambda}}$ can only be linear combinations of $ \iota \cos
\Phi $ or $ \iota \sin \Phi $ (or equivalently $\iota e^{\pm i \Phi}$).
Couplings like $e^{-i (m+m') (\Phi + \alpha)} \cos^a \Phi \sin^b \Phi $, with
$a,b\in \mathbb{N} $, arise at higher orders making the dependence in $\Phi$
more complicated. A close inspection to the results below [see
  Eqs.~(\ref{h22})--(\ref{h50}) with $\Psi \rightarrow \Phi$] confirm these
  expectations. Beware that our mode normalization is tuned to factor out the
  exponential factors $e^{-i m \alpha}$.

The structure of the modes is much more complicated for precessing binaries than for non-precessing binaries. When the orbital angular momentum is aligned with the total angular momentum $(\iota=0,\alpha=\pi)$, note that a rotation by $\phi$ about the z-axis produces an offset in the orbital phase angle, so that

\begin{equation} \label{eq:h_phi_ns}
   \overline{h}(\theta, \phi, \Phi) = \overline{h}(\theta, 0,\Phi - \phi)\, . 
\end{equation}
This ensures that only terms proportional to $e^{-i m \Phi}$ contribute to the integral over $\phi$ to compute $h_{\ell m}$. In the precessing case, a rotation by $\phi$ produces an offset in the $\alpha$ angle, and so terms with different powers of $e^{-i \Phi}$ can contribute to the same $h_{\ell m}$ mode. As we will see below, these terms with different powers of $e^{-i \Phi}$ interfere to produce rather complicated modulations to the modes on the \emph{orbital} time scale. Since the precessional motion is typically much slower than the orbital motion (several orbital cycles are completed in any precessional cycle for the systems we consider), it may be surprising that the relatively slow precessional motion can produce such rapid oscillations in the modes. This is simply a breakdown of the nice structure (i.e. that $h_{\ell m} \propto e^{-i m \Phi}$) of the $h_{\ell m}$ modes in the precessing case. Note however, that what is actually observed are the gravitational wave \emph{polarizations}. In the polarizations, precessional effects are indeed on a slower time scale than the orbital motion. They modulate the ``envelope'' of the waveform, rather than create orbital timescale interference.

A useful property of $ \overline{h} $ coming from the arbitrariness of the
body labeling is that it must be invariant in the exchange of particles 1 and
2: $m_1\leftrightarrow m_2 $, $ \chi_1 \leftrightarrow \chi_2 $, $\mathbf{n}
\rightarrow -\mathbf{n} $, $\mathbf{v} \rightarrow -\mathbf{v}$. Under this
transformation, the direction of the angular momentum $\mathbf{\hat{L}} $
remains invariant hence $ \boldsymbol{\lambda} \rightarrow -
\boldsymbol{\lambda}$. The orbital frequency $\omega_{\rm orb} = (\mathbf{v}
\cdot \boldsymbol{\lambda}) $ is unchanged as well as the direction of the
total angular momentum, due to its structure and parity. Therefore,
the phase $\Phi$ becomes $ (\hat{\mathbf{x}}_L, -\hat{\mathbf{n}}) = \Phi +
\pi$ whereas the angles $ \alpha $ and $ \iota $ are unaffected. This yields
the relation
\beq
  \overline{h}(\theta, \phi, \iota,
\alpha,\Phi, \chi_1^{x}, \chi_2^{x}, \chi_1^{y}, \chi_2^{y}, \chi_1^{z},
\chi_2^{z}) = [\overline{h}(\theta, \phi, \iota, \alpha, \Phi + \pi, \chi_2^{x},
\chi_1^{x}, \chi_2^{y}, \chi_1^{y}, \chi_2^{z}, \chi_1^{z})]_{\delta
   \rightarrow -\delta} \, .
\eeq
The previous identity may be supplemented by another one which originates from
the classical parity invariance of physics: for any given time instant $ t $,
the waveform resulting from the stress-energy tensor parametrized by the
world-lines $ \mathbf{x}_n $ and the spins $ \boldsymbol{\chi}_n $ must have
the same value at point $ \mathbf{x} $ as the waveform resulting from $
-\mathbf{x}_n $ and $ + \boldsymbol{\chi}_n $ at point $ -\mathbf{x} $. Taking
into account the transformation of the polarization vectors under parity, this
means for the function $ \overline{h} $:
\beq
\label{eq:parity}
  \overline{h}(\theta, \phi, \iota, \alpha,\Phi, \chi_n^x, \chi_n^y, \chi_n^z)
  \\= {\overline{h}^*}(\pi-\theta, \phi+\pi, \iota, \alpha,\Phi+\pi, \chi_n^x,
  \chi_n^y, \chi_n^z) \, .
\eeq
The above formula allows us to express the modes $ h_{\ell m} $ in terms of
the modes $ h_{\ell \, -m} $ by performing the change of variable $ \theta
\rightarrow \pi - \theta$ and $ \phi \rightarrow \phi + \pi $ in
Eq.~\eqref{eq:hlm}. The first factor of the integrand can be then rewritten as
$ {\overline{h}^*}(\theta, \phi, \iota, \alpha,\Phi+\pi, \chi_n^x, \chi_n^y,
\chi_n^z) $ making use of Eq.~\eqref{eq:parity}. The second factor
${}_{-2}Y^{\ell m*}(\pi-\theta, \phi+\pi) $ may be transformed by means of two
important symmetry properties of the spin-weighted spherical harmonics: $
{}_{-2}Y^{\ell m}(\pi - \theta, \phi+\pi)= (-1)^\ell {}_{+2}Y^{\ell m}(\theta,
\phi) $ and $ {}_{+2} Y^{\ell m}(\theta, \phi) = (-1)^m {}_{-2}{Y}^{\ell\, - m
  *}(\theta, \phi) $, which leads to the new expression $ (-1)^{m+\ell}
{}_{-2}Y^{\ell\, - m}(\theta, \phi)$ for this factor. As a consequence, the
link between $h_{\ell m}(\Phi) \equiv \int d \Omega \, \overline{h} (\theta,
\phi,\iota, \alpha, \Phi, \chi_n^x, \chi_n^y, \chi_n^z) \, {}_{-2}{Y}^{\ell\,
  m *}(\theta,\phi)$ and $h_{\ell \, -m}(\Phi)$ is given by
\begin{equation} \label{eq:hl-m}
  h_{\ell m}(\Phi) = (-1)^{m+\ell} \int d \Omega \,
  {\overline{h}^*}(\theta, \phi,\iota, \alpha, \Phi+\pi, \chi_n^x, \chi_n^y,
   \chi_n^z) \, {}_{-2}{Y}^{\ell\, - m}(\theta,\phi)
   =(-1)^{\ell + m} {h}^*_{\ell\, -m}(\Phi+\pi) \, 
\end{equation}
The explicit expressions for the modes $ h_{\ell m}$ are obtained by inserting
Eq.~\eqref{eq:h_phi} into the surface integral \eqref{eq:surface_integral}. We
normalize them in such a way that the leading order mode starts with
coefficient 1. Posing
\beq \label{eq:hlmhat}
h_{\ell m} = -\frac{(2 M \nu
  v^2)}{D_L}\,\sqrt{\frac{16\pi}{5}}\,e^{-im(\Psi+\alpha)}\,\hat{h}_{\ell
  m}\,,
\eeq 
and expanding in the inclination angle $\iota$, we arrive at\footnote{In the
  case of spins 
  aligned or anti-aligned with the Newtonian angular momentum, the modes
  $(2,2)$, $(2,1)$ and $(3,2)$ were also computed in Ref.~\cite{BCGSB07}. We
  fully agree with their results.}
\begin{subequations} \label{eq:modes}
\begin{align}
\label{h22}
    \hat{h}_{22} &= 1+\frac{1}{3} e^{i \Psi } \delta v \iota + v^2
    \bigg\{\frac{1}{42}(-107+55 \nu) -\frac{1}{2} e^{i (\alpha +\Psi )}
      \Big[\chi _a^x- i \chi _a^y +\delta (\chi _s^x - i
      \chi_s^y)\Big] \bigg\} 
    - \frac{\iota^2}{2} + v^3 \bigg[2 \pi -\frac{4\delta \chi _a^z}{3}
      \nonumber \\ & + \frac{4}{3} (-1+\nu ) \chi _s^z\bigg] + {\cal O}\left
    (\frac{1}{c^4} \right )\, ,\\ 
    \label{h21}
\hat{h}_{21} &= -\frac{\delta v }{3}+ e^{-i \Psi } \iota +\frac{v^2}{2}
      (\chi_a^z+\delta \chi _s^z ) + \delta\, \iota^2\,v\,
      \bigg(\frac{5}{12}- \frac{1}{4} e^{2 i \Psi } \bigg) 
      + v^3 \bigg\{-e^{i (\alpha +\Psi
      )} \bigg[\delta\,(\chi_a^x - i \chi_a^y) 
      + \bigg(1-\frac{\nu}{2}\bigg)
      (\chi_s^x - i \chi_s^y)\bigg] \nonumber \\ & +e^{-i (\alpha +\Psi )}
      \bigg[\delta (\chi_a^x + i \chi_a^y) + \bigg(1+\frac{5}{6} \nu \bigg)
      (\chi _s^x + i \chi_s^y)\bigg] + \frac{\delta}{84} (17-20 \nu ) \bigg\}
      + v^2 \iota \bigg\{\frac{e^{-i \Psi }}{42} (-107+55 \nu ) \nonumber \\ &
      + \frac{1}{4} \Big(\chi_a^x- i \chi_a^y +\delta (\chi _s^x - i
      \chi_{s}^y)\Big) e^{i\alpha } \Big(-1 + e^{2 i \Psi }\Big) \bigg\} +
      \frac{\iota ^3}{4} \bigg(-\frac{5}{3} e^{-i \Psi} - e^{3 i \Psi }\bigg)
      + {\cal O}\left (\frac{1}{c^4} \right )\, ,\\ \label{h20} \hat{h}_{20}
      &= \frac{1}{2} \sqrt{\frac{3}{2}} \bigg\{\frac{v^2}{3} \Big[- e^{i
      (\alpha +\Psi )} \Big(\chi_a^x - i \chi_a^y + \delta (\chi_s^x - i
      \chi_{s}^y)\Big) + e^{-i (\alpha +\Psi )} \Big(\chi _a^x + i
      \chi_a^y 
      +\delta (\chi_s^x + i \chi_{s}^y) \Big)\Big] \nonumber \\ & +
      \frac{4 
      i}{3} v\, \iota\, \delta \sin \Psi + 2 \iota^2 \cos 2\Psi -
    \frac{4 i}{3} v^2 
      \iota \sin \Psi (\chi _a^z + \delta \chi _s^z) \bigg\} + {\cal O}\left
      (\frac{1}{c^4} \right )\, , \\ \label{h33} \hat{h}_{33} &= -\frac{3}{4}
      \sqrt{\frac{15}{14}} \bigg\{ \delta v +v^3 \bigg[2 \delta (-2+\nu )
      +\frac{16}{9} e^{i (\alpha +\Psi)} \nu (\chi_s^x - i \chi_s^y)\bigg] -
      \frac{4}{9} e^{i \Psi } \iota v^2 (-1+3 \nu ) \nonumber \\ & +
      \frac{\delta v \iota^2}{4} \bigg(-3 +\frac{e^{2 i \Psi }}{9} \bigg)
      \bigg\} + {\cal O}\left (\frac{1}{c^4} \right ) \, ,\\
      \label{h32}\hat{h}_{32} &= -\frac{9}{8} \sqrt{\frac{5}{7}}
      \bigg[\frac{8}{27} v^2 (-1+3 \nu )+\delta v \iota \bigg(e^{-i \Psi }
      -\frac{ e^{i \Psi }}{27} \bigg) -\frac{32}{27} v^3 \nu \chi _s^z\bigg] +
      {\cal O}\left (\frac{1}{c^4} \right )\, ,\\\label{h31} \hat{h}_{31} &=
      -\frac{1}{12 \sqrt{14}} \bigg\{\delta v + v^3 \bigg[-\frac{2}{3} \delta
      (4 + \nu) -16 e^{-i (\alpha +\Psi)} \nu (\chi_s^x + i \chi_s^y)\bigg] +
      20 v^2 \iota (-1+3\nu ) e^{-i \Psi } \nonumber \\ & + \frac{\delta v
      \iota^2}{2} \bigg(-\frac{11}{2} + \frac{135}{2} e^{-2 i \Psi } - 3 e^{2
      i \Psi } \bigg) \bigg\} + {\cal O}\left (\frac{1}{c^4} \right )\,
      ,\\\label{h30} \hat{h}_{30} &= -\frac{1}{2 \sqrt{42}} \delta v \iota
      \cos \Psi + {\cal O}\left (\frac{1}{c^4} \right )\, ,\\\label{h44}
      \hat{h}_{44} &= \frac{8}{9} \sqrt{\frac{5}{7}} v^2 (1-3 \nu ) + {\cal
      O}\left (\frac{1}{c^4} \right )\, ,\\\label{h43} \hat{h}_{43} &=
      \frac{8}{9} \sqrt{\frac{10}{7}} \bigg[\frac{81}{320} v^3 \delta (-1+2
      \nu ) +v^2 \iota (1-3 \nu ) \left(e^{-i \Psi}-\frac{e^{i \Psi }}{16}
      \right) \bigg] + {\cal O}\left (\frac{1}{c^4} \right )\, ,\\\label{h42}
      \hat{h}_{42} &= \frac{\sqrt{5}}{63} v^2 (1-3 \nu ) + {\cal O}\left
      (\frac{1}{c^4} \right )\, ,\\\label{h41} \hat{h}_{41} &= \frac{1}{21}
      \sqrt{\frac{5}{2}} \bigg[\frac{\delta v^3}{20} (-1+2 \nu ) + v^2 \iota
      (1-3 \nu ) e^{-i \Psi } \bigg] + {\cal O}\left (\frac{1}{c^4} \right )\,
      ,\\\label{h40} \hat{h}_{40} &= {\cal O}\left (\frac{1}{c^4} \right )\,
      ,\\ \hat{h}_{55} &= -\frac{625}{96 \sqrt{66}} \delta v^3 (1-2 \nu ) +
      {\cal O}\left (\frac{1}{c^4} \right )\, ,\\ \hat{h}_{54} &= {\cal
      O}\left (\frac{1}{c^4} \right )\, ,\\ \hat{h}_{53} &= -\frac{9}{32}
      \sqrt{\frac{3}{110}} v^3 \delta (1-2 \nu ) + {\cal O}\left
      (\frac{1}{c^4} \right )\, ,\\ \hat{h}_{52} &= {\cal O}\left
      (\frac{1}{c^4} \right ) \, ,\\ \hat{h}_{51} &= -\frac{1}{288 \sqrt{385}}
      \delta v^3 (1-2 \nu ) + {\cal O}\left (\frac{1}{c^4} \right )\, ,\\
      \hat{h}_{50} &= {\cal O}\left (\frac{1}{c^4} \right ) \, .
\label{h50}
\end{align}
\end{subequations}
In Appendix \ref{appD} we display the modes $h_{22}$, $h_{33}$, and $h_{21}$
for generic inclination angle $\iota$. The modes $ \hat{h}_{\ell m} $ for $ m
< 0$ are derived from Eq.~\eqref{eq:modes} by means of the relation 
$\hat{h}_{\ell \, -m}(\Phi)= (-1)^\ell {\hat{h}}^*_{\ell m}(\Phi + \pi)$. The
non-precessing expressions are obtained by setting $ \iota = 0 $ and $ \alpha
= \pi $. Notice that when $ h_{\ell \, m}(\Phi) $ does not depend on $\Phi$,
we have simply $\hat{h}_{\ell \, -m}= (-1)^\ell {\hat{h}}^*_{\ell m}$. For
comparison with the modes of Refs.~\cite{K07, BFIS08} in the non-spinning
case, it is important to be aware that the origin of the azimuthal angle there
differs from ours by $-\pi/2 $, which produces an extra factor $(-i)^m $
(respectively $i^m$) with respect to us in the modes (respectively in the
spin-weighted spherical harmonics).

Finally, let us emphasize that the $(\ell,m)$ modes defined by Eq.
(\ref{eq:mode_definition}) depend on the particular choice of the source
frame. In fact, they are functions of the spin and angular momentum components
with respect to the $(\mathbf{\hat{x}},\mathbf{\hat{y}},\mathbf{\hat{z}})$
basis introduced in Sec.~\ref{secI}. As there is no canonical way to fix the
reference frame for precessing binaries because of the secular but perpetual
variation of the direction $\mathbf{J}/|\mathbf{J}|$, it is important to be
able to relate the $\hat{h}_{\ell m}$'s given in Eq.~\eqref{eq:modes} to the
polarization modes $h'_{\ell m}$ computed in another frame with different
polarization vectors. Under a passive rotation
  $$R^i_{~j}(A,B,\Gamma)=\left(\begin{array}{ccc} \cos A & -\sin 
      A & 0 \\ \sin A & \cos A & 0 \\ 0 & 0 & 1 
    \end{array}\right) 
  \left(\begin{array}{ccc} \cos B &  0 & \sin B \\ 0 & 1 & 0 \\ 
      -\sin B & 0 & \cos B \end{array}\right)
\left(\begin{array}{ccc} \cos \Gamma & -\sin 
      \Gamma & 0 \\ \sin \Gamma & \cos \Gamma & 0 \\ 0 & 0 & 1 
    \end{array}\right) \, ,$$ 
  the $(\ell,m)$ modes transform in the same way as they would
  in the case of a standard spherical harmonics decomposition
  \cite{GBCS,CLNZ08}. In fact, the spin-weighted $-2$ spherical harmonics are
  precisely devised to ensure this property for the modes of a spin-weighted 
  $-2$ object \cite{GMNRS67}. The law of transformation for the $h_{\ell m}$'s
  is given by
\begin{equation}
h'_{\ell m'}(\Phi', \alpha', \iota', \chi'^{x'}_n, \chi'^{y'}_n, \chi'^{z'}_n) = 
\sum_{m=-\ell}^\ell D^{* \ell}_{m m'}(A,B,\Gamma)\,  h_{\ell m}(\Phi, \alpha, \iota,
\chi_n^x, \chi_n^y, \chi_n^z) \, ,
\end{equation}
where the primed quantities refer to the new frame and where $D_{m'm}^\ell$ is the
  unitary Wigner matrix \cite{GMNRS67}
 \begin{equation}
  D_{m'm}^\ell(A,B,\Gamma)=(-1)^{m'} \sqrt{\frac{4\pi}{2\ell+1}}
    {}_{-m'} Y^{\ell m}(B, A) \, e^{i m' \Gamma}
  \end{equation}
 with the convention of Landau-Lifchitz \cite{LLvol3}. The new angles read
\begin{subequations}
\begin{align}
  &\iota'= \arccos \Big[\cos B \cos \iota - \cos (\Gamma + \alpha) \sin B \sin
  \iota
  \Big]\, ,\\
  &\alpha'= \arccos \cos \alpha' = \arccos \frac{\cos A \big[\cos \iota \sin B
    + \cos B \cos (\Gamma + \alpha) \sin \iota \big] - \sin A \sin \iota \sin
    (\Gamma + \alpha)}{\sqrt{1-\big(\cos B \cos \iota - \cos (\Gamma + \alpha)
      \sin B \sin \iota\big)^2}} \nonumber \\ & \qquad \qquad \qquad \qquad
  \text{if } \cos \iota \sin A \sin B + \sin \iota \big[\cos B \cos (\Gamma +
  \alpha) \sin A + \cos A \sin (\Gamma + \alpha)\big] \ge 0 \, , \\
  &\alpha'= 2\pi - \arccos \cos \alpha' \quad \text{otherwise} \, ,\\
  & \Phi'= \arccos \cos \Phi'\nonumber \\ &\quad =\arccos \frac{\cos B \cos \Phi
    \sin \iota - \sin B \sin \Gamma \, (\cos \iota \cos \Phi \sin \alpha + \cos
    \alpha \sin \Phi) + \cos \Gamma \sin B \, (\cos \iota \cos \alpha \cos
    \Phi - \sin \alpha \sin \Phi )}{\sqrt{1-\big(\cos B \cos \iota - \cos
      (\Gamma + \alpha) \sin B \sin
      \iota\big)^2}} \nonumber \\
  & \qquad \qquad \qquad \qquad \text{if } \cos \Phi \sin B \sin (\Gamma +
  \alpha) + (\cos \iota \cos (\Gamma + \alpha) \sin B + \cos B \sin \iota)
  \sin \Phi  \ge 0\, ,
  \label{eq:Psip1} \\ & \label{eq:Psip2} \Phi' = 2\pi - \arccos \cos \Phi' \quad
  \text{otherwise} \, .
\end{align}
\end{subequations}
When the direction of the total angular momentum used to built the new frame
coincides with that of $\mathbf{J}_0$, which results in the equality 
$\mathbf{\hat{z}}'=\mathbf{\hat{z}} $, the Euler angle $B$ vanishes. 
Then, it can be checked from Eqs.~\eqref{eq:Psip1} and~\eqref{eq:Psip2} that
$\Phi' = \Phi$ as expected.

\section{Features of gravitational-wave modes for precessing binaries on
  nearly circular orbits}
\label{secIIIbis}

We now study how spin effects change the waveform modes for generic precessing
binaries. We consider two maximally spinning configurations with mass ratios
1:1 and 4:1. We label the spin configurations with the angles
$\{\theta_1,\phi_1,\theta_2,\phi_2\}$, where $\{\theta_i,\phi_i\}$ describe
the orientation of the spin vector of the $i^\text{th}$ body relative to
the orbital angular momentum in the initial configuration, which we take to be
a circular orbit with $M\omega_{\rm orb} = 0.001$. We use the full expressions
for the $h_{\ell m}$'s (i.e., the expressions that have not been expanded in
$\iota$) as given in Appendix \ref{appD} and normalized following
  Eq.~\eqref{eq:hlmhat}, but we replace $\Psi$ with $\Phi$. 

\begin{figure}
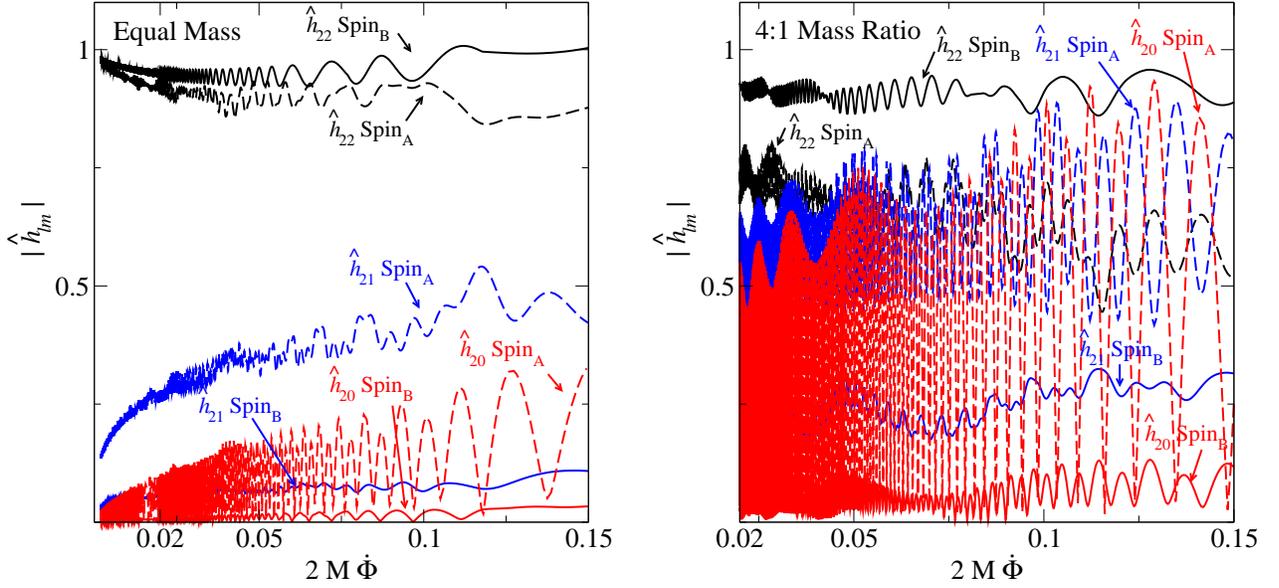

\includegraphics[width=0.45\linewidth]{Leq2equalPlot.eps} \hspace{0.3cm}
\includegraphics[width=0.45\linewidth]{Leq2unequalPlot.eps}
\caption{We plot the modulus of the $\ell = 2$ modes for mass ratios 1:1 (left
  panel) and 4:1 (right panel) with the spin configurations described in
  Fig.~\ref{figure:h22}. The computations use waveforms accurate to 1.5PN
  order in amplitude and phase evolved with precession equations at 1.5PN
  order. The dashed lines are the larger $\iota$ configuration (${\rm
    Spin}_{\rm A}$) and the solid lines are the smaller $\iota$ configuration
  (${\rm Spin}_{\rm B}$). We see that as $\iota$ increases, the modulus of
  $\hat{h}_{22}$ decreases while the modulus of the other $\ell = 2$ modes
  increases. This effect becomes more pronounced when the mass ratio is
  more extreme.
\label{figure:Leq2}}
\end{figure}

\begin{figure}
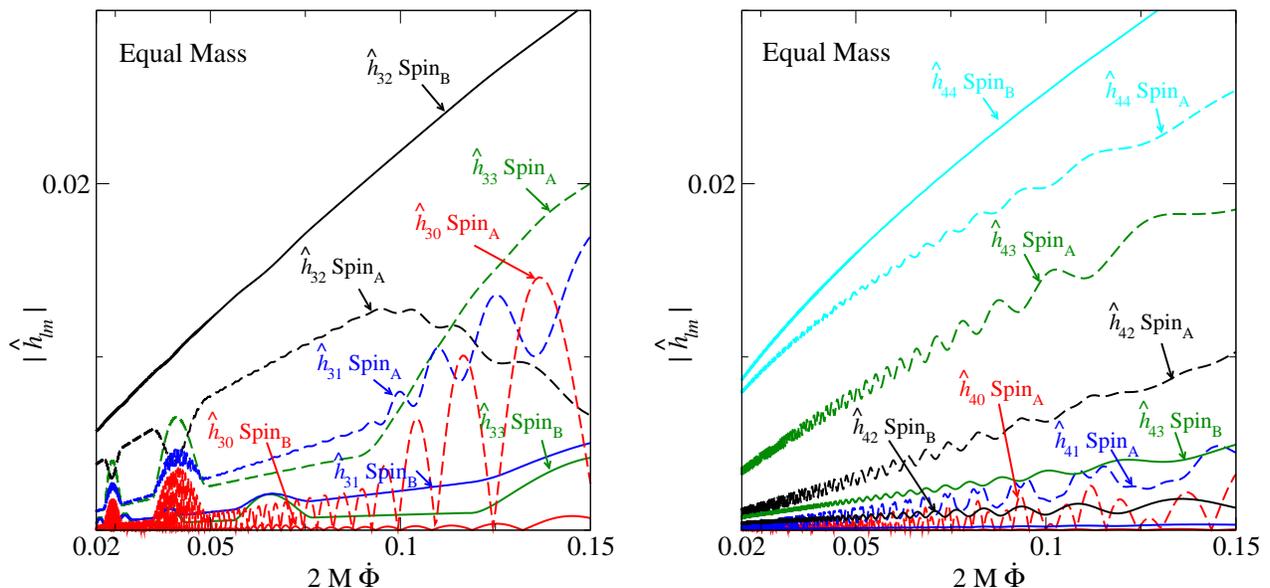

\includegraphics[width=0.45\linewidth]{Leq3equalPlot.eps} \hspace{0.3cm}
\includegraphics[width=0.45\linewidth]{Leq4equalPlot.eps}
\caption{We plot the modulus of the $\ell = 3$ modes (left panel) 
and $\ell = 4$ modes (right panel) for equal masses with the spin 
configurations described in Fig.~\ref{figure:h22}. The computations use 
waveforms accurate to 1.5PN order in both amplitude and phase evolved by means
of the 1.5PN precession equations. The dashed lines refer to the 
larger $\iota$ configuration (${\rm Spin}_{\rm A}$), the solid lines to the 
smaller iota configuration (${\rm Spin}_{\rm B}$). We see a redistribution of 
power among the modes similar to Fig.~\ref{figure:Leq2}. As $\iota$ increases,
the largest modes for the non-precessing cases ($|\hat{h}_{32}|$ and
$|\hat{h}_{44}|$) become smaller, while the other modes become larger.
\label{figure:Leq34}}
\end{figure}

After evolving through 1.5PN order all dynamical quantities they depend on 
[see Eqs.~(\ref{phiorb}), (\ref{preceq}), (\ref{angmomdot}), and
(\ref{omegadot})], we compute the modulus --- more 
often referred to as the absolute value --- of a
sample of modes. Considering the complicated structure of the $h_{\ell
    m}$'s, their qualitative behavior in the presence of spins is discussed
  here in terms of the $\iota$-expanded formulae~\eqref{eq:modes}. Let
us focus on two spin configurations. The configuration ${\rm Spin_A} =
\{{\pi}/{2},0,{\pi}/{2},{\pi}/{2}\}$ has both spin vectors in the orbital
plane, meaning a relatively large inclination angle. The configuration ${\rm
  Spin_B} = \{{\pi}/{6},{\pi}/{4},{\pi}/{6},\pi\}$ has a smaller component of
total spin transverse to the orbital angular momentum, hence a smaller
inclination angle. In Figs.~\ref{figure:Leq2} and \ref{figure:Leq34} we plot
the amplitude of the $\hat{h}_{\ell m}$ over the frequency range 
$2\,M\,\dot{\Phi} = 0.02 \mbox{--} 0.15$, the upper frequency being reached
roughly 2 cycles before merger, for an equal-mass, non-spinning
binary~\cite{UMD-CC08}. For a binary of total mass $16\,M_\odot$, the dominant
second harmonic varies over the frequency range 40--300 Hz. For a
$6.5 \times 10^6\,M_\odot$ binary, this
range is shifted to $10^{-4} \mbox{--} 7.5\times 10^{-4}$ Hz.

The left panel of Fig.~\ref{figure:h22} shows the inclination angle $\iota$ as
a function of the dimensionless frequency $2\, M\,\dot{\Phi}$. We see that the
inclinations are much larger in the case of 4:1 mass-ratio than for equal
masses. Moreover, the inclination increases monotonically in the equal-mass
case, whereas the 4:1 mass ratio exhibits nutation, since the inclination
oscillates, but grows on average. These observations can be explained as
follows. At early times, when the binary has a large orbital separation, we
have $|\mathbf{L}| \gg |\mathbf{S}|$, where
$\mathbf{S}=\mathbf{S}_1+\mathbf{S}_2$ is the total spin, so that $\mathbf{J}
= \mathbf{L} + \mathbf{S}$ and $\mathbf{L}$ are nearly aligned. Radiation
reaction causes $|\mathbf{L}|$ to decrease, making $\mathbf{J}$ move away from
$\mathbf{L}$ and toward $\mathbf{S}$. This is why the inclination angle,
$\iota$, grows on average as the frequency increases. The absence of
oscillations for the inclination angle $\iota$ in the equal-mass case can be
explained by the fact that we are evolving the dynamics, in particular the
precession equations (\ref{preceq}), through 1.5PN order, i.e., we are
neglecting spin-spin effects. Due to the equality $\boldsymbol{\Omega}_1 =
\boldsymbol{\Omega}_2$ at this accuracy level, the precession equations
simplify then to a single equation of the form ${d \mathbf{S}}/{d t} =
\mathbf{\Omega}\times \mathbf{S}$. In the absence of radiation reaction,
$\mathbf{S}$ precesses around a fixed direction with a constant frequency, and
the inclination is constant (apart from the increase produced by radiation
reaction). For unequal masses this symmetry does not exist, with the
consequence that one must solve two coupled equations for $\mathbf{S}_1$,
$\mathbf{S}_2$ instead of a single equation for $\mathbf{S}$. The motion of the
spin vectors is thus more complicated. Schematically, they rotate about a fixed
direction while also bobbing up and down~\cite{ER08}.

The right panel of Fig.~\ref{figure:h22} plots, for the case of equal masses,
the absolute value of the $h_{22}$ mode normalized to its Newtonian order
expression, $\hat{h}_{22}$, for both precessing spin configurations as well as
the non-spinning, aligned and anti-aligned cases. One interesting feature is
that the aligned and anti-aligned cases do not bound the absolute value of the
modes for generic spin configurations. This and other features of the plot can
be understood from the $\iota$ expansion~(\ref{h22}) of $\hat{h}_{22}$, which
contains four spin corrections. The first correction, $-\iota^2/2$, is zero
for the aligned and anti-aligned cases, while it decreases the absolute value
of the $\hat{h}_{22}$ mode for all other spin configurations. If $\iota$ is
comparable to 1 radian, it can be a significant correction. The second
correction, $({1}/{3}) e^{i \Psi } \delta v \iota$, vanishes for equal masses.
For unequal masses, it interferes with the non-spinning terms and, because it
has a different dependence on the orbital phase, produces oscillations in the
absolute value of $\hat{h}_{22}$. Next, the 1PN order spin correction,
$-(v^2/{2}) e^{i (\alpha +\Psi )}[\chi _a^x- i \chi _a^y +\delta (\chi _s^x
- i \chi_s^y)]$, generates oscillations that depend on the spin vector
components transverse to the total angular momentum. Finally, the 1.5PN order
spin correction, $v^3 [-4\delta \chi _a^z/{3}+({4}/{3}) (-1+\nu )
\chi _s^z]$, lowers (raises) the absolute value of $\hat{h}_{22}$ for spins
aligned (anti-aligned) with the total angular momentum. It is solely
responsible for the spread between the aligned and anti-aligned cases, as the
other corrections all vanish then.

A similar analysis can be applied to understand the behavior of the other
modes. As an illustration, we plot in Fig.~\ref{figure:Leq2} all of the
$\ell=2$ modes for mass ratios 1:1 and 4:1. The $\hat{h}_{21}$ mode
(\ref{h21}) is zero for non-spinning equal mass binaries. However, it contains 
several spin corrections and can have significant amplitude for precessing 
binaries, particularly for large $\iota$. It can exhibit complicated 
modulation, as its different spin corrections interfere with one another.
The $\hat{h}_{20}$ mode (\ref{h20}) also has several spin corrections, most
notably $2\, \iota^2\, \cos \Psi$. This correction is responsible for the
large oscillations in the absolute value of $\hat{h}_{20}$. Note that in the
late stages of the inspiral evolution for the 4:1 mass-ratio case, where
$\iota \sim$ 1 radian, these oscillations in the absolute value of
$\hat{h}_{20}$ peak near the absolute value of $\hat{h}_{22}$. The other spin
corrections in $\hat{h}_{20}$ are responsible for the further modulations of
the absolute value.

Figure~\ref{figure:Leq34} plots the absolute value of the $\ell=3$ and
$\ell=4$ modes for equal masses. Note that these modes are about two orders of
magnitude smaller than the $\ell=2$ modes. This remains true in the
non-spinning case for the mass ratios we consider. In the
case of non-spinning, equal mass binaries, for $\ell=3$ only the
$\hat{h}_{32}$ mode (\ref{h32}) is non-zero. The 1.5PN order spin correction
decreases (increases) this mode's absolute value if the spins are aligned
(anti-aligned) with the total angular momentum, in a similar way as in the
$\hat{h}_{22}$ mode. For unequal masses, the $\hat{h}_{32}$ mode
(\ref{h32}) also has an interference term proportional to $\iota$. The other
$\ell=3$ modes are non-zero for generic precessing binaries, and generally
have larger absolute values for larger inclinations. In the left panel of
Fig.~\ref{figure:Leq34}, we indeed observe that $|\hat{h}_{33}|$,
$|\hat{h}_{31}|$, and $|\hat{h}_{30}|$ are greater for the configuration ${\rm
  Spin}_{\rm A}$ than for ${\rm Spin}_{\rm B}$. For $\ell=4$, only the
$\hat{h}_{44}$ and $\hat{h}_{42}$ modes (\ref{h44}), (\ref{h42}) are non-zero
for non-spinning, equal-mass binaries, with $\hat{h}_{44}$ being the largest.
Though the $\iota$-expanded form of the $\hat{h}_{44}$, $\hat{h}_{42}$ and
$\hat{h}_{40}$ modes in Eq.~(\ref{eq:modes}) do not have any spin corrections
through ${\cal O}({1}/{c^3})$, we do see spin effects in all of the $\ell=4$
modes when we plot the full expressions accurate through $v^3$. The $v^2$ and
$v^3$ coefficients in the full expressions for the $\hat{h}_{44}$,
$\hat{h}_{42}$ and $\hat{h}_{40}$ modes depend on the inclination $\iota$ but
\emph{not} on the spin vector components and this dependence is such that if
we treat $\iota$ as a ${1}/{c}$ correction by performing a Taylor expansion in
powers of $\iota$, then the spin terms are proportional to $v^2\,\iota^2$,
$v^3\,\iota$ and higher order in $\iota$. They are thus considered as higher
order corrections in the $\iota$ expansion, though they are present when we
expand only in powers of $v$. None of the $\ell=4$ modes contain any spin
corrections proportional to the spin vector components through order $v^3$.
Those corrections would appear only at higher order in $v$.

While we have plotted the $\ell=3$ and $\ell=4$ modes solely for equal masses,
we have also studied these modes for 4:1 mass-ratio binaries. We find that
they are affected by the change of mass ratio in much the same way as the
$\ell=2$ modes: the redistribution of signal among the $\ell=3$ and $\ell=4$
modes is more pronounced for asymmetric binaries than for equal mass binaries.
However, even for 4:1 binaries with large $\iota$ (${\rm Spin_A}$) spin
configuration, all of the $\ell=3$ and $\ell=4$ modes are still one or two
orders of magnitude smaller than $\hat{h}_{22}$, whereas the $\hat{h}_{21}$
and $\hat{h}_{20}$ modes can be comparable to $\hat{h}_{22}$. The reason is
essentially that $\hat{h}_{21}$ and $\hat{h}_{20}$ have $\iota$
corrections at leading order in $v$, while $\iota$ corrections to the $\ell=3$
and $\ell=4$ modes all appear at higher order, and so they do not have
as strong an effect as for $\hat{h}_{21}$ and $\hat{h}_{20}$.

By examining the absolute values of the $\hat{h}_{\ell m}$ modes for
  precessing binaries, we see that they can be significantly altered by the
  motion of the orbital plane relative to the frame used to perform the mode
  decomposition and the signal may be redistributed among the modes as
  observed in Ref.~\cite{CLNZ08}. This suggests that \emph{all} modes, not
just the dominant ones for non-spinning binaries, are needed to accurately
describe the waveforms emitted from precessing binary systems,
especially for asymmetric binaries and binaries with large inclinations
where this redistribution of signal among the modes is most dramatic.

Let us close this section with a few comments about the applicability of the
full and $\iota$-expanded expressions for the modes and the possibility of
combining them with higher order non-spinning corrections. We find that the
absolute values of the full and $\iota$-expanded $\hat{h}_{\ell m}$ modes are
often quite close to each other for relatively small $\iota$. For inclinations
less than half a radian $(30^\circ)$, the difference in $|\hat{h}_{22}|$ is
typically of a few percent. For $\iota$ comparable to or larger than a radian
($60^\circ$), significant differences between the full and $\iota$-expanded
modes develop and the absolute values may differ by $\sim 10\mbox{--}100\%$
when $\iota \geq 1$ radian. Nonetheless, the $\iota$-expanded modes are very
useful in understanding the qualitative behavior of precessing binaries, even
for inclinations $\sim 1$ radian, albeit they should not be used for precise
quantitative studies of binaries with large inclination angles.

In Refs.~\cite{K07,BFIS08}, expressions of the modes are given to 3PN order
for non-spinning binaries. We have compared their absolute values to that of
the corresponding quantities truncated at 1.5PN order, which has shown us that
they typically differ by $\sim 1 \mbox{--} 10\%$. For example, the absolute
value of the 1.5PN and 3PN order $h_{22}$ modes for equal-mass binaries differ
by less than $1\%$ at $2M\,\dot{\Phi} = 0.01$ and by about $3\%$ at
$2M\,\dot{\Phi} = 0.05$ or $2 M\,\dot{\Phi} = 0.12$. The other modes typically
have a larger difference. The 1.5PN order and highest known order absolute
values for $h_{32}$, $h_{44}$ and $h_{42}$ differ by about $5 \mbox{--} 15\%$
over this same frequency range for equal mass binaries. For 4:1 mass ratio
binaries, the differences in absolute value are similar.

Known higher-order non-spinning terms can actually be included in the
amplitude if enough care is taken. In the non-precessing case, the modes
$h_{\ell m}$ are proportional to $e^{-i\,m\,\Psi}$, as in the non-spinning
case, because $\mathbf{\hat{n}}$ and $\boldsymbol{\hat{\lambda}}$ appearing in
the strain tensor $h_{ij}$ are trigonometric functions of the orbital phase
$\Phi$. In constructing the $h_{\ell m}$'s from $h_{ij}$, they generate an
exponential dependence on multiples of $\Phi$. However, for the case of a
precessing binary with small $\iota$, $\mathbf{\hat{n}}$ and
$\boldsymbol{\hat{\lambda}}$ are trigonometric functions of $\Phi+\alpha$.
Thus, the $h_{\ell m}$'s contain then all of the non-spinning terms, but with
the substitution $\Psi \rightarrow \Psi+\alpha$. The situation is different
for the general precessing case. The vectors $\mathbf{\hat{n}}$ and
$\boldsymbol{\hat{\lambda}}$ depend on $\Phi$, $\alpha$, and $\iota$, and the
resulting $h_{\ell m}$ have a complicated dependence on all three of these
quantities that cannot be simply related to the non-spinning case. These
considerations show that it is only for binaries with a small inclination (or
no inclination) that we can readily construct the $h_{\ell m}$'s with spin
effects up to 1.5PN order and non-spinning corrections up to 3PN order. For
spins (anti-)aligned it is trivial to add the higher-order non-spinning
corrections of Refs.~\cite{K07,BFIS08} to the $h_{\ell m}$ given in
Eq.~(\ref{eq:modes}). For precessing binaries with small inclinations, they
can be added to our expressions in Eq.~(\ref{eq:modes}) with the substitution
$\Psi \rightarrow \Psi+\alpha$. For general precessing binaries, it is not so
simple to include higher-order non-spinning corrections to the full
expressions for the $h_{\ell m}$ given in Appendix~\ref{appD}. To do this
properly, we would need the spin terms at the same order as the
non-spinning terms and repeat the derivation of the $h_{\ell m}$ to a higher
order.

\section{Ready-to-use frequency-domain templates for spinning, 
non-precessing binaries}
\label{secIV}

\subsection{Gravitational-wave polarizations in time domain}
\label{subsecIVa}

In the non-precessing case, the orbital angular momentum points in a fixed
direction which we take to be the $z$-axis (see Fig.~\ref{figure:SourceFrame})
and the spins are either aligned or anti-aligned with it. The basis vectors of
the orbital plane, $\mathbf{\hat{x}}_L(t)$ and $\mathbf{\hat{y}}_L(t)$, are
constant in time. They can be freely chosen to be any pair of orthogonal
unit vectors in the $x\mbox{--}y$ plane. Here, following the convention of
Ref.~\cite{ABIQ04}, we choose $\mathbf{\hat{x}}_L = \mathbf{\hat{P}} =
{\mathbf{\hat{N}} \times \mathbf{\hat{L}}_{\rm N}}/{|\mathbf{\hat{N}} \times
  \mathbf{\hat{L}}_{\rm N}|}$, so that the phase is zero at the ascending node
(where the orbital separation vector crosses the plane of the sky from below).
This is equivalent to setting $\iota = 0$ and $\alpha = \pi$ in
Eqs.~\eqref{h}, \eqref{eq:Hplus_prec}, and \eqref{eq:Hcross_prec}. Note also
that since the orbital plane remains fixed, the phase $\Phi$ defined through
Eq.~\eqref{omegaorb} coincides with the standard definition of the orbital
phase, that is
\begin{equation}
\label{omeganonprecess}
\omega_{\rm{orb}} = \dot{\Phi}\,.
\end{equation}
In the non-precessing case, the vectors in terms of which the GW polarizations
are expressed originally take the simpler form 
\begin{align}
\label{nvecNP}
\mathbf{\hat{n}} &= \left(\sin \Phi\, ,\, -\cos \Phi\, ,\,0 \right)\,,\\
\label{lvecNP}
\boldsymbol{\hat{\lambda}} &= \left(\cos \Phi\, ,\,\sin \Phi\, ,\,0\right)\,,\\
\label{NvecNP}
\mathbf{\hat{N}} &= \left(\sin \theta\, ,\, 0\, ,\,\cos \theta\right)\,,\\
\label{LvecNP}
\mathbf{\hat{L}}_{\rm N} &= \left(0\, ,\,0\, ,\,1\right)\,.
\end{align}
By plugging the expressions (\ref{nvecNP})--(\ref{LvecNP}) into
Eq.~(\ref{hij}) and taking the combinations given in Eq.~(\ref{poldefs}), we
obtain an equation similar to Eq.~(\ref{h}). The spin-dependent 1PN, 1.5PN and
2PN order polarization coefficients read

\begin{subequations}
\begin{align}
\label{Hplus1}
H_{+}^{(1,{\rm SO})} &= v^2{s_\theta}\,\cos \Psi\,\left[
\boldsymbol{\chi}_a\cdot\mathbf{\hat{L}}_{\rm N} 
+ \delta\,\boldsymbol{\chi}_s\cdot\mathbf{\hat{L}}_{\rm N}\right]\,,\\
\label{Hplus15}
H_{+}^{(3/2,{\rm SO})} & =  v^3 \cos 2\Psi\,\frac{4}{3}\,\left[(1 + c_\theta^2)
\left(\boldsymbol{\chi}_s\cdot\mathbf{\hat{L}}_{\rm N} 
+ \delta\,\boldsymbol{\chi}_a\cdot\mathbf{\hat{L}}_{\rm N}\right) + \nu\,(1 -
5 c_\theta^2)\,\boldsymbol{\chi}_s\cdot\mathbf{\hat{L}}_{\rm N}\right]\,,\\ 
\label{Hplus2}
H_{+}^{(2,{\rm SS})}&=  - v^4\, 2\,\nu\,(1 +c_\theta^2)
\left(\boldsymbol{\chi}_s^2 - \boldsymbol{\chi}_a^2\right)\,\cos 2 \Psi\,,\\
\label{Hcross1}
H_{\times}^{(1,{\rm SO})} &=  v^2 {s_\theta\,c_\theta}\,\sin \Psi\,\left[
\boldsymbol{\chi}_a\cdot\mathbf{\hat{L}}_{\rm N} 
+ \delta\,\boldsymbol{\chi}_s\cdot\mathbf{\hat{L}}_{\rm N}\right]\,,\\
\label{Hcross15}
H_{\times}^{(3/2,{\rm SO})} &=  v^3 \sin 2\Psi\,\frac{4}{3}\,c_\theta\, 
\left[2 \left(\boldsymbol{\chi}_s\cdot\mathbf{\hat{L}}_{\rm N}
+ \delta\,\boldsymbol{\chi}_a\cdot\mathbf{\hat{L}}_{\rm N}\right)
- \nu\,(1 + 3 c_\theta^2)\,\boldsymbol{\chi}_s\cdot\mathbf{\hat{L}}_{\rm N}
\right]\,,\\
\label{Hcross2}
H_{\times}^{(2,{\rm SS})} &= - v^4\,4\,\nu\,c_\theta\,
\left(\boldsymbol{\chi}_s^2 - \boldsymbol{\chi}_a^2\right)\,\sin 2 \Psi\,,
\end{align}
where SS labels the spin(1)-spin(2) contributions.\footnote{We remind that
  spin(1)-spin(1) and spin(2)-spin(2) effects in the waveform polarizations
  are currently unknown.} In the equations above, we use the shorthand
$c_\theta = \cos\theta$ and $s_\theta = \sin\theta$. Note also that the phase
$\Psi$ is the shifted orbital phase that relates to $\Phi$ at our accuracy
level as
\begin{equation}
\label{psidef}
\Psi = \Phi - 2\,v^3\,\left(1 -
\frac{\nu}{2}\,v^2\,\right)\,\ln\left(\frac{\omega_{\rm
    orb}}{\omega_0}\right)\,,
\end{equation}
\end{subequations}
where $\omega_0$ can be chosen arbitrarily. Expressed in terms of the orbital
phase $\Phi$, the GW polarizations would contain terms logarithmic in
$\omega_{\rm orb}$, arising from the propagation of the tails. However,
introducing the phase (\ref{psidef}), they are all absorbed (up to the 2.5PN
order we are considering) into the phase variable~\cite{ABIQ04}.

The spin-dependent polarizations (\ref{Hplus1})--(\ref{Hcross2}) were derived
in Ref.~\cite{WWi96} [see Eqs.~(F24a)--(F25c) in that paper], although the
1.5PN and 2PN order cross polarizations had an erroneous sign, which is
corrected here.

\subsection{Spin-orbit effects at 1.5PN order and spin-spin effects at 2PN 
order in the frequency-domain gravitational-wave amplitude}
\label{subsecIVb}

Writing $h(t) = h_+\,F_+ + h_\times\,F_\times$ and collecting terms
by PN order and by sines or cosines of harmonics of the
orbital frequency, we can write the time-domain strain in the compact form:
\bea h(t) &=& \frac{2\,M\,\nu}{D_L}\, \sum_{n=0}^5\ \sum_{k=1}^7
V_k^{2+n}\,\left [\alpha_k^{(n)}\,\cos(k\,\Psi(t)) +
  \beta_k^{(n)}\,\sin(k\,\Psi(t)) \right
] + {\cal O}(v^8)\,,\nonumber\\ &=&\frac{2\,M\,\nu}{D_L}\ \sum_{n=0}^5
\sum_{k=1}^7\ V_k^{2+n}\,\left [\alpha_k^{(n)}\,\cos(k\,\Psi(t)) +
  \beta_k^{(n)}\,\cos(k\,\Psi(t)-\frac{\pi}{2}) \right ]+{\cal
  O}(v^8)\,,\label{hoft} \eea
where ${n}/{2}$ is the PN order and $k$ labels the
harmonics of the orbital phase. The PN expansion
parameter is defined as $V = (2\pi M F)^{1/3}$, with $F=\omega_{\rm
  orb}/(2\pi)$. We shall denote the GW frequency by $f$.  
For the $k^{\rm th}$ harmonic, we have then the relation $f=f_k \equiv k F$,
so that  
\begin{equation}
\label{vk}
V_k = \left (2\,\pi\,M\,\frac{f_k}{k} \right )^{1/3}\,.
\end{equation}
Given a function of the form $h(t) = A(t)\,\cos \phi(t)$, where $\phi(t)$ is a
monotonically increasing function satisfying ${d\rm{ln} A(t)}/{dt} \ll {d
  \phi(t)}/{dt}$, we can compute its Fourier transform by applying the
stationary-phase approximation (SPA):
\begin{equation}
\tilde{h}(f) \simeq \frac{1}{2}\, A(t(f))\,
\sqrt{\frac{2\,\pi}{\ddot{\phi}(t(f))}}\,e^{i(2\,\pi\,f\,t(f) - \phi(t(f)) -
  \pi/4)}\,,\label{SPAformula} 
\end{equation}
$t(f)$ being defined here for each frequency $f$  as the value of $t$ for which
$(d\phi/dt)(t) = 2 \pi f$. In a similar manner, we apply the SPA to each term in
the sum of Eq.~(\ref{hoft}). Moreover, for each harmonic of the orbital phase,
we expand the factor inversely proportional to the second time derivative of 
the orbital phase entering Eq.~(\ref{SPAformula}) in a PN series of the form
\begin{equation}
\sqrt{\frac{2\, \pi}{k\,\ddot{\phi}}} = \left(k\,\frac{dF}{dt}\right)^{-1/2} =
\sqrt{\frac{5\,\pi}{k\,48\,\nu}}\,M\,V_k^{-11/2}\,\left [1 +
  \mathcal{S}_2\,V_k^2 + \mathcal{S}_3\,V_k^3 + \mathcal{S}_4\,V_k^4 +
  \mathcal{S}_5\,V_k^5 + \mathcal{O}(V_k^6) \right]\,,\label{derivexpansion}
\end{equation}
with
\bea
\mathcal{S}_2 & = & \frac{743}{672}\,+\,\frac{11}{8}\,\nu\,,\nonumber\\
\mathcal{S}_3 & = & -2\pi+\left(\frac{113}{24} - \frac{19}{6}\,\nu\right)
\,\boldsymbol{\chi}_s\cdot\mathbf{\hat{L}}_{\rm N} + \frac{113}{24}
\,\delta\,\boldsymbol{\chi}_a\cdot\mathbf{\hat{L}}_{\rm N}\,, \nonumber\\
\mathcal{S}_4 & = & \frac{7266251}{8128512}\,+\,\frac{18913}{16128}\,\nu\,
+\,\frac{1379}{1152}\,\nu^2 
- \nu\, \left(\frac{721}{96}\,\left(\left(\boldsymbol{\chi}_s
\cdot\mathbf{\hat{L}}_{\rm N}\right)^2 - 
\left(\boldsymbol{\chi}_a\cdot\mathbf{\hat{L}}_{\rm N}\right)^2\right) 
- \frac{247}{96}\left(\boldsymbol{\chi}_s^2 
- \boldsymbol{\chi}_a^2\right)\right)\,, \nonumber\\
\mathcal{S}_5 & = &
\pi\,\left(-\frac{4757}{1344}\,+\,\frac{57}{16}\,\nu\right)\,.
\label{S-expansion}
\eea 
The expansion (\ref{derivexpansion}) {\it without} spin corrections in the
amplitude was first given in Ref.~\cite{ChrisAnand06b}. We have added to it
the leading order SO corrections through 1.5PN order and the spin(1)-spin(2) SS
corrections appearing at 2PN order. In principle,  SO corrections at 2.5PN
order and spin(1)-spin(1), spin(2)-spin(2) SS corrections at 2PN arising from
the spin contribution to the orbital frequency are also present. However, when
calculating spin terms in the frequency-domain amplitude, we neglect them
because they have not been calculated yet beyond the 1.5PN order in the 
time domain amplitude. The spin contribution at 2PN and 2.5PN order to the 
Fourier domain amplitude is not complete unless we take both into account.

Defining the frequency-dependent SPA phase as 
\begin{equation}
\Psi_{\rm SPA}(f) = 2 \pi\,f\,t(f) - \Psi(f)\,,\label{SPAphase}
\end{equation}
the frequency domain waveform with amplitude corrections containing SO effects
through 1.5PN order and spin(1)-spin(2) effects through 2PN order is
\bea \tilde{h}(f) &=& \frac{M\,\nu}{D_L} \sum_{n=0}^5 \sum_{k=1}^7
V_k^{2+n}\,\left(k\,\frac{dF}{dt}\right)^{-1/2}
\left(\alpha_k^{(n)}\,e^{i\,\left(2\,\pi\,f\,t(F)-k\,\Psi(F)\,-\,\pi/4\right)}
+
\beta_k^{(n)}\,e^{i\left(2\,\pi\,f\,t(F)\,-(k\,\Psi(F)-\pi/2)-\pi/4\right)}
\right)\,, 
\nonumber\\ &=& 
\frac{M\,\nu}{D_L}\sum_{n=0}^5\sum_{k=1}^7 V_k^{n-\frac{7}{2}}\,
\sqrt{\frac{5\,\pi}{k\,48\,\nu}}\,M\, \left(1+ {\cal S}_2\,V_k^2 + {\cal
  S}_3\,V_k^3 + {\cal S}_4\,V_k^4 + {\cal S}_5\,V_k^5\right) (\alpha_k^{(n)} +
e^{i\,\pi/2}\,\beta_k^{(n)}) e^{i\,\left(k\,\Psi_{\rm
    SPA}(f/k)-\pi/4\right)}\,,\nonumber\\ &=& \frac{M^2}{D_L}\,
\sqrt{\frac{5\,\pi\,\nu}{48}} \sum_{n=0}^5
\sum_{k=1}^7\ V_k^{n-\frac{7}{2}}\,{\cal C}_k^{(n)} e^{i\,\left(k\,\Psi_{\rm
    SPA}(f/k)-\pi/4\right)}\,.\label{hoff} \eea
with
\beq \mathcal{C}_k^{(n)}= \frac{1}{\sqrt{k}}\, \left(\alpha_k^{(n)} +
i\,\beta_k^{(n)}\right) + \sum_{m=2}^{n}
\frac{\mathcal{S}_{m}}{\sqrt{k}}\,\left(\alpha_k^{(n-m)} +
i\,\beta_k^{(n-m)}\right)\,,
\label{Cceff}
\eeq
where the index $n$ denotes the PN order and the index $k$ the harmonics.
Explicit expressions for the $\mathcal{C}_k^{(n)}$ can be found in
Appendix~\ref{appC}. The non-spinning terms in the amplitude agree with
Ref.~\cite{ChrisAnand06}, although we have written them in a different, more
explicit manner. Notice that recently the non-spinning amplitude corrections
were calculated through 3PN order~\cite{BFIS08}, but in this paper we
restricted the computation to 2.5PN order.

\subsection{Spin-orbit effects at 2.5PN order in the frequency-domain 
gravitational-wave phase}
\label{subsecIVc}

For matched filtering, it is best to know the GW phasing at the highest PN
order. We now derive the SO contributions to the SPA phase through 2.5PN
and the SS contributions (\emph{including} spin(1)-spin(1) and 
spin(2)-spin(2) contributions) to the SPA phase through 2PN order.

The PN expansion of the SPA phase $\Psi_{\rm SPA}(F)$ can be obtained from the
PN expansions of the binary center-of-mass energy, $E$, and GW flux, ${F}$,
via the energy balance equation 
\beq -\frac{dE}{dt} = {\cal F}\,.
\label{EB}
\eeq 
Using ${d\Psi/dt}=2\,\pi\,F = {v^3}/{M}$, we can re-write  
the energy balance equation as the differential equations
\begin{equation}
dt = -\frac{dE}{dv}\,\frac{1}{\cal F} \, dv\,,\label{dt}
\end{equation}
and 
\bea 
d\Psi= -\frac{dE}{dv}\,\frac{1}{\cal F}\,\frac{v^3}{M}\, dv\,.\label{dphi}
\eea 
The quantities $E$ and ${\cal F}$ are known as power series in
$v=(2\pi\,M\,F)^{1/3}$. The non-spinning terms in the expansions of $E$ and
${\cal F}$ have been calculated by Refs.~\cite{JS99,JaraS00,Andrade01,
  Blanchet01a,Damour01,FBBu06}, while the spin contributions to these
quantities through 2.5PN order were derived by 
Refs.~\cite{K95,WWi96,Poisson98,BFIJ02,BDEI04,BBuF06}. 
The center-of-mass energy and the flux read 
\bea
\label{Energy}
E(v) &=& E_{\rm Newt}\, v^2\, \left(1 + \sum_{i=2}^6 E_i v^i \right)\,,\\
{\cal F}(v) &=& F_{\rm Newt}\,v^{10}\, \left ( 1 + \sum_{i=2}^7 F_i v^i \right
)\,, 
\label{Flux}
\eea 
where the coefficients $E_i$ and $F_i$ are explicitly given in
Appendix~\ref{appA}. By inserting Eqs.~(\ref{Energy}), (\ref{Flux}) into
Eqs.~(\ref{dt}), (\ref{dphi}), we obtain rational function approximations to
the integrands. We then find the Taylor series of the rational functions and
integrate up to some reference frequency, often chosen to be the time of
coalescence, when the orbital frequency formally diverges. Thus, we obtain PN
approximations of the form
\bea t(v) & = & t_c - \int_{v_c}^v \frac{E_{\rm Newt}}{F_{\rm Newt}}\,\left (2
v^{-9}+\sum_{j=2}^7 t_j\, v^{j-9} \right )\,,\label{t}\\ \Psi(v) & = & \Psi_c
- \frac{1}{M}\,\int_{v_c}^v \frac{E_{\rm Newt}}{F_{\rm Newt}}\,\left (2
v^{-6}+\sum_{j=2}^7 t_j\, v^{j-6}\right )\,,\label{phi} \nonumber \\ \eea
where the $t_j$ coefficients are linear combinations of products of the $E_i$
and $F_i$. Plugging Eqs.~(\ref{t}), (\ref{phi}) into Eq.~(\ref{SPAphase}), we
obtain the following expression for the SPA phase through 2.5PN
order\footnote{The non-spinning terms in the SPA phase through 3.5PN order
  can be found in Ref.~\cite{AISS05}.}
\bea \Psi_{\rm SPA}(F)&=&2\pi\, F\, t_c-\Psi_c+ \frac{3}{256}\,(2\pi {\cal
  M}F)^{-5/3}\, \left\{1+\left(\frac{3715}{756}+\frac{55}{9}\nu\right)\,(2\pi
M F)^{2/3}+(4\, \beta-16\pi)\,(2\pi M
F)\right.\nonumber\\ &+&\left.\left(\frac{15293365}{508032}+
\frac{27145}{504}\nu+\frac{3085}{72}\nu^2 
-10\, \sigma \right)\,(2\pi M F)^{4/3}+\left( \frac{38645}{756}\pi -
\frac{65}{9}\pi\,\nu \, - \gamma \right)\,\left(1+3\log(v) \right) \times
\right . \nonumber \\ && \left . (2\pi M F)^{5/3} \right \}\,, \label{SPA} \eea
where the 1.5PN SO phase corrections are contained in $\beta$, the 2PN SS
corrections are contained in $\sigma$, and the 2.5PN SO corrections are
contained in $\gamma$. Note that $\beta$ and the spin(1)-spin(2) contributions
to $\sigma$ were previously known~\cite{CF94,PW95}, while we have calculated
the spin(1)-spin(1) and spin(2)-spin(2) contributions to $\sigma$ and the
2.5PN SO corrections to the SPA phase using the results for the center-of-mass
energy and GW flux of Refs.~\cite{Poisson98,BBuF06,FBBu06,RBK08}. Explicitly,
these corrections are
\bea
\beta&=& \left(\frac{113}{12} -\frac{19}{3}\,\nu\right)
\,\boldsymbol{\chi}_s\cdot\mathbf{\hat{L}}_{\rm N} + \frac{113}{12}
\,\delta\,\boldsymbol{\chi}_a\cdot\mathbf{\hat{L}}_{\rm N}\,,\label{beta}\\
\sigma&=& \nu\, \left \{\frac{721}{48}\,\left[\left(\boldsymbol{\chi}_s
\cdot\mathbf{\hat{L}}_{\rm N}\right)^2 - 
\left(\boldsymbol{\chi}_a\cdot\mathbf{\hat{L}}_{\rm N}\right)^2\right] 
- \frac{247}{48}\left(\boldsymbol{\chi}_s^2 
- \boldsymbol{\chi}_a^2\right)\right \}\nonumber\\ 
 &+& \left(1-2\nu\right)\,\left \{\frac{719}{96}\,
\left[\left(\boldsymbol{\chi}_s \cdot\mathbf{\hat{L}}_{\rm N}\right)^2 + 
\left(\boldsymbol{\chi}_a\cdot\mathbf{\hat{L}}_{\rm N}\right)^2\right]
- \frac{233}{96}\,\left(\boldsymbol{\chi}_s^2 
+ \boldsymbol{\chi}_a^2\right)\right \}\nonumber\\
 &+& \delta\,\left[\frac{719}{48}\,
\left(\boldsymbol{\chi}_s\cdot\mathbf{\hat{L}}_{\rm N}\right)\,
\left(\boldsymbol{\chi}_a\cdot\mathbf{\hat{L}}_{\rm N}\right)
- \frac{233}{48}\,\boldsymbol{\chi}_s\cdot\boldsymbol{\chi}_a\right]
\,,\label{sigma}\\
\gamma&=&\left(\frac{732985}{2268} - \frac{24260}{81}\,\nu -
  \frac{340}{9}\,\nu^2 \right)\,
\boldsymbol{\chi}_s\cdot\widehat{\mathbf{L}}_{\rm N} + \left(\frac{732985}{2268} +
  \frac{140}{9}\,\nu\right)\,\delta\,
\boldsymbol{\chi}_a\cdot\widehat{\mathbf{L}}_{\rm N}\,.\label{gamma} \eea
We note that these expressions are only valid when both component spins 
are aligned or anti-aligned with the orbital angular momentum. The 
spin(1)-spin(1) and spin(2)-spin(2) contributions to $\sigma$ were also 
derived in Refs.~\cite{MVG05,RBK08} and we found full agreement with them.

\subsection{Features of frequency-domain non-precessing waveforms with higher 
harmonics}
\label{subsecIVd}

We now discuss some interesting features of the {\it spinning, non-precessing}
waveforms derived in Sec.~\ref{subsecIVb}. Several papers have studied the
effect of higher harmonics in the amplitude corrections of {\it non-spinning}
binaries observable by ground- and space-based detectors~\cite {ChrisAnand06,
  ChrisAnand06b, AISS07, AISSV07,TriasSintes07,PorterCornish08}. 

One important feature of the higher harmonics in the waveform amplitude is
that they can increase the mass reach of a detector~\cite{AISS07}. This is
because high-mass binaries whose dominant second harmonic is not in the
detector's sensitive band can have higher harmonics in band and therefore
become visible to the detector. A closer look at Eq.~(\ref{Cceff}) and
Appendix~\ref{appC} shows that spin corrections through 2PN order appear only
in the first and second harmonics. In particular, the only SPA amplitude 
coefficients with spin dependence are
$\mathcal{C}^{(2)}_1$, $\mathcal{C}^{(3)}_2$ and $\mathcal{C}^{(4)}_2$ given
in Appendix~\ref{appC}. Thus, in the non-precessing case 
spin corrections through 2PN order in the waveform amplitude do not affect the
mass reach of the detector, and only affect binaries whose second harmonic
appears in band.

Another general feature of the higher harmonics is that they interfere with
one another, typically destructively~\cite{ChrisAnand06,AISS07,TriasSintes07}.
For binaries that would be visible with Newtonian waveforms, this effect tends
to decrease the signal to noise ratio (SNR). As we shall study in detail in
this section, spin effects can play a role in this interference, either
raising or lowering the SNR depending on the spin orientations.

\begin{figure}
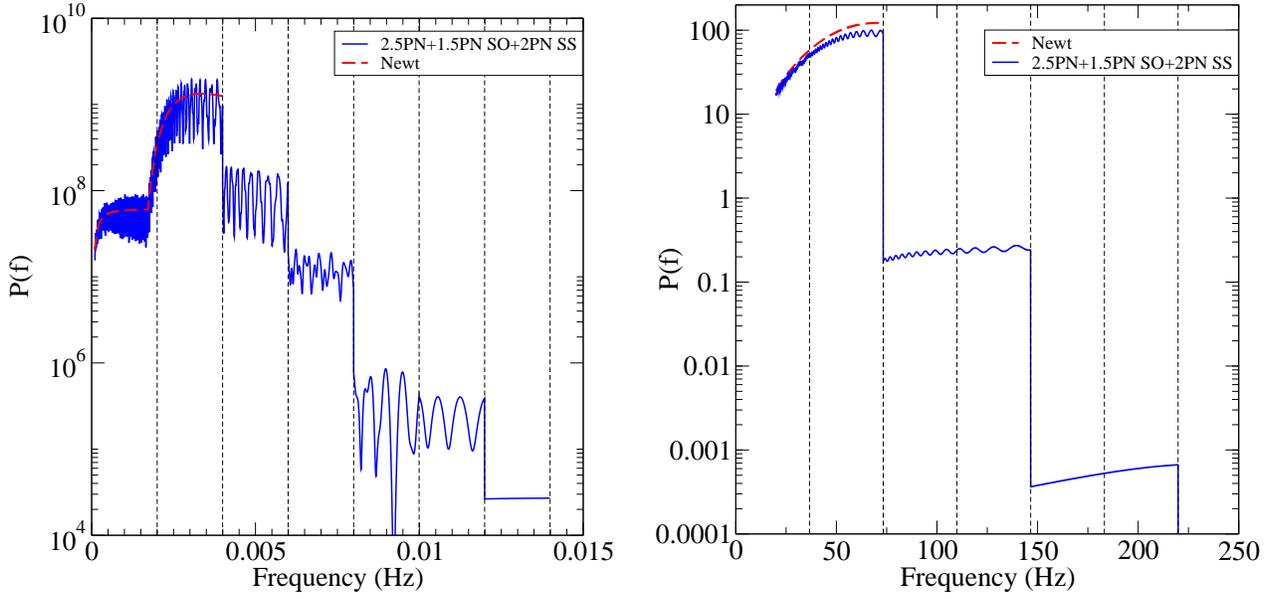

\includegraphics[width=0.45\linewidth]{PowerSpecLISA.eps} \hspace{0.3cm}
\includegraphics[width=0.45\linewidth]{PowerSpecADVLIGO.eps}
\caption{We compare the power spectra computed with the Newtonian amplitude 
  waveform (red dashed line) and the 2.5PN waveform 
  with 1.5PN SO and 2PN SS effects included (blue, continuous line). 
  In the left panel
  we consider a typical source for LISA, a binary with total mass $(10^6 +
  10^5) M_\odot$, and spins maximal and aligned with the orbital angular
  momentum. In the right panel we consider a typical source for Advanced
  LIGO, a binary of total mass $(30 + 30) M_\odot$ with spins $\chi_1 = 1$,
  $\chi_2 = 0.5$ aligned with the orbital angular momentum. Note that the
  $k^{\rm th}$ harmonic ends at $k\, F_{\rm LSO}$, and these frequencies are
  marked by the vertical dashed lines on the graph. The spectrum of the 2.5PN 
  waveform is much simpler in the equal-mass case than unequal mass case
  because in the former case all non-spinning odd harmonics are suppressed.
\label{figure:PowerSpecLISAAdvLIGO}}
\end{figure}

We define the power spectrum, $P(f)$, as
\begin{equation}
\label{Poff}
P(f) = \frac{|\tilde{h}(f)|^2}{S_n(f)}\,,
\end{equation}
and the {\it optimal} SNR, $\rho$, as 
\begin{equation}
\label{rho}
\rho^2 = 4 \int_{f_s}^{7 F_{\rm LSO}} \frac{|\tilde{h}(f)|^2}{S_n(f)} df\,,
\end{equation}
where $f_s$ is the low frequency seismic cutoff of the detector, and the 
upper frequency cutoff is taken to be the highest harmonic 
of the orbital frequency at the last stable orbit (LSO) which 
for simplicity we choose to be the LSO of a test particle in Schwarzschild, 
\begin{equation}
\label{Flso}
F_{\rm LSO} = \frac{1}{2\pi 6^{3/2} M}\,.
\end{equation}
Note that the $k^{\rm th}$ harmonic ends at $k F_{\rm LSO}$ as enforced by a
step function $\theta(k F_{\rm LSO} - f)$ [see Eqs.~(\ref{b1})--(\ref{b21}) in
Appendix~\ref{appC}]. In Eqs.~(\ref{Poff}), (\ref{rho}), we denote with
$S_n(f)$ the noise power spectral density of the detector. For Advanced LIGO,
we take the spectral density to be Eq.~(4.3) of Ref. ~\cite{ChrisAnand06b} and
fix $f_s = 20$ Hz. For LISA, we use the so-called \emph{effective}
non-sky-averaged spectral density given in Eqs.~(2.28)--(2.32) of
Ref.~\cite{BBW05a}. We do not consider the orbital motion of the LISA
spacecraft~\cite{Cutler98} and consider only the single detector configuration
\footnote{It should be noted that in our model, though we do not perform an
  average over the antenna pattern functions, we do not account for the
  orbital motion of LISA either. In this sense, our model falls in between the
  {\it pattern averaged} and {\it non-pattern averaged} cases described in
  Ref.~\cite{BBW05a}}. In the presence of higher harmonics, the lower and
upper cut-off frequencies are chosen following Sec. IIIA of
Ref.~\cite{AISS07}. For LISA we assume an observation time of one year, and
the orbital frequency at the beginning of observation to be Eq.~(3.3) of
Ref.~\cite{AISS07}. As explained in Ref.~\cite{AISS07}, this can be
implemented by multiplying the $k^{\rm th}$ harmonic by the step function
$\theta(f - k F_{in})$ where $F_{in}$ is the orbital frequency at the
beginning of observation. Finally, because of the $60^\circ$ angle between
LISA's arms, we use $\tilde{h}(f) \rightarrow (\sqrt{3} / 2)\, \tilde{h}(f)$
in Eqs.~ (\ref{Poff}), (\ref{rho}) in the case of LISA.

All tables and figures in this section, refer to a binary with orbital angular
momentum inclined relative to the line of sight by $\theta = \pi / 3$, sky
location $\bar{\theta} = \bar{\phi} = \pi / 6$ and polarization angle
$\bar{\psi} = \pi / 4$ [see Eqs.~(\ref{antennaresponse1}),
(\ref{antennaresponse2})]. We have verified, by considering random values for
the four angles, that the qualitative trends reported in this section are
generic and do not depend on the specific values of them (see a detailed
discussion at the end of this section). Regardless of the PN order of the
amplitude, all waveforms use the SPA phase with non-spinning terms up to 3.5PN
order~\cite{AISS05}, and spin terms up to 2.5PN order, as given in
Eqs.~(\ref{SPA})--(\ref{gamma}). In the case of Advanced LIGO (LISA) we
consider binaries at a distance of 100 Mpc (3 Gpc). Moreover, all masses and
distances refer to the redshifted quantities.

\begin{figure}
\includegraphics[width=0.45\linewidth]{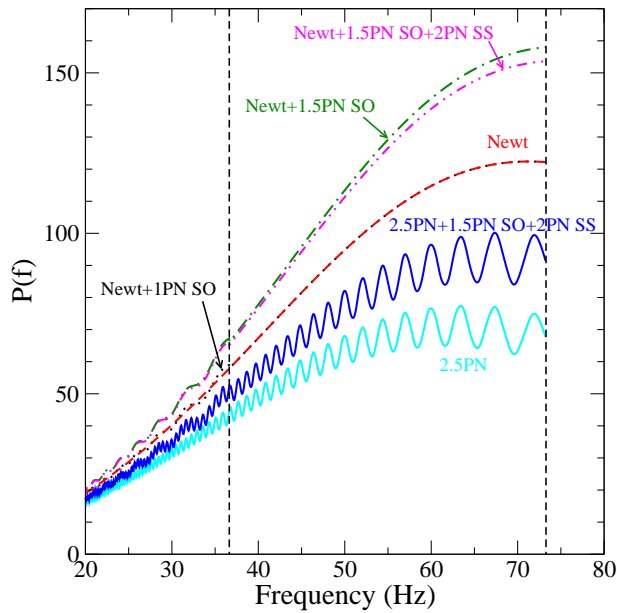}
\caption{For a binary of total mass $(30+30) M_\odot$ with spins $\chi_1=1,
  \chi_2=0.5$ aligned with the orbital angular momentum (the same binary of
  the right panel of Fig.~\ref{figure:PowerSpecLISAAdvLIGO}),  
we show the power spectra up to $2 F_{\rm LSO}$. We plot the power spectrum
for the waveform through 2.5 PN order with no spin corrections (cyan solid
line) and with SO corrections through 1.5PN (that is, 1PN and 1.5PN) and SS
corrections at 2PN order (dark blue solid line). We also plot power spectra
for the waveform with Newtonian amplitude (red dashed line), Newtonian
amplitude plus the 1PN SO correction (black dotted line), Newtonian amplitude
plus SO effects through 1.5PN (green,dot dashed line), and Newtonian amplitude
plus SO corrections through 1.5PN and the 2PN SS correction 
  (magenta, double-dot-dashed line). 
  The 1.5PN SO and 2PN SS effects raise and lower the power in
  the dominant harmonic while the 1PN SO effect merely changes the modulation
  pattern up to its cutoff frequency of $F_{\rm LSO}$. Vertical dashed lines
  mark the frequencies $F_{\rm LSO}$ and $2\, F_{\rm LSO}$.\\
  \label{figure:PowerSpecZoom}}
\end{figure}

The destructive interference between different harmonics can be seen in
Fig.~\ref{figure:PowerSpecLISAAdvLIGO}. The Newtonian waveform's power
spectrum is simply proportional to $f^{-7/3} / S_n(f)$. The higher harmonics
present in the 2.5PN waveform introduce oscillatory cross terms that on
average lower the power. Notice that although the higher harmonics extend the
observable frequency band significantly, the power beyond the cutoff of the
second harmonic, being at a higher PN order, is suppressed by one or several
orders of magnitude. These features explain why the SNR listed in
Tables~\ref{table:advLIGOSNR}, \ref{table:LISASNR} tends to decrease as the PN
order increases for the range of masses we consider. For equal-mass binaries,
all non-spinning odd harmonic corrections are suppressed because the latter
are proportional to $\delta$ which is zero for equal masses [see
Eqs.~(\ref{b1})--(\ref{b21}) in Appendix~\ref{appC}]. This is not true of
spin-dependent amplitude corrections. For example, the first harmonic has a
spin dependent amplitude correction at 1PN order which does not vanish for
equal mass systems unless spins are equal and aligned with one another [see
Eq.~(\ref{Hplus1})].

Tables~\ref{table:advLIGOSNR}, \ref{table:LISASNR} show the SNR for the case
of maximal spins both aligned or anti-aligned with the orbital angular
momentum. From the bottom three rows of
Tables~\ref{table:advLIGOSNR}, \ref{table:LISASNR}, we see that, depending on
the spin orientation, the 2.5PN amplitude corrections with spins can have SNR
$\sim 10\%$ higher or lower than the 2.5PN amplitude corrections without spins.
We caution that this $\sim 10\%$ change in the SNR from spin corrections
is only meant as a bound on spin effects for spinning, \emph{non-precessing} 
binaries. As we have seen in Sec.~\ref{secIIIbis}, the affect of spin 
corrections on precessing binaries is not bounded by the cases of maximal 
spins aligned and anti-aligned with the orbital angular momentum.

\begin{table}[ht]
\centering
\begin{tabular}{|l| c | c | c | c |}
\hline
 & \multicolumn{4}{c|}{Advanced LIGO SNR} \\ 
 & \multicolumn{2}{c|}{$(50 + 5) M_\odot$}
& \multicolumn{2}{c|}{$(30 + 30) M_\odot$} \\ 
 & $(1,1)$ & $(-1,-1)$ & $(1,1)$ & $(-1,-1)$ \\ 
 & & & & \\ \hline 
Newt & 76.4& 76.4& 131.1& 131.1 \\
0.5PN & 84.9& 82.3& 131.1& 131.1 \\
1PN & 74.2& 71.9& 116.9& 115.8 \\
1PN + 1PN SO & 74.1& 72.1& 116.9& 115.8 \\
1.5PN & 69.2& 67.6& 116.9& 115.8\\
1.5PN + 1.5PN SO & 79.7& 58.1& 134.2& 98.8 \\
2PN + 1.5PN SO & 75.8& 55.1& 123.1& 88.3 \\
2PN + 1.5PN SO + 2PN SS & 75.5& 54.8& 121.4& 86.7 \\
2.5PN & 64.0 & 62.6& 106.3 & 105.3 \\
2.5PN + 1.5PN SO & 74.2& 53.6& 123.5& 88.5 \\
2.5PN + 1.5PN SO + 2PN SS & 73.9& 53.4& 121.7& 86.9 \\
\hline
\end{tabular}
\caption{For several binary configurations observable by Advanced LIGO we list 
the SNR as the PN order of the amplitude corrections is varied. 
In each column we show the component 
spins $(\boldsymbol{\chi}_1 \cdot \mathbf{\hat{L}}_{\rm N},\boldsymbol{\chi}_2 
\cdot \mathbf{\hat{L}}_{\rm N})$. 
We include all non-spinning, spin-orbit (SO) and spin-spin (SS) corrections 
up to the orders given in the first column. For example, 
2.5PN + 1.5PN SO + 2PN SS means we include non-spinning amplitude corrections 
from Newtonian to 2.5PN order, 1PN and 1.5PN SO corrections, and the 
2PN SS correction. Regardless of the PN order of the amplitude, we always use
the SPA phase with non-spinning terms up to 3.5PN order, and spin terms up to
2.5PN order, as given in Eqs.~(\ref{SPA})--(\ref{gamma}).
The binary is at a distance of 100 Mpc with orbital angular
momentum inclined relative to the line of sight by $\theta = \pi / 3$, 
sky location $\bar{\theta} = \bar{\phi} = \pi / 6$ 
and polarization angle $\bar{\psi} = \pi / 4$ 
[see Eqs.~(\ref{antennaresponse1}),(\ref{antennaresponse2})].
\label{table:advLIGOSNR}}
\end{table}

\begin{table}[ht]
\centering
\begin{tabular}{|l| c | c | c | c | c | c | c | c |}
\hline
 & \multicolumn{4}{c|}{LISA SNR}\\ 
 & \multicolumn{2}{c|}{$(2 \times 10^6 + 10^4) M_\odot$}
& \multicolumn{2}{c|}{$(10^6 + 10^6) M_\odot$}\\ 
 & $(1,1)$ & $(-1,-1)$ & $(1,1)$ & $(-1,-1)$ \\ 
 & & & & \\ \hline
Newt & 382.6& 382.6& 2764.4& 2764.4\\
0.5PN & 598.5& 598.6& 2764.4& 2764.4\\
1PN & 620.1& 621.5& 2510.0& 2469.7\\
1PN + 1PN SO & 619.9& 621.7& 2510.0& 2469.7\\
1.5PN & 512.2& 517.0& 2510.0& 2469.7\\
1.5PN + 1.5PN SO & 551.9& 484.1& 2875.9& 2118.2\\
2PN + 1.5PN SO & 523.6& 457.5& 2608.5& 1870.5\\
2PN + 1.5PN SO + 2PN SS & 523.5& 457.4& 2570.2& 1836.1\\
2.5PN & 479.4& 481.1& 2280.3& 2242.0\\
2.5PN + 1.5PN SO & 516.5& 451.6& 2639.2& 1901.9\\
2.5PN + 1.5PN SO + 2PN SS & 516.4& 451.6& 2601.4& 1868.4\\
\hline
\end{tabular}
\caption{For several binary configurations observable by LISA we list
the SNR as the PN order of the amplitude corrections is varied. In each 
column we show the component spins $(\boldsymbol{\chi}_1 \cdot 
\mathbf{\hat{L}}_{\rm N},\boldsymbol{\chi}_2 \cdot \mathbf{\hat{L}}_{\rm N})$. 
We include all non-spinning, spin-orbit (SO) and spin-spin (SS) corrections 
up to the orders given in the first column.
Regardless of the PN order of the amplitude, we always use the SPA phase with
non-spinning terms up to 3.5PN order, and spin terms  
up to 2.5PN order, as given in Eqs.~(\ref{SPA})--(\ref{gamma}).
The binary is at a distance of 3 Gpc with the same orientation as in
Table~\ref{table:advLIGOSNR}. 
The binary masses and distances refer to the redshifted quantities.
\label{table:LISASNR}}
\end{table}

Quite interestingly, the 1.5PN SO and 2PN SS corrections are far more important 
than the 1PN SO correction in terms of their effect on 
the power spectrum and the SNR. Notice that in Tables ~\ref{table:advLIGOSNR}, 
\ref{table:LISASNR} the 1PN SO term always has little or no effect, while  
the 1.5PN SO term changes the SNR by $\sim 10\%$, and the 2PN SS term changes 
the SNR for the equal-mass binary. The reason the 1PN SO term is less
important is that the 1.5PN SO and 2PN SS terms are corrections to the second 
harmonic, so they increase or decrease the power in the dominant term.
On the other hand, the 1PN SO term is a correction to the first harmonic. 
Thus, it is merely a perturbation to the dominant 
signal, and only in the lowest part of the spectrum where the first harmonic is 
observable. This is illustrated in Fig.~\ref{figure:PowerSpecZoom}, where we 
plot the power spectrum as a function of frequency (up to $2F_{\rm LSO}$) 
for different spin contributions for the $(30+30) M_\odot$ binary system. 
We see that the 1.5PN SO and 2PN SS corrections add or subtract 
their power coherently with the dominant second harmonic. 
Their net effect is to shift the power spectrum of the full waveform 
upward without changing its shape. On the other hand, the 1PN SO correction,
which is proportional to the (sine or cosine of) half the dominant harmonic, 
simply changes the modulation pattern of the full waveform up to 
$F_{\rm LSO}$ (37 Hz). It should however be noted that the structures in
the power spectra could be more complicated for asymmetric systems 
where the non-spinning terms proportional to $\cos \Psi$ and $\sin \Psi$ 
are not suppressed.

\begin{table}[ht]
\centering
\begin{tabular}{|l| c | c | c | c |}
\hline
 & \multicolumn{4}{c|}{Advanced LIGO SNR}\\
 & \multicolumn{4}{c|}{$(60 + 40) M_\odot$}\\
 & $(1,-1)$ & $(0.8,-0.8)$ & $(0.5,-0.5)$ & $(0.2,-0.2)$ \\ 
 & & & & \\ \hline 
2.5PN & 81.0& 80.5& 80.8& 81.8\\
2.5PN + 1.5PN SO & 84.4& 83.3& 82.5& 82.5\\
2.5PN + 1.5PN SO + 2PN SS & 85.8& 84.2& 82.8& 82.5\\
\hline
\end{tabular}
\caption{For a typical binary observable by Advanced LIGO, we compare the SNR
obtained using the 2.5PN amplitude corrected waveform without spin effects,
with spin-orbit effects, and with spin-orbit and spin-spin effects. In each
column we show the component spins $(\boldsymbol{\chi}_1 \cdot
\mathbf{\hat{L}}_{\rm N},\boldsymbol{\chi}_2 \cdot \mathbf{\hat{L}}_{\rm N})$. 
In all cases we use the SPA phase with non-spinning terms up to 3.5PN order, and
spin terms up to 2.5PN order, as given in Eqs.~(\ref{SPA})--(\ref{gamma}). 
The binary is at a distance of 100 Mpc with the same orientation as in
Table~\ref{table:advLIGOSNR}. 
\label{table:advLIGOVarySpin}}
\end{table}

The 1.5PN SO term is typically the most important of the spin terms. This term
is linearly proportional to the spins of the two bodies, as can be seen in
Eq.~(\ref{b8}). If the spins are aligned with the orbital angular momentum it
increases the SNR. If the spin terms are anti-aligned with the orbital
angular momentum it decreases the SNR. If one spin is aligned with
$\mathbf{\hat{L}}_{\rm N}$ and the other anti-aligned, the body with the
greater spin $\mathbf{S}_i = m_i^2\, \boldsymbol{\chi}_i$, which is typically
the larger body, dominates. Thus, the large body dictates whether the SO
effect increases or decreases the SNR, unless the spin of the smaller body is
much greater than the spin of the large body. This is illustrated in Table
~\ref{table:LISAVarySpin}, where the mass ratio $m_1:m_2= 10:1$. The spin of
the larger body is aligned with $\mathbf{\hat{L}}_{\rm N}$ and tends to
increase the SNR while the spin of the smaller body is anti-aligned with
$\mathbf{\hat{L}}_{\rm N}$ and tends to decrease the SNR. For a spin ratio
$\chi_1:\chi_2 = 1:1$ there is a large increase in SNR due the larger BH. For
a spin ratio 1:10, the larger BH still dominates and we get a small increase
in SNR. For the spin ratios of 1:100 and 1:1000, the smaller BH is now able to
overcome the larger BH and produce a net decrease in the SNR.

\begin{table}[ht]
\centering 
\begin{tabular}{|l| c | c | c | c |}
\hline
 & \multicolumn{4}{c|}{LISA SNR}\\
 & \multicolumn{4}{c|}{$(10^6 + 10^5) M_\odot$}\\ 
 & $(1,-1)$ & $(0.1,-1)$ & $(0.01,-1)$ & $(0.001,-1)$ \\ 
 & & & & \\ \hline
2.5PN & 2538.7& 2570.4& 2522.2& 2572.4\\
2.5PN + 1.5PN SO & 2917.5& 2583.8& 2500.6& 2546.7\\
2.5PN + 1.5PN SO + 2PN SS & 2930.4& 2585.0& 2500.8& 2546.7\\
\hline
\end{tabular}
\caption{For a typical binary observable by LISA, we compare the SNR obtained
using the 2.5PN waveform without spin effects, with spin-orbit effects, and
with spin-orbit and spin-spin effects. In each column we
show the component spins $(\boldsymbol{\chi}_1 \cdot \mathbf{\hat{L}}_{\rm N},
\boldsymbol{\chi}_2 \cdot \mathbf{\hat{L}}_{\rm N})$. 
In all cases we use the SPA phase with non-spinning terms up to 3.5PN order, and
spin terms up to 2.5PN order, as given in Eqs.~(\ref{SPA})--(\ref{gamma}). 
The binary is at a distance of 3 Gpc with the same orientation as in
Table~\ref{table:advLIGOSNR}. 
The binary masses and distances refer to the redshifted quantities.
\label{table:LISAVarySpin}}
\end{table}

The 2PN SS term decreases the power spectrum and SNR when the 
component spins are aligned with one another, and increases the 
power spectrum and SNR when they are anti-aligned with one another. 
The 2PN SS term has a greater effect on the SNR and 
power spectrum than the 1PN SO term, but is less important than the 
1.5PN SO term. This is because it is suppressed relative to the 
1.5PN SO term by a factor of $v/c$ and it is quadratic in the spins 
and proportional to the symmetric mass-ratio $\nu$. 
Thus, the 2PN SS term are most important for binaries 
with two large component spins and comparable masses. 
From Tables ~\ref{table:advLIGOSNR}, \ref{table:advLIGOVarySpin}, 
we can see that the 2PN SS term has little or 
no effect on binaries with a mass ratio greater than 10:1. In Tables 
~\ref{table:advLIGOSNR}, \ref{table:LISASNR}, for the columns with equal 
masses and spins aligned with one another, the 2PN SS term decreases the SNR 
by a few percent. For the binary in Table~\ref{table:advLIGOVarySpin}, 
we see that the 2PN SS term increases the SNR by an amount comparable to the 
SO terms when the spins are maximal. As we decrease the spin magnitude, the 
SS effect is suppressed faster than the SO effect because it is quadratic in 
the spins while the SO effect is linear. 

Before ending this section we study how different values of the source
position and inclination angle can affect the SNR trends shown in
Table~\ref{table:advLIGOSNR}. For $(5 + 50) M_\odot$ and $(30 + 30)M_\odot$
systems we calculated the SNRs at different PN orders in amplitude for various
random realizations of $\bar\theta,\bar\phi,\bar\psi$ and ${\theta}$ and for
the spinning and non-spinning cases. For the spinning cases, when all the
known spin effects are included at different PN orders, the trends across
different orders remains the same for all the random realizations except
between the Newtonian and 0.5PN order. Though on most of the occasions, the
SNR increases from Newtonian to 0.5PN order, there are cases when it
decreases, albeit slightly. All these cases where the SNR decreases have
inclination angle $\theta$ very close to zero or $\pi$. For these cases, the
third harmonic, which is proportional to $\sin \theta$, is largely suppressed
and the spin-dependent interference accounts for the small drop in SNR. This
drop is observed for systems for which $\chi_1=\chi_2=-1$ whereas the
non-spinning and $\chi_1=\chi_2=1$ cases consistently showed the increase in
SNR between Newtonian and 0.5PN order. To further assert this, we fix the
inclination angle to a value very close to zero and $\pi$ and randomly varied
the other three angles. We find that for all the realizations the SNR
decreases in going from Newtonian to 0.5PN order. In brief, the trends shown
in Table~\ref{table:advLIGOSNR} is quite general except for inclination angles
close to zero or $\pi$. We however note that the trends of
Table~\ref{table:advLIGOSNR} need not be same for much higher masses when the
leading harmonic approaches the lower cut-off frequency of the detector
($2F_{\rm LSO}\simeq f_s$). We have not done a thorough analysis for the whole
mass range.

\section{Conclusions}
\label{concl}

The ongoing search for GWs from compact binaries with the network of 
interferometers LIGO, Virgo and GEO, and the work at the 
interface between analytical and numerical relativity aimed at 
providing accurate templates for the search, has made it urgent 
to include higher-order PN effects in the theoretical predictions of 
the waveforms. This paper is a step forward in this direction. 

We provided ready-to-use time-domain waveforms for spinning, precessing
binaries moving on nearly circular orbits through 1.5PN order and decompose
those waveforms in spin-weighted $-2$ spherical harmonics [see
Appendices~\ref{appB} and \ref{appD}]. Neglecting radiation-reaction effects
and assuming $S \ll L$, we found that the inclination angle $\iota$ between
the total angular momentum and the Newtonian orbital angular momentum (see
Fig.~\ref{figure:SourceFrame}) is a 0.5PN correction. Motivated by this, we
expanded the GW polarizations and spin-weighted spherical harmonic modes in a
Taylor series in $\iota$ [see Eqs.~\eqref{eq:Hplus_prec_iotaexpanded},
\eqref{eq:Hcross_prec_iotaexpanded} and Eqs.~(\ref{h22})--(\ref{h50})]. Their
expressions become much simpler and allow one to extract interesting physical
features of the gravitational waves from precessing binaries.

We found that, in contrast to what happens in the non-spinning case, the
$h_{\ell m}$'s are not in general proportional to $ e^{-i m \Psi}$. They also
depend on the angles $\iota$ and $\alpha$, where $\iota$ is the inclination
angle of the Newtonian orbital momentum relative to the total angular momentum
and $\alpha$ is the angle between the $x$-axis and the projection of the
Newtonian orbital angular momentum onto the $x\mbox{--}y$ plane (see
Fig.~\ref{figure:SourceFrame}). For example, the terms independent of $\iota$
are proportional to $e^{-i m (\Psi+\alpha)}$, the terms that are linear in
$\iota$ are proportional to $e^{-i (m+m') (\Psi + \alpha)} \cos \Psi $ or
$e^{-i (m+m') (\Psi + \alpha)}\sin \Psi$, while higher-order contributions in
$\iota$ involve terms of the form $e^{-i (m+m') (\Psi + \alpha)} \cos^a
\Psi\,\sin^b \Psi$, where $a,b \in \mathbb{N}$ and $m' \in {-1,0,1}$. In the
presence of precession, the angles $\iota$ and $\alpha$ vary in time and the
different harmonics present in each of the modes interfere, causing a strong 
modulation of the mode amplitudes. We also found that, in contrast to what 
happens in the non-spinning case, the signal can be largely distributed among
modes $(\ell,m)$ other than the $(2,2)$ mode. With our choice of the source
frame, when spins are maximal and the binary system has significant mass 
asymmetry and/or a large inclination angle, we found that the amplitude of the 
$(2,0)$ and $(2,1)$ modes can be comparable to the amplitude of the $(2,2)$ 
mode, especially during the last stages of inspiral. For the mass ratios we 
considered, we found that the $\ell=3$ and $\ell=4$ modes are 
generally one or two orders of magnitude smaller than the $\ell = 2$ modes. 
These results are summarized in Figs.~\ref{figure:h22}, \ref{figure:Leq2} and
\ref{figure:Leq34}, for binaries with mass ratio 1:1 and 4:1, and for two
maximal spin configurations having a small or large inclination angle $\iota$.
The ready-to-use time-domain waveforms for spinning, precessing binaries can
be employed for accurate comparisons with numerical simulations of binary BHs
~\cite{Campanelli2006c,Campanelli2006d,Campanelli2007b,Herrmann2007,
Herrmann2007c,HHBG07,CLNZ08} and for designing
time-domain~\cite{BuonEOB07,Damour2007a,DN2008,UMD-CC08} analytical templates.

Restricting ourselves to spinning, non-precessing binaries, we computed
ready-to-use frequency-domain waveforms in the stationary-phase approximation.
We derived 1PN and 1.5PN order spin-orbit effects, and 2PN order spin-spin
(spin(1)-spin(2) only) effects in the frequency-domain GW amplitude [see Eq.
(\ref{Cceff}), and Eqs.(\ref{b1})--(\ref{b21}) in Appendix~\ref{appC}]. We
also calculated the 2PN spin-spin (\emph{including} spin(1)-spin(1) and
spin(2)-spin(2) effects), and the 2.5PN order spin-orbit effects in the
frequency-domain GW phase [see Eqs.~(\ref{SPA}), (\ref{gamma}) and
(\ref{sigma})]. For the 2PN spin-spin terms, we found agreement with
Refs.~\cite{MVG05,RBK08}. We wrote the frequency-domain waveforms in a rather
compact way, so that they can be easily used for data analysis and for
building analytical frequency-domain~\cite{AjithNR07a,Pan07comparison}
templates.

In the non-precessing case, we found that, through 2PN order, spin effects in
the amplitude affect only the PN corrections to the first and second
harmonics. Thus, through 2PN order, spin effects do not {\it yet} extend the
mass reach of GW detectors. However, as seen in
Figs.~\ref{figure:PowerSpecLISAAdvLIGO}, \ref{figure:PowerSpecZoom}, they can
interfere with other harmonics and, depending on the spin orientation, lower
or raise the signal-to-noise ratio of ground-based (see
Tables~\ref{table:advLIGOSNR}, \ref{table:advLIGOVarySpin}) and space-based
detectors (see Tables~\ref{table:LISASNR}, \ref{table:LISAVarySpin}). We also
expect that those spin terms will help in localizing the binary source in the
sky. We leave to a future publication the use of the waveforms derived in this
paper to extend parameter-estimation predictions~\cite{AISS05,
  SinVecc00a,BBW05a,Chris06,ChrisAnand06,
  ChrisAnand06b,SinVecc00b,MH02,HM03,AISS07,AISSV07,TriasSintes07,Vecchio04,
  LangHughes06,LangHughes07, Sluys07MCMC,BHHS} of ground-based and space-based
detectors to spinning, precessing binaries.

Finally, we notice that the gravitational polarizations computed in this paper do not 
include the modification of the orbital phase evolution at the relative 2.5PN order 
induced by the flow of energy into the black hole horizons as explicitly computed 
in Ref.~\cite{KA}. As summarized in Table IV of Ref.~\cite{KA}, this effect can cause 
a variation of the number of GW cycles at the Schwarzschild ISCO 
of $3\% \mbox{--} 24 \%$ depending on the binary mass ratio. We postpone to a future 
publication the inclusion of those effects.

\begin{acknowledgments}
We thank Luc Blanchet, Bala Iyer, Yi Pan and Etienne Racine for useful discussions. 
K.G.A is a VESF fellow of the European Gravitational Observatory (EGO). 
A.B. and E.O. acknowledge support from NSF grant PHY-0603762, 
and A.B. also acknowledges support from the Alfred P Sloan Foundation. 
\end{acknowledgments}

\appendix

\section{Ready-to-use gravitational-wave polarizations for precessing binaries
on nearly circular orbits through 1.5PN order: generic inclination angles}
\label{appB}

In Sec.~\ref{secII} we wrote the GW polarizations 
\begin{equation}
h_{+,\times}\ =\
\frac{2\,M\,\nu\,v^2}{D_L}\,\left[\hat{h}_{+,\times}^{(0)}+ \left(\hat{h}_{+,\times}^{(1/2)} +
\hat{h}_{+,\times}^{(1/2,{\rm SO})}\right)\,v +  \left(\hat{h}_{+,\times}^{(1)} 
+ \hat{h}_{+,\times}^{(1,{\rm SO})}\right)\,v^2 +
\left(\hat{h}_{+,\times}^{(3/2)} + \hat{h}_{+,\times}^{(3/2,{\rm SO})}\right)\,v^3  \right]\,,
\label{hv}
\end{equation}
expanded in the inclination angle $\iota$. Here we give the full expressions.  
The Newtonian, 0.5PN and 1PN order coefficients were computed explicitly 
in Ref.~\cite{K95}, the 1.5PN order coefficients, are 
computed for the first time in this paper. They read
\begin{subequations} \label{eq:Hplus_prec}

\end{subequations}
where $s_X$ and $c_X$ are shorthand for $\sin X$ and $\cos X$, respectively, 
with $X = \theta, \iota, \dots$. 

\section{Gravitational-wave modes for precessing binaries on nearly circular
orbits  through 1.5PN order: generic inclination angles}
\label{appD}

In Sec.~\ref{secIII} we wrote the gravitational-wave modes expanded in the 
inclination angle $\iota$. Here we give the full expression of the ${h}_{22}$, 
${h}_{33}$ and ${h}_{21}$ modes. They read 

\begin{eqnarray}
  h_{22}&=& -\frac{2\,M\,\nu\,v^2}{D_L}\,e^{-2i\,(\alpha+\Psi)}\frac{3}{2}
  \sqrt{\frac{\pi }{5}} \left\{\frac{1}{6} e^{-2 i \iota } \left(e^{4 i \Psi }
      \left(-1+e^{i \iota}\right)^4+\left(1+e^{i \iota
        }\right)^4\right)+\frac{1}{9} i e^{-i (2 \iota -\Psi )} \left[e^{2 i
        \Psi } \left(-1+e^{i \iota}\right)^3 \left(1+e^{i \iota}
      \right)\right.\right.\nonumber\\  
  &&-\left.\left.\left(-1+e^{i \iota }\right) \left(1+e^{i
          \iota}\right)^3\right] v \,\delta \right.\nonumber\\  
  &&+\left.v^2
    \left[\frac{1}{252} e^{-2 i \iota } \left(-107 e^{4 i \Psi } \left(-1+e^{i
            \iota }\right)^4-107 \left(1+e^{i \iota 
          }\right)^4\right)+\frac{55}{252} e^{-2 i \iota } \left(e^{4 i \Psi }
        \left(-1+e^{i \iota }\right)^4+\left(1+e^{i \iota}\right)^4\right)
      \eta\right.\right. \nonumber\\ 
  &&+\left.\left.\frac{1}{3} e^{i (\alpha -\iota +\Psi )} \left(e^{2 i \Psi }
        \left(-1+e^{i \iota}\right)^2-\left(1+e^{i \iota }\right)^2\right)
      \chi _a^x+\frac{1}{3} e^{i (\alpha -\iota +\Psi )} \left(e^{2 i \Psi} 
        \left(-1+e^{i \iota }\right)^2-\left(1+e^{i \iota }\right)^2\right)
      \delta  \chi _s^x\right.\right.\nonumber\\ 
  &&+\left.\left.i \left(\frac{1}{3} e^{i (\alpha -\iota
          +\Psi )} \left(-e^{2 i \Psi } \left(-1+e^{i \iota
            }\right)^2+\left(1+e^{i \iota }\right)^2\right) \chi
        _a^y+\frac{1}{3} e^{i (\alpha -\iota +\Psi )} \left(-e^{2 i \Psi }
          \left(-1+e^{i \iota }\right)^2+\left(1+e^{i \iota }\right)^2\right)
        \delta  \chi_s^y\right)\right]\right.\nonumber\\
  &&+\left.v^3 \left[\frac{1}{3} e^{-2 i \iota } \left(e^{4 i \Psi }
        \left(-1+e^{i \iota }\right)^4+\left(1+e^{i \iota}\right)^4\right) \pi
      +\delta  \left(\frac{1}{36} e^{-i (\alpha +3 \iota )} \left(-5
          \left(-1+e^{i \iota }\right) \left(1+e^{i \iota}
          \right)^5\right.\right.\right.\right.\nonumber\\  
  &&+\left.\left.\left.\left.e^{2 i \alpha } \left(1+e^{i \iota }\right)^3
          \left(-5+7 e^{i \iota }-7 e^{2 i \iota }+5 e^{3 i \iota} \right) +
          e^{4 i \Psi } \left(-1+e^{i \iota }\right)^3 \left(1+e^{i
              \iota}\right) \left(-5+5 e^{2 i \alpha }+5 e^{2 i \iota }   
     \left(-1+e^{2 i \alpha
       }\right)\right.\right.\right.\right.\right.\nonumber\\ 
  &&+\left.\left.\left.\left.\left.2 e^{i \iota } \left(5+e^{2 i \alpha}
            \right)\right)+6 e^{2 i \Psi } \left(\left(-1+e^{2 i \iota}
            \right)^3 - e^{2 i \alpha } \left(-1-5 e^{2 i \iota }+5 e^{4 i
                \iota }+e^{6 i \iota }\right)\right)\right)
        \chi_a^y\right.\right.\right.\nonumber\\ 
  && + \left.\left.\left.\frac{1}{18} e^{-3 i \iota } \left(6 e^{2 i \Psi } 
    \left(-1+e^{2 i \iota }\right)^2 \left(1+e^{2 i \iota} \right) - \left(1 +
      e^{i \iota }\right)^4 \left(5-6 e^{i \iota }+5 e^{2 i \iota}\right)
  \right.\right.\right.\right.\nonumber\\  
  &&-\left.\left.\left.\left.e^{4 i \Psi } \left(-1+e^{i \iota} \right)^4
          \left(5+6 e^{i \iota }+5 e^{2 i \iota }\right)\right) \chi
        _a^z\right)\right.\right.\nonumber\\ 
  && +\left.\left.\frac{1}{36} e^{-i (\alpha +3 \iota )}
      \left(-5 \left(-1+e^{i \iota }\right) \left(1+e^{i \iota }\right)^5
     +e^{2 i \alpha } \left(1+e^{i \iota }\right)^3 \left(-5+7 
          e^{i \iota }-7 e^{2 i \iota }+ 5 e^{3 i \iota}
        \right)\right.\right.\right. \nonumber\\ 
  &&+\left.\left.\left.e^{4 i \Psi } \left(-1+e^{i \iota }\right)^3 \left(1 +
          e^{i \iota }\right) 
        \left(-5+5 e^{2 i \alpha }+5 e^{2 i \iota } \left(-1 + e^{2 i \alpha}
          \right) + 2 e^{i \iota } \left(5+e^{2 i \alpha} \right) \right) + 6
        e^{2 i \Psi } \left(\left(-1+e^{2 i \iota}\right)^3 \right.
      \right.\right.\right.\nonumber\\  
  &&-\left.\left.\left.\left.e^{2 i \alpha } \left(-1-5 e^{2 i \iota }+5 e^{4 i
              \iota } + e^{6 i \iota } \right) \right) \right) \chi _s^y +
      \frac{1}{18} e^{-3 i \iota } \left(6 e^{2 i \Psi } \left(-1 + e^{2 i \iota
          }\right)^2 \left(1 + e^{2 i \iota } \right) - \left( 1 + e^{i \iota
          } \right)^4 \left( 5 - 6 e^{i \iota } + 5 e^{2 i \iota } \right)
      \right. \right.\right.\nonumber\\
  &&-\left.\left.\left.e^{4 i
          \Psi } \left(-1+e^{i \iota }\right)^4 \left(5+6 e^{i \iota } + 5
          e^{2 i \iota } \right)\right) \chi _s^z + i \left(\delta 
        \left( \frac{1}{252} e^{-i (2 \iota -\Psi )} \left(-17 e^{2 i \Psi }
            \left(-1 + e^{i \iota }\right)^3 \left( 1 + e^{i \iota
              } \right) \right. \right. \right.\right.\right.\nonumber\\
  && + \left.\left.\left.\left. \left. 17 \left( -1 + e^{i \iota } \right)
            \left(1 + e^{i \iota }\right)^3 \right) + \frac{1}{63} e^{-i (2
            \iota -\Psi)} \left(5 
           e^{2 i \Psi } \left(-1 + e^{i \iota } \right)^3 \left(1 + e^{i
               \iota }\right) - 5 \left(-1 + e^{i \iota }\right) \left(1 +
             e^{i \iota 
              }\right)^3\right) \eta \right.\right. \right.\right.\nonumber\\
  &&+\left.\left.\left.\left.\frac{1}{36} e^{-i (\alpha +3 \iota )} \left(5
            \left(-1 + e^{i \iota }\right) \left(1+e^{i \iota
              }\right)^5 + e^{2 i \alpha } \left(1+e^{i \iota }\right)^3
            \left(-5+7 e^{i \iota }-7 e^{2 i \iota } + 5 e^{3 i \iota
              }\right)\right. \right.\right.\right.\right.\nonumber\\
  &&+\left.\left.\left.\left.\left.e^{4
              i \Psi } \left(-1 + e^{i \iota }\right)^3 \left(1 + e^{i \iota
              }\right) \left(5 \left(-1 + e^{i \iota }\right)^2 + e^{2 i
                \alpha } 
              \left(5 + 2 e^{i \iota } + 5 e^{2 i \iota }\right)\right) + 6
            e^{2 i \Psi } \left(-\left(-1 + e^{2 i \iota }\right)^3 \right.
          \right. \right.\right.\right.\right.\nonumber\\
  &&-\left.\left.\left.\left.\left.\left.e^{2 i \alpha }
              \left(-1-5 e^{2 i \iota } + 5 e^{4 i \iota } + e^{6 i \iota
                }\right) \right) \right) \chi _a^x \right) + \frac{1}{36}
        e^{-i (\alpha +3 
          \iota )} \left(5 \left(-1 + e^{i \iota }\right) \left(1 + e^{i \iota
            }\right)^5 \right.\right.\right.\right.\nonumber\\
  &&+ \left.\left.\left.\left. e^{2 i \alpha } \left(1 + e^{i \iota } \right)^3
          \left(-5 + 7 e^{i \iota } - 7 e^{2 i \iota } + 5 e^{3 i \iota
            }\right) + e^{4 i \Psi } \left(-1+e^{i \iota }\right)^3 \left(1 +
            e^{i \iota }\right)\right.\right. \right. \right.\nonumber\\
  &&\times \left.\left.\left.\left. \left(5 \left(-1+e^{i \iota }\right)^2 +
            e^{2 i \alpha } \left(5 + 2 e^{i \iota }+5 e^{2 i \iota } \right)
          \right) \right.\right.\right.\right.\nonumber\\
  &&+\left.\left.\left.\left. 6
          e^{2 i \Psi } \left(-\left(-1 + e^{2 i \iota }\right)^3 - e^{2 i
              \alpha } \left(-1 - 5 e^{2 i \iota }+5 e^{4 i \iota }+e^{6 i \iota
              }\right) \right) \right) \chi _s^x + \frac{1}{72} e^{-i (\alpha
          +3 \iota )} \left(\left(-1 + e^{i \iota }\right) \left(1 + e^{i \iota
            }\right)^3 \left(1 + e^{2 i \alpha } \right. \right. \right.
    \right. \right.\nonumber\\
  &&+ \left.\left.\left. \left. \left. e^{2 i \iota } \left(1 + e^{2 i \alpha
              }\right) - 2 e^{i \iota } \left(-1+9 e^{2 i \alpha
              }\right)\right) + e^{4 i \Psi } \left(-1 + e^{i \iota }\right)^3
          \left(1 + e^{i \iota }\right) \right. \right. \right.
  \right.\nonumber\\ 
  &&\times\left.\left.\left. \left. \left(1 + e^{2 i \alpha } + e^{2 i \iota
            } \left(1 + e^{2 i \alpha }\right) + 2 e^{i \iota } \left(-1 + 9
              e^{2 i \alpha } \right)\right) \right.\right.\right.
  \right. \nonumber\\ 
  &&+ \left.\left.\left. \left. 6 e^{2 i \Psi } \left(\left(-1 + e^{2 i
                \iota }\right)^3 + e^{2 i \alpha } \left(-1 - 5 e^{2 i \iota }
              + 5 e^{4 i \iota }+e^{6 i \iota } \right)\right)\right) \eta  \chi
        _s^x \right) + \eta  \left(\frac{1}{72} e^{-i (\alpha +3 \iota )}
        \left(-\left(-1+e^{i \iota } \right) \left(1 + e^{i \iota
            } \right)^5 \right.\right.\right.\right. \nonumber\\
  &&+\left.\left.\left.\left. e^{2 i \alpha } \left(1 + e^{i \iota }\right)^3
          \left(-1 + 19 e^{i \iota } - 19 e^{2 i \iota } + e^{3 i \iota
            }\right) + e^{4
            i \Psi } \left(-1 + e^{i \iota }\right)^3 \left(1 + e^{i \iota
            }\right) \left(-1 + e^{2 i \alpha } + e^{2 i \iota } \left(-1 +
  e^{2 i \alpha }\right) \right.\right.\right.\right.\right.\nonumber\\
  &&+ \left.\left.\left.\left.\left. 2 e^{i \iota } \left(1 + 9 e^{2 i \alpha
              } \right)\right) + 6 e^{2 i \Psi } \left( -\left(-1 + e^{2 i \iota
              }\right)^3 + e^{2 i \alpha } \left(-1 - 5 e^{2 i \iota } + 5
              e^{4 i \iota } + e^{6 i \iota }\right)\right)\right) \chi
        _s^y\right.\right.\right.\nonumber\\
  &&+\left.\left.\left.\frac{1}{36} e^{-3 i \iota } \left(-6 e^{2 i \Psi }
          \left(-1 + e^{2 i \iota }\right)^2 \left(1+e^{2 i \iota
            }\right)-\left(1 + e^{i \iota }\right)^4 \left(1 - 10 e^{i \iota
            }+e^{2 i \iota }\right)\right.\right.\right.\right.\nonumber\\ 
  &&-\left.\left.e^{4 i \Psi } \left(-1+e^{i \iota
        }\right)^4 \left(1+10 e^{i \iota } + e^{2 i \iota }\right)\right) \chi
    _s^z \right)\biggr] + {\cal O}(v^4) \biggr\} \, ,
\end{eqnarray}

\begin{eqnarray}
h_{21}&=&-\frac{2\,M\,\nu\,v^2}{D_L}\,e^{ - i\,(\alpha+\Psi)}
i \sqrt{\frac{\pi }{5}} \left\{\frac{1}{2} e^{ - i (2 \iota + \Psi )} \left( -
    e^{4 i \Psi } \left( - 1 + e^{i \iota }\right)^3\left(1 + e^{i \iota
      }\right) - \left( - 1 + e^{i \iota }\right) \left(1 + e^{i \iota
      }\right)^3\right)\right. \nonumber\\ 
&& + \left. \frac{1}{3} i e^{ - 2 i \iota }   \biggl[1 + e^{i \iota } + e^{3 i
    \iota } + e^{4 i \iota } 
 - e^{2 i \Psi } \left( - 1 + e^{i \iota }\right)^2 \left(1 + e^{i \iota } +
   e^{2 i\iota }\right)\biggr] v \delta \right. \nonumber\\ 
&& + \left. v^2 \biggl[\frac{107}{84}   e^{ - i (2 \iota + \Psi )} \left( - 1
    + e^{i \iota }\right) \left(1 + e^{i \iota }\right) \left(e^{4 i \Psi }
    \left( - 1 + e^{i \iota   }\right)^2 + \left(1 + e^{i \iota
      }\right)^2\right) + \frac{1}{84} e^{ - i (2 \iota + \Psi )} \left( - 55
    e^{4 i \Psi } \left( - 1 + e^{i   \iota }\right)^3 \right. \right.
\nonumber\\ 
&&\times\left. \left. \left. \left(1 + e^{i \iota }\right) - 55 \left( - 1 +
        e^{i \iota }\right) \left(1 + e^{i \iota }\right)^3\right) \eta   -
    \frac{1}{2} e^{i (\alpha - \iota )} \left( - 1 + e^{2 i \iota }\right)
    \left( - 1 + e^{2 i \Psi }\right) \chi _a^x\right. \right. \nonumber\\ 
&& - \left. \left. \frac{1}{2} e^{i   (\alpha - \iota )} \left( - 1 + e^{2 i
        \iota }\right) \left( - 1 + e^{2 i \Psi }\right) \delta  \chi _s^x + i
    \left(\frac{1}{2} e^{i   (\alpha - \iota )} \left( - 1 + e^{2 i \iota
        }\right) \left( - 1 + e^{2 i \Psi }\right) \chi _a^y\right. \right.
\right. \nonumber\\ 
&& + \left. \left. \left. \frac{1}{2} e^{ - i \iota } \left(e^{2   i \Psi }
        \left( - 1 + e^{i \iota }\right)^2 - \left(1 + e^{i \iota
          }\right)^2\right) \chi _a^z + \delta  \left(\frac{1}{2} e^{i (\alpha
          - \iota )} \left( - 1 + e^{2 i \iota }\right) \left( - 1 + e^{2 i
            \Psi }\right) \chi _s^y\right. \right. \right. \right. \nonumber\\ 
&&\left. \left. \left.  + \frac{1}{2} e^{ - i \iota } \left(e^{2 i \Psi   }
        \left( - 1 + e^{i \iota }\right)^2 - \left(1 + e^{i \iota
          }\right)^2\right) \chi _s^z\right)\right)\biggr]\right. \nonumber\\ 
&& + \left. v^3 \biggl[e^{ - i (2 \iota + \Psi )} \left( - e^{4 i \Psi }
    \left( - 1 + e^{i \iota }\right)^3\left(1 + e^{i \iota }\right) \pi -
    \left( - 1 + e^{i \iota }\right) \left(1 + e^{i \iota }\right)^3 \pi
  \right)\right. \nonumber\\ 
&& + \left. \left. \delta\left(\frac{1}{12} e^{ - i (\alpha + 3 \iota + \Psi
        )} \left(\left(1 + e^{i \iota }\right)^4 \left(5 - 7 e^{i \iota } + 5
          e^{2 i \iota   }\right) - e^{2 i \alpha } \left( - 1 + e^{2 i \iota
          }\right)^2 \left(5 + e^{i \iota } + 5 e^{2 i \iota }\right)\right.
    \right. \right. \right. \nonumber\\ 
&& + \left. \left. \left. \left. 6 e^{2 i \Psi }\left(1 + e^{2 i \iota
          }\right) \left( - \left( - 1 + e^{2 i \iota }\right)^2\right.
      \right. \right. \right. \right. \nonumber\\ 
&& + \left. \left. \left. \left. \left. e^{2 i \alpha } \left(1 + e^{2 i
              \iota}\right)^2\right) + e^{4 i \Psi } \left( - e^{2 i \alpha }
          \left( - 1 + e^{2 i \iota }\right)^2 \left(5 - e^{i \iota } + 5 e^{2
              i \iota}\right) + \left( - 1 + e^{i \iota }\right)^4 \left(5 + 7
            e^{i \iota } + 5 e^{2 i \iota }\right)\right)\right) \right.
  \right. \right. \chi _a^y\nonumber\\ 
&& + \left. \left. \left. \frac{1}{6}e^{ - i (3 \iota + \Psi )} \left(\left(1
          + e^{i \iota }\right)^3 \left( - 5 + 8 e^{i \iota } - 8 e^{2 i \iota
          } + 5 e^{3 i \iota}\right) + e^{4 i \Psi } \left( - 1 + e^{i \iota
          }\right)^3 \left(5 + 8 e^{i \iota } + 8 e^{2 i \iota } + 5 e^{3 i
            \iota }\right)\right. \right. \right. \right. \nonumber\\ 
&& - \left. \left. \left. \left. 6 e^{2i \Psi } \left( - 1 + e^{2 i \iota } -
          e^{4 i \iota } + e^{6 i \iota }\right)\right) \chi _a^z\right) +
    \frac{1}{12} e^{ - i (\alpha + 3\iota + \Psi )} \left(\left(1 + e^{i \iota
        }\right)^4 \left(5 - 7 e^{i \iota } + 5 e^{2 i \iota }\right)\right.
  \right. \right. \nonumber\\ 
&& - \left. \left. \left. e^{2 i \alpha }\left( - 1 + e^{2 i \iota }\right)^2
      \left(5 + e^{i \iota } + 5 e^{2 i \iota }\right) + 6 e^{2 i \Psi }
      \left(1 + e^{2 i \iota }\right) \left( - \left( - 1 + e^{2 i \iota
          }\right)^2 + e^{2 i \alpha } \left(1 + e^{2 i \iota
          }\right)^2\right)\right. \right. \right. \nonumber\\ 
&& + \left. \left. \left. e^{4 i \Psi } \left( - e^{2 i 
   \alpha } \left( - 1 + e^{2 i \iota }\right)^2 \left(5 - e^{i \iota } + 5
   e^{2 i \iota }\right) + \left( - 1 + e^{i \iota }\right)^4 \left(5 + 7 e^{i
     \iota } + 5 e^{2 i \iota }\right)\right)\right) \chi _s^y\right. \right.
\nonumber\\ 
&& + \left. \left. \frac{1}{6} e^{ - i (3 \iota + \Psi )} \left(\left(1 + e^{i
          \iota }\right)^3 \left( - 5 + 8 e^{i \iota } - 8 e^{2 i \iota } + 5
        e^{3 i \iota }\right) + e^{4 i \Psi } \left( - 1 + e^{i \iota
        }\right)^3 \left(5 + 8 e^{i \iota } + 8 e^{2 i \iota } + 5 e^{3 i
          \iota }\right)\right. \right. \right. \nonumber\\ 
&& - \left. \left. \left. 6 e^{2 i \Psi } \left( - 1 + e^{2 i \iota } - e^{4 i
          \iota } + e^{6 i \iota }\right)\right) \chi _s^z + i \left(\delta
      \left(\frac{1}{84} e^{ - 2 i \iota } \left( - 17 + 17 e^{2 i \Psi }
          \left( - 1 + e^{i \iota }\right)^2 \left(1 + e^{i \iota } + e^{2 i
              \iota }\right)\right. \right. \right. \right. \right.
\nonumber\\ 
&& - \left. \left. \left. \left. \left. 17 e^{i \iota } \left(1 + e^{2 i \iota
            } + e^{3 i \iota   }\right)\right) + \frac{5}{21} e^{ - 2 i \iota
        } \left(1 + e^{i \iota } + e^{3 i \iota } + e^{4 i \iota } - e^{2 i
            \Psi } \left( - 1 + e^{i   \iota }\right)^2 \left(1 + e^{i \iota }
            + e^{2 i \iota }\right)\right) \eta \right. \right. \right.
\right. \nonumber\\ 
&& + \left. \left. \left. \left. \frac{1}{12} e^{ - i (\alpha + 3 \iota + \Psi
          )}   \left( - \left(1 + e^{i \iota }\right)^4 \left(5 - 7 e^{i \iota
            } + 5 e^{2 i \iota }\right) - e^{2 i \alpha } \left( - 1 + e^{2 i
              \iota   }\right)^2 \left(5 + e^{i \iota } + 5 e^{2 i \iota
            }\right)\right. \right. \right. \right. \right. \nonumber\\ 
&& + \left. \left. \left. \left. \left. 6 e^{2 i \Psi } \left(1 + e^{2 i \iota
            }\right) \left(\left( - 1 + e^{2 i   \iota }\right)^2 + e^{2 i
              \alpha } \left(1 + e^{2 i \iota }\right)^2\right) + e^{4 i \Psi
          } \left( - e^{2 i \alpha } \left( - 1 + e^{2 i   \iota }\right)^2
            \left(5 - e^{i \iota } + 5 e^{2 i \iota }\right)\right. \right.
      \right. \right. \right. \right. \nonumber\\ 
&& - \left. \left. \left. \left. \left. \left. \left( - 1 + e^{i \iota
              }\right)^4 \left(5 + 7 e^{i \iota } + 5 e^{2 i   \iota
              }\right)\right)\right) \chi _a^x\right) + \frac{1}{12} e^{ - i
        (\alpha + 3 \iota + \Psi )} \left( - \left(1 + e^{i \iota   }\right)^4
        \left(5 - 7 e^{i \iota } + 5 e^{2 i \iota }\right)\right. \right.
  \right. \right. \nonumber\\ 
&& - \left. \left. \left. \left. e^{2 i \alpha } \left( - 1 + e^{2 i \iota
          }\right)^2 \left(5 + e^{i \iota   } + 5 e^{2 i \iota }\right) + 6
        e^{2 i \Psi } \left(1 + e^{2 i \iota }\right) \left(\left( - 1 + e^{2
              i \iota }\right)^2 + e^{2 i \alpha   } \left(1 + e^{2 i \iota
            }\right)^2\right)\right. \right. \right. \right. \nonumber\\ 
&& + \left. \left. \left. \left. e^{4 i \Psi } \left( - e^{2 i \alpha } \left(
            - 1 + e^{2 i \iota }\right)^2 \left(5 - e^{i   \iota } + 5 e^{2 i
              \iota }\right) - \left( - 1 + e^{i \iota }\right)^4 \left(5 + 7
            e^{i \iota } + 5 e^{2 i \iota }\right)\right)\right)   \chi
      _s^x\right. \right. \right. \nonumber\\ 
&& + \left. \left. \left. \frac{1}{24} e^{ - i (\alpha + 3 \iota + \Psi )}
      \left( - e^{2 i \alpha } \left( - 1 + e^{2 i \iota }\right)^2 \left(1 -
          13   e^{i \iota } + e^{2 i \iota }\right) - \left(1 + e^{i \iota
          }\right)^4 \left(1 + 3 e^{i \iota } + e^{2 i \iota }\right)\right.
    \right. \right. \right. \nonumber\\ 
&& + \left. \left. \left. \left. 6 e^{2 i \Psi }   \left(1 + e^{2 i \iota
          }\right) \left( - \left( - 1 + e^{2 i \iota }\right)^2 - e^{2 i
            \alpha } \left(1 + e^{2 i \iota   }\right)^2\right) + e^{4 i \Psi
        } \left( - \left( - 1 + e^{i \iota }\right)^4 \left(1 - 3 e^{i \iota }
            + e^{2 i \iota }\right)\right. \right. \right. \right. \right.
\nonumber\\ 
&& - \left. \left. \left. \left. \left. e^{2 i   \alpha } \left( - 1 + e^{2 i
              \iota }\right)^2 \left(1 + 13 e^{i \iota } + e^{2 i \iota
            }\right)\right)\right) \eta  \chi   _s^x\right) + \eta
    \left(\frac{1}{24} e^{ - i (\alpha + 3 \iota + \Psi )} \left( - e^{2 i
          \alpha } \left( - 1 + e^{2 i \iota }\right)^2   \left(1 - 13 e^{i
            \iota } + e^{2 i \iota }\right)\right. \right. \right. \right.
\nonumber\\ 
&& + \left. \left. \left. \left. \left(1 + e^{i \iota }\right)^4 \left(1 + 3
          e^{i \iota } + e^{2 i \iota }\right) + 6   e^{2 i \Psi } \left(1 +
          e^{2 i \iota }\right) \left(\left( - 1 + e^{2 i \iota }\right)^2 -
          e^{2 i \alpha } \left(1 + e^{2 i \iota   }\right)^2\right)\right.
    \right. \right. \right. \nonumber\\ 
&& + \left. \left. \left. \left. e^{4 i \Psi } \left(\left( - 1 + e^{i \iota
            }\right)^4 \left(1 - 3 e^{i \iota } + e^{2 i \iota }\right) - e^{2
            i   \alpha } \left( - 1 + e^{2 i \iota }\right)^2 \left(1 + 13
            e^{i \iota } + e^{2 i \iota }\right)\right)\right) \chi
      _s^y\right. \right. \right. \nonumber\\ 
&& + \left. \left. \left. \frac{1}{12}   e^{ - i (3 \iota + \Psi )}
      \left(\left(1 + e^{i \iota }\right)^3 \left( - 1 + 6 e^{i \iota } - 6
          e^{2 i \iota } + e^{3 i \iota   }\right) + e^{4 i \Psi } \left( - 1
          + e^{i \iota }\right)^3 \left(1 + 6 e^{i \iota } + 6 e^{2 i \iota }
          + e^{3 i \iota }\right)\right. \right. \right. \right. \nonumber\\ 
&& + \left. \left. 6 e^{2 i   \Psi } \left( - 1 + e^{2 i \iota } - e^{4 i
        \iota } + e^{6 i \iota }\right)\right) \chi _s^z\right)\biggr] + {\cal
  O}(v^4)\biggr\} \, ,
\end{eqnarray}


\begin{eqnarray}
h_{33}&=&  \frac{2\,M\,\nu\,v^2}{D_L}\,e^{ - 3i\,(\alpha+\Psi)}\frac{1}{64}
\sqrt{\frac{\pi }{42}} \left\{e^{ - 3 i \iota } \left(9 \,e^{6 i \Psi } \left(
- 1 + e^{i \iota }\right)^6 - e^{4 i \Psi }\left( - 1 + e^{i \iota }\right)^4
\left(1 + e^{i \iota }\right)^2\right. \right. \nonumber\\ 
&& - \left. \left. e^{2 i \Psi } \left( - 1 + e^{i \iota }\right)^2 \left(1 +
e^{i \iota}\right)^4 + 9 \left(1 + e^{i \iota }\right)^6\right)\, v
\,\delta\,\right. \nonumber\\ 
&& + \left. i \,v^2 \left[e^{ - i (3 \iota - \Psi )} \left(8 \,e^{4 i \Psi
}\left( - 1 + e^{i \iota }\right)^5 \left(1 + e^{i \iota }\right) - 8 \left( -
1 + e^{i \iota }\right) \left(1 + e^{i \iota}\right)^5\right)\right. \right.
\nonumber\\ 
&& + \left. \left. e^{ - i (3 \iota - \Psi )} \left( - 24 \,e^{4 i \Psi }
\left( - 1 + e^{i \iota }\right)^5 \left(1 + e^{i \iota}\right)\right. \right.
\right. \nonumber\\ 
&& + \left. \left. \left. 24 \left( - 1 + e^{i \iota }\right) \left(1 + e^{i
\iota }\right)^5\right) \eta \right] + v^3\,
\biggl[\,\delta\,\left(\frac{1}{3} \,e^{ - 3 i \iota } \left( - 108 \,e^{6 i
\Psi } \left( - 1 + e^{i \iota }\right)^6 + 8 \,e^{4 i \Psi } \left( - 1 +
e^{i \iota}\right)^4 \left(1 + e^{i \iota }\right)^2\right. \right. \right.
\nonumber\\ 
&& + \left. \left. \left. \left. 8 \,e^{2 i \Psi } \left( - 1 + e^{i \iota
}\right)^2 \left(1 + e^{i \iota }\right)^4 - 108\left(1 + e^{i \iota
}\right)^6\right) + \frac{1}{3} \,e^{ - 3 i \iota } \left(54 \,e^{6 i \Psi }
\left( - 1 + e^{i \iota }\right)^6 + 2 \,e^{4 i \Psi } \left( - 1 + e^{i \iota
}\right)^4 \left(1 + e^{i \iota }\right)^2\right. \right. \right. \right.
\nonumber\\ 
&& + \left. \left. \left. \left. 2 \,e^{2 i \Psi } \left( - 1 + e^{i \iota
}\right)^2\left(1 + e^{i \iota }\right)^4 + 54 \left(1 + e^{i \iota
}\right)^6\right) \eta \right) + 64 \,e^{i (\alpha - 2 \iota + \Psi )}\left( -
e^{4 i \Psi } \left( - 1 + e^{i \iota }\right)^4 + \left(1 + e^{i \iota
}\right)^4\right) \eta  \chi _s^x\right. \right. \nonumber\\ 
&& + \left. 64 i \,e^{i (\alpha - 2\iota + \Psi )} \left(e^{4 i \Psi } \left(
- 
1 + e^{i \iota }\right)^4 - \left(1 + e^{i \iota }\right)^4\right) \eta
\chi_s^y\biggr] + {\cal O}(v^4)\right\} \, .
\end{eqnarray}
\section{Center-of-mass energy  and gravitational-wave energy flux}
\label{appA}

For nearly circular orbits, the center-of-mass energy is known through 2PN
order, when spins are present and 3PN order when spins are neglected. The
coefficients entering Eq.~(\ref{Energy})
are~\cite{K95,WWi96,Poisson98,JS99,JaraS00,
Andrade01,Blanchet01a,Damour01,FBBu06}
\bea
E_{\rm Newt} &=& - {M \over 2}\,\nu\,,\\
E_2 &=& - {3 \over 4} -{1 \over 12}\nu\,,\\
E_3 &=& \left[\left(\frac{8}{3} - \frac{4}{3}\,\nu\right)
\boldsymbol{\chi}_s\cdot\mathbf{\hat{L}}_{\rm N} + \frac{8}{3}\,\delta\,
\boldsymbol{\chi}_a\cdot\mathbf{\hat{L}}_{\rm N}\right]\,, \\
E_4 &=& -{27 \over 8}+ {19 \over 8}\nu -{1 \over 24}\nu^2 + \nu\,
\left \{\left(\boldsymbol{\chi}_s^2 - \boldsymbol{\chi}_a^2\right) 
- 3\,\left[\left(\boldsymbol{\chi}_s
\cdot\mathbf{\hat{L}}_{\rm N}\right)^2 - \left(\boldsymbol{\chi}_a
\cdot\mathbf{\hat{L}}_{\rm N}\right)^2\right]\right \}\nonumber\\
 &+ & \left(\frac{1}{2}-\nu\right)\,\left \{\boldsymbol{\chi}_s^2 
+ \boldsymbol{\chi}_a^2 - 3 \left[\left(\boldsymbol{\chi}_s
\cdot\mathbf{\hat{L}}_{\rm N}\right)^2 + 
\left(\boldsymbol{\chi}_a\cdot\mathbf{\hat{L}}_{\rm N}\right)^2\right]\right \}
\nonumber\\
 &+& \delta\,\left \{\boldsymbol{\chi}_s\cdot\boldsymbol{\chi}_a 
- 3\,\left[\left(\boldsymbol{\chi}_s\cdot\mathbf{\hat{L}}_{\rm N}\right)\,
\left(\boldsymbol{\chi}_a\cdot\mathbf{\hat{L}}_{\rm N}\right)\right]\right \}
\,, \\
E_5 &=& \left[\left(8 - \frac{121}{9}\,\nu + \frac{2}{9}\,\nu^2\right)
  \boldsymbol{\chi}_s\cdot\mathbf{\hat{L}}_{\rm N} + \left(8 -
    \frac{31}{9}\,\nu\right)\,\delta\,
  \boldsymbol{\chi}_a\cdot\mathbf{\hat{L}}_{\rm N}\right]\,, \\
E_6 &=& -{675 \over 64}+\left({34445 \over 576}-{205 \over
  96}\pi^2\right)\nu-{155 \over 96}\nu^2-{35 \over 5184}\nu^3\,. 
\eea
The GW energy flux is known through 2.5PN order for spin
effects~\cite{BBuF06,FBBu06,MVG05,RBK08}, and 3.5PN order when spin effects are
absent~\cite{BDEI04}. The coefficients in Eq.~(\ref{Flux}) read 
\bea
  F_{\rm Newt} &=& {32 \over 5}\nu^2\,,\\
  F_2 &=& -{1247 \over 336} -{35 \over 12}\nu\,,\\
  F_3 &=& 4\pi - \left[\left(\frac{11}{4} - 3\nu\right)
\boldsymbol{\chi}_s\cdot\mathbf{\hat{L}}_{\rm N} + \frac{11}{4}\,\delta\,
\boldsymbol{\chi}_a\cdot\mathbf{\hat{L}}_{\rm N}\right]\,,\\
  F_4 &=& -\frac{44711}{9072}+\frac{9271}{504}\nu + \frac{65}{18}\nu^2 +
\left(\frac{287}{96} + \frac{\nu}{24}\right)\,
\left(\boldsymbol{\chi}_s\cdot\mathbf{\hat{L}}_{\rm N}\right)^2  - 
\left(\frac{89}{96} + \frac{7 \nu}{24}\right)\,\boldsymbol{\chi}_s^2 \nonumber\\
& & + \left(\frac{287}{96} - 12\,\nu\right)\,
\left(\boldsymbol{\chi}_a\cdot\mathbf{\hat{L}}_{\rm N}\right)^2 +
\left(-\frac{89}{96} + 4\,\nu\right)\,\boldsymbol{\chi}_a^2 +
\frac{287}{48}\,\delta\, \left(\boldsymbol{\chi}_s\cdot\mathbf{\hat{L}}_{\rm
    N}\right)\,
\left(\boldsymbol{\chi}_a\cdot\mathbf{\hat{L}}_{\rm N}\right) \nonumber \\
&& - \frac{89}{48}\,\delta\,
\left(\boldsymbol{\chi}_s\cdot\boldsymbol{\chi}_a\right)
\,, \\
F_5 &=& \left(-\frac{8191}{672}-\frac{583}{24}\nu\right)\pi + \left[
  \left(-\frac{59}{16} + \frac{227}{9}\,\nu - \frac{157}{9}\,\nu^2\right)
  \boldsymbol{\chi}_s\cdot\mathbf{\hat{L}}_{\rm N} + \left(-\frac{59}{16} + \frac{701}{36}\,\nu\right)\,\delta\,
  \boldsymbol{\chi}_a\cdot\mathbf{\hat{L}}_{\rm N}\right]\,\\
  F_6 &=& {6643739519 \over 69854400}+{16 \over 3}\pi^2-{1712 \over
    105}\gamma_E-{856 \over 105} \log(16 v^2)+\left (-{134543 \over 7776}+{41
    \over 48}\pi^2 \right )\nu-{94403 \over 3024}\nu^2-{775 \over
    324}\nu^3\,,\\ 
  F_7 &=& \left(-{16285 \over 504}+{214745 \over 1728}\nu + {193385
      \over 3024}\nu^2\right)\,\pi\,.\label{fi} 
\eea

\section{Frequency-domain amplitude corrections}
\label{appC}

We give here the complex coefficients $\mathcal{C}_k^{(n)}$ appearing in the
frequency domain non-precessing waveform (\ref{hoff}). The lower index in
$\mathcal{C}_k^{(n)}$ denotes the harmonic of the orbital phase, and the upper
index denotes the (half) PN order. Since the different harmonics end at
different GW frequencies, the $k^{\rm th}$ harmonic ends at $k$ times the
orbital frequency cutoff. Thus, we introduce step functions
$\Theta(k\,F_{\rm{cut}} - f)$ to ensure each harmonic ends at its proper
frequency. We derive
\bea
\label{b1}
\mathcal{C}_2^{(0)} & = & \frac{1}{\sqrt{2}}\,\left[-(1+c_\theta^2)\,F_+ - 2
  i\,c_\theta\,F_\times\right] \Theta(2\,F_{\rm{cut}} - f)\,, \eea \bea
\label{b2}
\mathcal{C}_1^{(1)} & = & s_\theta\,
\delta \,\left[-\left(\frac{5}{8}+\frac{c_\theta^2}{8}\right)\,F_+ -
  \frac{3}{4}i\,c_\theta\,F_\times\right]\Theta(F_{\rm{cut}} - f)\,, \eea \bea
\label{b3}
\mathcal{C}_3^{(1)} & = &
\frac{1}{\sqrt{3}}\,\left[\frac{9}{8}\,s_\theta\,\delta\,\left((1+c_\theta^2)\,
F_+ + i\,2\,c_\theta\,F_\times\right)\right]\Theta(3\,F_{\rm{cut}} - f)\,, \eea
\bea
\label{b4}
\mathcal{C}_1^{(2)} & = &
s_\theta\left(\delta\,\boldsymbol{\chi}_s\cdot\mathbf{\hat{L}}_{\rm N} +
\boldsymbol{\chi}_a\cdot\mathbf{\hat{L}}_{\rm N}\right)\,\left(F_+ + 
i\,c_\theta\,F_\times\right)\Theta(F_{\rm{cut}} - f)\,, \eea \bea
\label{b5}
\mathcal{C}_2^{(2)} & = & \frac{1}{\sqrt{2}}\,\left[\left(\frac{1385}{672} -
  \frac{109\nu}{24}+\left(\frac{265}{672}+\frac{11\nu}{24}\right)\,c_\theta^2
  + \left(-\frac{1}{3}+\nu\right)\,c_\theta^4 \right)\,F_+\right.\nonumber\\ &
  & \left.  + i\,\left(\left(\frac{387}{112} -\frac{85}{12}\nu\right)\,c_\theta
  + \left(-\frac{4}{3} +
  4\nu\right)\,c_\theta^3\right)\,F_\times\right]\Theta(2\,F_{\rm{cut}}-f)\,,
\eea \bea
\label{b6}
\mathcal{C}_4^{(2)} & = &
\frac{1}{2}s_\theta^2\,\left[\frac{4}{3}(3\nu-1)\left((1+c_\theta^2)\,F_+ +
  i\,2\,c_\theta\,F_\times\right)\right]\Theta(4\,F_{\rm{cut}}-f)\,, \eea\bea
\label{b7}
\mathcal{C}_1^{(3)} & = & s_\theta\,\delta\,\left[\left(-\frac{2119}{5376} -
  \frac{263\nu}{192}+
  \left(\frac{937}{5376}-\frac{3\nu}{64}\right)\,c_\theta^2 +
  \left(-\frac{1}{192}+\frac{\nu}{96}\right)
  \,c_\theta^4\right)\,F_+\right.\nonumber\\ & & \left.+ i\,
  \left(-\left(\frac{155}{896}+\frac{145\nu}{96}\right)\,c_\theta +
  \frac{5}{96}(2\nu-1)\,c_\theta^3\right)\,F_\times\right]\Theta(F_{\rm{cut}}-f)
\,, \eea \bea
\label{b8}
\mathcal{C}_2^{(3)} & = & \frac{1}{\sqrt{2}}\,\left[\left(-\frac{27}{8} 
\left(1+c_{\theta}^2\right) \delta 
\left( \boldsymbol{\chi}_a\cdot\mathbf{\hat{L}}_{\rm N}\right)
+\left(-\frac{27}{8} \left(1+c_{\theta}^2\right)+\frac{1}{2} \left(9-7
   c_{\theta}^2\right) \nu \right) \left(\boldsymbol{\chi}_s\cdot
\mathbf{\hat{L}}_{\rm N}\right)\right) F_{+}\right.\nonumber\\
&&+\left.i\,c_{\theta}\left(-\frac{27}{4}\,\delta
\left(\boldsymbol{\chi}_a\cdot\mathbf{\hat{L}}_{\rm N}\right)
+\left(-\frac{27}{4}+ \left(5-4 c_\theta^2\right) \nu \right) 
\left(\boldsymbol{\chi}_s\cdot\mathbf{\hat{L}}_{\rm N}\right)\right)
F_\times\right]\,\Theta(2\,F_{\rm{cut}} - f)\,, \eea \bea
\mathcal{C}_3^{(3)} & = &
\frac{1}{\sqrt{3}}\,s_\theta\,\delta\,\left[\left(-\frac{6969}{1792}+
  \frac{81\nu}{16} +
  \left(-\frac{2811}{1792}+\frac{27\nu}{64}\right)\,c_\theta^2 +
  \frac{81}{128}(1-2\nu)\,c_\theta^4\right)\,F_+\right.\nonumber\\ & & \left.
  + i\,\left(\left(-\frac{6213}{896} + \frac{135\nu}{16}\right)\,c_\theta +
  \frac{135}{64}(1-2\nu)c_\theta^3\right)\,F_\times\right]
\Theta(3\,F_{\rm{cut}}-f)\,,
\eea \bea
\label{b10}
\mathcal{C}_5^{(3)} & = & \frac{1}{\sqrt{5}}\,s_\theta^3\,\delta\,(1-2\nu)
\left[\frac{625}{384} \left((1+c_\theta^2)\,F_+ +
  i\,2\,c_\theta\,F_\times\right)\right]\Theta(5\,F_{\rm{cut}}-f)\,, \eea \bea
\label{b11}
\mathcal{C}_1^{(4)} & = & s_\theta\,\delta\,\left[\left(\frac{5\pi}{8} +
  i\,\left(\frac{11}{40}+\frac{5\ln 2}{4}\right)
  +\left(\frac{\pi}{8}+i\,\left(\frac{7}{40}+\frac{\ln
    2}{4}\right)\right)\,c_\theta^2\right)\,F_+\right.\nonumber\\ & & \left. +
  \left(-\left(\frac{9}{20}+\frac{3\ln 2}{2}\right) +
  i\,\frac{3\pi}{4}\right)\,c_\theta\,F_\times\right]\Theta(F_{\rm{cut}}-f)\,,
\eea \bea
\label{b12}
\mathcal{C}_2^{(4)} & = & \frac{1}{\sqrt{2}}\,
\left[\left(\frac{113419241}{40642560}+
  \frac{152987\nu}{16128}-\frac{11099\nu^2}{1152} +
  \left(\frac{165194153}{40642560}-\frac{149\nu}{1792} +
  \frac{6709\nu^2}{1152}\right)\,c_\theta^2 +
  \left(\frac{1693}{2016}-\frac{5723\nu}{2016}+
  \frac{13\nu^2}{12}\right)\,c_\theta^4\right.\right.\nonumber\\ & &
  \left.\left.+ \left(-\frac{1}{24}+\frac{5\nu}{24}-
  \frac{5\nu^2}{24}\right)\,c_\theta^6\right)\,F_+ + i\,
  \left(\left(\frac{114020009}{20321280}+\frac{133411\nu}{8064}-
  \frac{7499\nu^2}{576}\right)\,c_\theta\right.\right.\nonumber\\ & &
  \left.\left. + \left(\frac{5777}{2520}-\frac{5555\nu}{504}+
  \frac{34\nu^2}{3}\right)\,c_\theta^3 +
  \left(-\frac{1}{4}+\frac{5\nu}{4}-\frac{5\nu^2}{4}\right)
  \,c_\theta^5\right)\,F_\times\right. \nonumber\\ & & \left. +
  \nu\,\left(\frac{721}{96}\left(\left(\boldsymbol{\chi}_s
  \cdot\mathbf{\hat{L}}_{\rm N}\right)^2 - \left(\boldsymbol{\chi}_a
  \cdot\mathbf{\hat{L}}_{\rm N}\right)^2\right) -
  \frac{439}{96}\left(\boldsymbol{\chi}_s^2 - \boldsymbol{\chi}_a^2\right)
  \right)\left((1+c_\theta^2)\,F_+ + i\,2\,c_\theta\,F_\times\right)\right]
\Theta(2\,F_{\rm{cut}}-f)\,, \eea \bea
\label{b13}
\mathcal{C}_3^{(4)} & = &
\frac{1}{\sqrt{3}}\,s_\theta\,\delta\,\left[\left(\frac{9\pi}{8}+i\,
  \left(-\frac{189}{40}+\frac{27\ln(3/2)}{4}\right)\right)
  \left((1+c_\theta^2)\,F_+ +
  i\,2\,c_\theta\,F_\times\right)\right]\Theta(3\,F_{\rm{cut}}-f)\,, \eea \bea
\label{b14}
\mathcal{C}_4^{(4)} & = &
\frac{1}{2}\,\left[\left(\frac{16109}{2520}-\frac{13367\nu}{504} +
  \frac{39\nu^2}{2} +
  \left(\frac{16}{15}-\frac{16\nu}{3}+\frac{16\nu^2}{3}\right)\,c_\theta^2\,
  \left(c_\theta^4-3\right)
  +\left(-\frac{10733}{2520}+\frac{7991\nu}{504}-\frac{53\nu^2}{6}\right)
  \,c_\theta^4\right)\,F_+\right.\nonumber\\ & & \left. +
  i\,\left(\left(\frac{2953}{252}-\frac{12023\nu}{252}+
  \frac{101\nu^2}{3}\right)\,c_\theta +
  \left(-\frac{18797}{1260}+\frac{16055\nu}{252}-\frac{149\nu^2}{3}\right)
  \,c_\theta^3\right.\right.\nonumber\\ & & \left.\left. +
  \left(\frac{16}{5}-16\,\nu+16\,\nu^2\right)\,c_\theta^5\right)
  \,F_\times\right]\Theta(4\,F_{\rm{cut}}-f)\,, \eea \bea
\label{b15}
\mathcal{C}_6^{(4)} & = & \frac{1}{\sqrt{6}}\,s_\theta^4\,
\left[\left(-\frac{81}{40}+\frac{81\nu}{8}-\frac{81\nu^2}{8}\right)
  \left((1+c_\theta^2)\,F_+ +
  i\,2\,c_\theta\,F_\times\right)\right]\Theta(6\,F_{\rm{cut}}-f)\,, \eea \bea
\label{b16}
\mathcal{C}_1^{(5)} & = & s_\theta\,\delta\,
\left[\left(\frac{37533829}{325140480} + \frac{76171\nu}{43008} -
  \frac{8407\nu^2}{4608}+c_\theta^2\,\left(-\frac{29850823}{325140480} +
  \frac{56543\nu}{129024} + \frac{139\nu^2}{576}\right)
  \right.\right.\nonumber\\ & & \left.\left.+
  c_\theta^4\,\left(\frac{255}{14336}-\frac{2659\nu}{64512}+
  \frac{127\nu^2}{9216}\right) +
  c_\theta^6\,\left(-\frac{1}{9216}+\frac{\nu}{2304}-
  \frac{\nu^2}{3072}\right)\right)\,F_+ \right.\nonumber\\ & & \left. +
  i\,\left(c_\theta\,\left(-\frac{3453823}{54190080} + \frac{163015\nu}{64512}
  - \frac{4237\nu^2}{2304}\right) +
  c_\theta^3\,\left(\frac{34373}{322560}-\frac{11755\nu}{32256} +
  \frac{631\nu^2}{2304}\right) \right.\right.\nonumber\\ & & \left.\left.+
  c_\theta^5\,\left(-\frac{7}{4608}+\frac{7\nu}{1152}-
  \frac{7\nu^2}{1536}\right)\right)\,F_\times\right]\Theta(F_{\rm{cut}}-f)\,,
\eea \bea
\label{b17}
\mathcal{C}_2^{(5)} & = & \frac{1}{\sqrt{2}}
\,\left[\left(\frac{85\pi}{64}(1-4\,\nu) + i\,\left(-\frac{9}{5} + 32\nu\right)
  + c_\theta^2\,\left(\frac{85\pi}{64}(1-4\,\nu) +
  i\,\frac{14}{5}(1+4\,\nu)\right) +
  i\,c_\theta^4\,\frac{7}{5}(1-4\,\nu)\right)\,F_+ \right.\nonumber\\ & &
  \left. + \left(c_\theta\,\left(2 - \frac{282\nu}{5} + i\,
  \frac{85\pi}{32}(1-4\,\nu)\right) +
  c_\theta^3\,\left(-\frac{22}{5}+\frac{94\nu}{5}\right)\right)\,F_\times\right]
\Theta(2\,F_{\rm{cut}}-f)\,, \eea \bea
\label{b18}
\mathcal{C}_3^{(5)} & = & \frac{1}{\sqrt{3}}\,s_\theta\,
\delta\,\left[\left(-\frac{8781361}{7225344}-\frac{366781\nu}{17920}+
  \frac{15193\nu^2}{1280} + c_\theta^2\,\left(-\frac{238136057}{36126720} +
  \frac{37829\nu}{71680} - \frac{7073\nu^2}{1280}\right)
  \right.\right.\nonumber\\ & & \left.\left. +
  c_\theta^4\,\left(-\frac{328347}{143360}+\frac{396009\nu}{71680} -
  \frac{10557\nu^2}{5120}\right) + c_\theta^6\,\left( \frac{729}{5120} -
  \frac{729\nu}{1280} + \frac{2187\nu^2}{5120}\right)\right)\,F_+
  \right.\nonumber\\ & & \left. +
  i\,\left(c_\theta\,\left(-\frac{63633869}{18063360} - \frac{89609\nu}{2560}
  + \frac{697\nu^2}{40}\right) + c_\theta^3\,\left(-\frac{508689}{71680} +
  \frac{812727\nu}{35840} - \frac{4707\nu^2}{320}\right)
  \right.\right.\nonumber\\ & & \left.\left. +
  c_\theta^5\,\left(\frac{1701}{2560} - \frac{1701\nu}{640} +
  \frac{5103\nu^2}{2560}\right)\right)\,F_\times\right]\Theta(3\,
F_{\rm{cut}}-f)\,, \eea \bea
\label{b19}
\mathcal{C}_4^{(5)} & = & \frac{1}{2}\,s_\theta^2\,\left[\left(
  \frac{8\pi}{3}(3\,\nu-1) + 
  i\,\left(\frac{56}{5} - \frac{1193\nu}{30} + \frac{32\,\ln
    2}{3}(3\,\nu-1)\right)\right)\left((1+c_\theta^2)\,F_+ +
  i\,2\,c_\theta\,F_\times\right)\right]\Theta(4\,F_{\rm{cut}}-f)\,, \eea \bea
\label{b20}
\mathcal{C}_5^{(5)} & = & \frac{1}{\sqrt{5}}\,s_\theta\,\delta
\,\left[\left(-\frac{854375}{86016} + \frac{3919375\nu}{129024} -
  \frac{160625\nu^2}{9216} + c_\theta^2\,\left(\frac{40625}{9216} -
  \frac{40625\nu}{2304} + \frac{40625\nu^2}{3072}\right)
  \right.\right.\nonumber\\ & & \left.\left. +
  c_\theta^4\,\left(\frac{1863125}{258048} - \frac{2519375\nu}{129024} +
  \frac{85625\nu^2}{9216}\right) +
  c_\theta^6\,\left(-\frac{15625}{9216}+\frac{15625\nu}{2304} -
  \frac{15625\nu^2}{3072}\right)\right)\,F_+ \right.\nonumber\\ & & \left. +
  i\,\left(c_\theta\,\left(-\frac{2388125}{129024}+
  \frac{3569375\nu}{64512}-\frac{141875\nu^2}{4608}\right) +
  c_\theta^3\,\left(\frac{3000625}{129024} - \frac{1598125\nu}{21504} +
  \frac{51875\nu^2}{1152}\right) \right.\right.\nonumber\\ & & \left.\left. +
  c_\theta^5\,\left(-\frac{21875}{4608}+ \frac{21875\nu}{1152} -
  \frac{21875\nu^2}{1536}\right)\right)\,F_\times\right]
\Theta(5\,F_{\rm{cut}}-f)\,, 
\eea 
\bea 
\mathcal{C}_7^{(5)} & = &
\frac{1}{\sqrt{7}}\,s_\theta^5\,\delta\,\left[ \left(\frac{117649}{46080} -
  \frac{117649\nu}{11520} +
  \frac{117649\nu^2}{15360}\right)\left((1+c_\theta^2)\,F_+ +
  i\,2\,c_\theta\,F_\times\right)\right]\Theta(7\,F_{\rm{cut}}-f)\,.
\label{b21}
\eea
\bibliography{ref-list}

\begin{thebibliography}{96}
\expandafter\ifx\csname natexlab\endcsname\relax\def\natexlab#1{#1}\fi
\expandafter\ifx\csname bibnamefont\endcsname\relax
  \def\bibnamefont#1{#1}\fi
\expandafter\ifx\csname bibfnamefont\endcsname\relax
  \def\bibfnamefont#1{#1}\fi
\expandafter\ifx\csname citenamefont\endcsname\relax
  \def\citenamefont#1{#1}\fi
\expandafter\ifx\csname url\endcsname\relax
  \def\url#1{\texttt{#1}}\fi
\expandafter\ifx\csname urlprefix\endcsname\relax\def\urlprefix{URL }\fi
\providecommand{\bibinfo}[2]{#2}
\providecommand{\eprint}[2][]{\url{#2}}

\bibitem[{lig()}]{ligo}
\bibinfo{howpublished}{\url{http://www.ligo.caltech.edu}}.

\bibitem[{vir()}]{virgo}
\bibinfo{howpublished}{\url{http://www.virgo.infn.it}}.

\bibitem[{geo()}]{geo}
\bibinfo{howpublished}{\url{http://www.geo600.uni-hannover.de}}.

\bibitem[{lis()}]{lisa}
\bibinfo{howpublished}{\url{http://lisa.jpl.nasa.gov}}.

\bibitem[{\citenamefont{Allen et~al.}(1999)}]{LIGO40m99}
\bibinfo{author}{\bibfnamefont{B.}~\bibnamefont{Allen}} \bibnamefont{et~al.},
  \bibinfo{journal}{Phys. Rev. Lett.} \textbf{\bibinfo{volume}{83}},
  \bibinfo{pages}{1498} (\bibinfo{year}{1999}).

\bibitem[{\citenamefont{Abbott et~al.}(2005)}]{LSCnsns}
\bibinfo{author}{\bibfnamefont{B.}~\bibnamefont{Abbott}} \bibnamefont{et~al.}
  (\bibinfo{collaboration}{LIGO Scientific Collaboration}),
  \bibinfo{journal}{Phys.\ Rev.\ D} \textbf{\bibinfo{volume}{72}},
  \bibinfo{pages}{082001} (\bibinfo{year}{2005}).

\bibitem[{\citenamefont{Abbott et~al.}(2006)}]{LSC05insp}
\bibinfo{author}{\bibfnamefont{B.}~\bibnamefont{Abbott}} \bibnamefont{et~al.}
  (\bibinfo{collaboration}{LIGO Scientific Collaboration}),
  \bibinfo{journal}{Phys. Rev.~D} \textbf{\bibinfo{volume}{73}},
  \bibinfo{pages}{062001} (\bibinfo{year}{2006}).

\bibitem[{\citenamefont{Blanchet}(2006)}]{Bliving}
\bibinfo{author}{\bibfnamefont{L.}~\bibnamefont{Blanchet}},
  \bibinfo{journal}{Living Rev. Rel.} \textbf{\bibinfo{volume}{9}},
  \bibinfo{pages}{4} (\bibinfo{year}{2006}).

\bibitem[{\citenamefont{Buonanno and Damour}(2000)}]{BuonD00}
\bibinfo{author}{\bibfnamefont{A.}~\bibnamefont{Buonanno}} \bibnamefont{and}
  \bibinfo{author}{\bibfnamefont{T.}~\bibnamefont{Damour}},
  \bibinfo{journal}{Phys. Rev. D} \textbf{\bibinfo{volume}{62}},
  \bibinfo{pages}{064015} (\bibinfo{year}{2000}).

\bibitem[{\citenamefont{{Buonanno} et~al.}(2005)\citenamefont{{Buonanno},
  {Chen}, {Pan}, {Tagoshi}, and {Vallisneri}}}]{BCPTV05}
\bibinfo{author}{\bibfnamefont{A.}~\bibnamefont{{Buonanno}}},
  \bibinfo{author}{\bibfnamefont{Y.}~\bibnamefont{{Chen}}},
  \bibinfo{author}{\bibfnamefont{Y.}~\bibnamefont{{Pan}}},
  \bibinfo{author}{\bibfnamefont{H.}~\bibnamefont{{Tagoshi}}},
  \bibnamefont{and}
  \bibinfo{author}{\bibfnamefont{M.}~\bibnamefont{{Vallisneri}}},
  \bibinfo{journal}{Phys.\ Rev.\ D} \textbf{\bibinfo{volume}{72}},
  \bibinfo{pages}{084027} (\bibinfo{year}{2005}).

\bibitem[{\citenamefont{Belczynski et~al.}(2007)\citenamefont{Belczynski, Taam,
  Rantsiou, and van~der Sluys}}]{BelczynskiSpinEvoln07}
\bibinfo{author}{\bibfnamefont{K.}~\bibnamefont{Belczynski}},
  \bibinfo{author}{\bibfnamefont{R.~E.} \bibnamefont{Taam}},
  \bibinfo{author}{\bibfnamefont{E.}~\bibnamefont{Rantsiou}}, \bibnamefont{and}
  \bibinfo{author}{\bibfnamefont{M.}~\bibnamefont{van~der Sluys}},
  \bibinfo{journal}{\apj} \textbf{\bibinfo{volume}{682}}, \bibinfo{pages}{474}
  (\bibinfo{year}{2007}).

\bibitem[{\citenamefont{Buonanno et~al.}(2003)\citenamefont{Buonanno, Chen, and
  Vallisneri}}]{BCV03b}
\bibinfo{author}{\bibfnamefont{A.}~\bibnamefont{Buonanno}},
  \bibinfo{author}{\bibfnamefont{Y.}~\bibnamefont{Chen}}, \bibnamefont{and}
  \bibinfo{author}{\bibfnamefont{M.}~\bibnamefont{Vallisneri}},
  \bibinfo{journal}{Phys. Rev. D} \textbf{\bibinfo{volume}{67}},
  \bibinfo{pages}{104025} (\bibinfo{year}{2003}), \bibinfo{note}{erratum-ibid.
  { D}~74, 029904(E) (2006)}.

\bibitem[{\citenamefont{Cutler and Flanagan}(1994)}]{CF94}
\bibinfo{author}{\bibfnamefont{C.}~\bibnamefont{Cutler}} \bibnamefont{and}
  \bibinfo{author}{\bibfnamefont{E.}~\bibnamefont{Flanagan}},
  \bibinfo{journal}{Phys. Rev. D} \textbf{\bibinfo{volume}{49}},
  \bibinfo{pages}{2658} (\bibinfo{year}{1994}).

\bibitem[{\citenamefont{Poisson and Will}(1995)}]{PW95}
\bibinfo{author}{\bibfnamefont{E.}~\bibnamefont{Poisson}} \bibnamefont{and}
  \bibinfo{author}{\bibfnamefont{C.}~\bibnamefont{Will}},
  \bibinfo{journal}{Phys. Rev. D} \textbf{\bibinfo{volume}{52}},
  \bibinfo{pages}{848} (\bibinfo{year}{1995}).

\bibitem[{\citenamefont{van~der Sluys et~al.}(2007)}]{Sluys07MCMC}
\bibinfo{author}{\bibfnamefont{M.~V.} \bibnamefont{van~der Sluys}}
  \bibnamefont{et~al.} (\bibinfo{year}{2007}), \eprint{arXiv:0710.1897
  [astro-ph]}.

\bibitem[{\citenamefont{{Pan} et~al.}(2004)\citenamefont{{Pan}, {Buonanno},
  {Chen}, and {Vallisneri}}}]{PBCV04}
\bibinfo{author}{\bibfnamefont{Y.}~\bibnamefont{{Pan}}},
  \bibinfo{author}{\bibfnamefont{A.}~\bibnamefont{{Buonanno}}},
  \bibinfo{author}{\bibfnamefont{Y.}~\bibnamefont{{Chen}}}, \bibnamefont{and}
  \bibinfo{author}{\bibfnamefont{M.}~\bibnamefont{{Vallisneri}}},
  \bibinfo{journal}{Phys.~Rev.~D} \textbf{\bibinfo{volume}{69}},
  \bibinfo{pages}{104017} (\bibinfo{year}{2004}), \bibinfo{note}{erratum-ibid.
  { D}~74, 029905(E) (2006)}.

\bibitem[{\citenamefont{{Buonanno} et~al.}(2004)\citenamefont{{Buonanno},
  {Chen}, {Pan}, and {Vallisneri}}}]{BCPV04}
\bibinfo{author}{\bibfnamefont{A.}~\bibnamefont{{Buonanno}}},
  \bibinfo{author}{\bibfnamefont{Y.}~\bibnamefont{{Chen}}},
  \bibinfo{author}{\bibfnamefont{Y.}~\bibnamefont{{Pan}}}, \bibnamefont{and}
  \bibinfo{author}{\bibfnamefont{M.}~\bibnamefont{{Vallisneri}}},
  \bibinfo{journal}{Phys.\ Rev.\ D} \textbf{\bibinfo{volume}{70}},
  \bibinfo{pages}{104003} (\bibinfo{year}{2004}), \bibinfo{note}{erratum-ibid.
  {D}~{\bf 74}, 029902(E) (2006)}.

\bibitem[{\citenamefont{{Buonanno} et~al.}(2006)\citenamefont{{Buonanno},
  {Chen}, and {Damour}}}]{Buonanno06}
\bibinfo{author}{\bibfnamefont{A.}~\bibnamefont{{Buonanno}}},
  \bibinfo{author}{\bibfnamefont{Y.}~\bibnamefont{{Chen}}}, \bibnamefont{and}
  \bibinfo{author}{\bibfnamefont{T.}~\bibnamefont{{Damour}}},
  \bibinfo{journal}{Phys.\ Rev.\ D} \textbf{\bibinfo{volume}{74}},
  \bibinfo{pages}{104005} (\bibinfo{year}{2006}).

\bibitem[{\citenamefont{Blanchet
  et~al.}(1995{\natexlab{a}})\citenamefont{Blanchet, Damour, and Iyer}}]{BDI95}
\bibinfo{author}{\bibfnamefont{L.}~\bibnamefont{Blanchet}},
  \bibinfo{author}{\bibfnamefont{T.}~\bibnamefont{Damour}}, \bibnamefont{and}
  \bibinfo{author}{\bibfnamefont{B.~R.} \bibnamefont{Iyer}},
  \bibinfo{journal}{Phys. Rev. D} \textbf{\bibinfo{volume}{51}},
  \bibinfo{pages}{5360} (\bibinfo{year}{1995}{\natexlab{a}}).

\bibitem[{\citenamefont{Blanchet
  et~al.}(1995{\natexlab{b}})\citenamefont{Blanchet, Damour, Iyer, Will, and
  Wiseman}}]{BDIWW95}
\bibinfo{author}{\bibfnamefont{L.}~\bibnamefont{Blanchet}},
  \bibinfo{author}{\bibfnamefont{T.}~\bibnamefont{Damour}},
  \bibinfo{author}{\bibfnamefont{B.~R.} \bibnamefont{Iyer}},
  \bibinfo{author}{\bibfnamefont{C.~M.} \bibnamefont{Will}}, \bibnamefont{and}
  \bibinfo{author}{\bibfnamefont{A.~G.} \bibnamefont{Wiseman}},
  \bibinfo{journal}{Phys. Rev. Lett.} \textbf{\bibinfo{volume}{74}},
  \bibinfo{pages}{3515} (\bibinfo{year}{1995}{\natexlab{b}}).

\bibitem[{\citenamefont{Blanchet}(1996)}]{B96}
\bibinfo{author}{\bibfnamefont{L.}~\bibnamefont{Blanchet}},
  \bibinfo{journal}{Phys. Rev. D} \textbf{\bibinfo{volume}{54}},
  \bibinfo{pages}{1417} (\bibinfo{year}{1996}), \bibinfo{note}{{erratum-ibid.
  D~{\bf 71}, 129904(E) (2005)}}.

\bibitem[{\citenamefont{Blanchet et~al.}(2002)\citenamefont{Blanchet, Faye,
  Iyer, and Joguet}}]{BFIJ02}
\bibinfo{author}{\bibfnamefont{L.}~\bibnamefont{Blanchet}},
  \bibinfo{author}{\bibfnamefont{G.}~\bibnamefont{Faye}},
  \bibinfo{author}{\bibfnamefont{B.~R.} \bibnamefont{Iyer}}, \bibnamefont{and}
  \bibinfo{author}{\bibfnamefont{B.}~\bibnamefont{Joguet}},
  \bibinfo{journal}{Phys. Rev. D} \textbf{\bibinfo{volume}{65}},
  \bibinfo{pages}{061501(R)} (\bibinfo{year}{2002}),
  \bibinfo{note}{{erratum-ibid~D.~{\bf 71}, 129902(E) (2005)}}.

\bibitem[{\citenamefont{Blanchet et~al.}(2004)\citenamefont{Blanchet, Damour,
  Esposito-Far{\`e}se, and Iyer}}]{BDEI04}
\bibinfo{author}{\bibfnamefont{L.}~\bibnamefont{Blanchet}},
  \bibinfo{author}{\bibfnamefont{T.}~\bibnamefont{Damour}},
  \bibinfo{author}{\bibfnamefont{G.}~\bibnamefont{Esposito-Far{\`e}se}},
  \bibnamefont{and} \bibinfo{author}{\bibfnamefont{B.~R.} \bibnamefont{Iyer}},
  \bibinfo{journal}{Phys. Rev. Lett.} \textbf{\bibinfo{volume}{93}},
  \bibinfo{pages}{091101} (\bibinfo{year}{2004}).

\bibitem[{\citenamefont{Blanchet et~al.}(1996)\citenamefont{Blanchet, Iyer,
  Will, and Wiseman}}]{BIWW96}
\bibinfo{author}{\bibfnamefont{L.}~\bibnamefont{Blanchet}},
  \bibinfo{author}{\bibfnamefont{B.~R.} \bibnamefont{Iyer}},
  \bibinfo{author}{\bibfnamefont{C.~M.} \bibnamefont{Will}}, \bibnamefont{and}
  \bibinfo{author}{\bibfnamefont{A.~G.} \bibnamefont{Wiseman}},
  \bibinfo{journal}{Class. Quantum Grav.} \textbf{\bibinfo{volume}{13}},
  \bibinfo{pages}{575} (\bibinfo{year}{1996}).

\bibitem[{\citenamefont{Arun et~al.}(2004)\citenamefont{Arun, Blanchet, Iyer,
  and Qusailah}}]{ABIQ04}
\bibinfo{author}{\bibfnamefont{K.~G.} \bibnamefont{Arun}},
  \bibinfo{author}{\bibfnamefont{L.}~\bibnamefont{Blanchet}},
  \bibinfo{author}{\bibfnamefont{B.~R.} \bibnamefont{Iyer}}, \bibnamefont{and}
  \bibinfo{author}{\bibfnamefont{M.~S.~S.} \bibnamefont{Qusailah}},
  \bibinfo{journal}{Class. Quantum Grav.} \textbf{\bibinfo{volume}{21}},
  \bibinfo{pages}{3771} (\bibinfo{year}{2004}), \bibinfo{note}{erratum-ibid.
  {\bf 22}, 3115 (2005)}.

\bibitem[{\citenamefont{Kidder et~al.}(2007)\citenamefont{Kidder, Blanchet, and
  Iyer}}]{KBI07}
\bibinfo{author}{\bibfnamefont{L.~E.} \bibnamefont{Kidder}},
  \bibinfo{author}{\bibfnamefont{L.}~\bibnamefont{Blanchet}}, \bibnamefont{and}
  \bibinfo{author}{\bibfnamefont{B.~R.} \bibnamefont{Iyer}},
  \bibinfo{journal}{Class. Quant. Grav.} \textbf{\bibinfo{volume}{24}},
  \bibinfo{pages}{5307} (\bibinfo{year}{2007}).

\bibitem[{\citenamefont{{Kidder}}(2007)}]{K07}
\bibinfo{author}{\bibfnamefont{L.~E.} \bibnamefont{{Kidder}}},
  \bibinfo{journal}{Phys.\ Rev.\ D} \textbf{\bibinfo{volume}{77}},
  \bibinfo{pages}{044016} (\bibinfo{year}{2007}).

\bibitem[{\citenamefont{Blanchet et~al.}(2008)\citenamefont{Blanchet, Faye,
  Iyer, and Sinha}}]{BFIS08}
\bibinfo{author}{\bibfnamefont{L.}~\bibnamefont{Blanchet}},
  \bibinfo{author}{\bibfnamefont{G.}~\bibnamefont{Faye}},
  \bibinfo{author}{\bibfnamefont{B.~R.} \bibnamefont{Iyer}}, \bibnamefont{and}
  \bibinfo{author}{\bibfnamefont{S.}~\bibnamefont{Sinha}},
  \bibinfo{journal}{Class.\ Quantum Grav.} \textbf{\bibinfo{volume}{25}},
  \bibinfo{pages}{165003} (\bibinfo{year}{2008}).

\bibitem[{\citenamefont{Faye et~al.}(2006)\citenamefont{Faye, Blanchet, and
  Buonanno}}]{FBBu06}
\bibinfo{author}{\bibfnamefont{G.}~\bibnamefont{Faye}},
  \bibinfo{author}{\bibfnamefont{L.}~\bibnamefont{Blanchet}}, \bibnamefont{and}
  \bibinfo{author}{\bibfnamefont{A.}~\bibnamefont{Buonanno}},
  \bibinfo{journal}{Phys. Rev.~D} \textbf{\bibinfo{volume}{{\bf 74}}},
  \bibinfo{pages}{104033} (\bibinfo{year}{2006}).

\bibitem[{\citenamefont{Blanchet et~al.}(2006)\citenamefont{Blanchet, Buonanno,
  and Faye}}]{BBuF06}
\bibinfo{author}{\bibfnamefont{L.}~\bibnamefont{Blanchet}},
  \bibinfo{author}{\bibfnamefont{A.}~\bibnamefont{Buonanno}}, \bibnamefont{and}
  \bibinfo{author}{\bibfnamefont{G.}~\bibnamefont{Faye}},
  \bibinfo{journal}{Phys. Rev.~D} \textbf{\bibinfo{volume}{{\bf 74}}},
  \bibinfo{pages}{104034} (\bibinfo{year}{2006}), \bibinfo{note}{erratum-ibid.
  { D} {\bf 75}, 049903 (E) (2007)}.

\bibitem[{\citenamefont{Kidder et~al.}(1993)\citenamefont{Kidder, Will, and
  Wiseman}}]{KWWi93}
\bibinfo{author}{\bibfnamefont{L.}~\bibnamefont{Kidder}},
  \bibinfo{author}{\bibfnamefont{C.}~\bibnamefont{Will}}, \bibnamefont{and}
  \bibinfo{author}{\bibfnamefont{A.}~\bibnamefont{Wiseman}},
  \bibinfo{journal}{Phys. Rev. D} \textbf{\bibinfo{volume}{47}},
  \bibinfo{pages}{R4183} (\bibinfo{year}{1993}).

\bibitem[{\citenamefont{{Mik{\'o}czi} et~al.}(2005)\citenamefont{{Mik{\'o}czi},
  {Vas{\'u}th}, and {Gergely}}}]{MVG05}
\bibinfo{author}{\bibfnamefont{B.}~\bibnamefont{{Mik{\'o}czi}}},
  \bibinfo{author}{\bibfnamefont{M.}~\bibnamefont{{Vas{\'u}th}}},
  \bibnamefont{and} \bibinfo{author}{\bibfnamefont{L.~{\'A}.}
  \bibnamefont{{Gergely}}}, \bibinfo{journal}{Phys.\ Rev.\ D}
  \textbf{\bibinfo{volume}{71}}, \bibinfo{pages}{124043}
  (\bibinfo{year}{2005}).

\bibitem[{\citenamefont{{Racine} et~al.}(2008)\citenamefont{{Racine},
  {Buonanno}, and {Kidder}}}]{RBK08}
\bibinfo{author}{\bibfnamefont{E.}~\bibnamefont{{Racine}}},
  \bibinfo{author}{\bibfnamefont{A.}~\bibnamefont{{Buonanno}}},
  \bibnamefont{and} \bibinfo{author}{\bibfnamefont{L.}~\bibnamefont{{Kidder}}}
  (\bibinfo{year}{2008}), \bibinfo{note}{in preparation}.

\bibitem[{\citenamefont{Kidder}(1995)}]{K95}
\bibinfo{author}{\bibfnamefont{L.}~\bibnamefont{Kidder}},
  \bibinfo{journal}{Phys. Rev. D} \textbf{\bibinfo{volume}{52}},
  \bibinfo{pages}{821} (\bibinfo{year}{1995}).

\bibitem[{\citenamefont{Will and Wiseman}(1996)}]{WWi96}
\bibinfo{author}{\bibfnamefont{C.}~\bibnamefont{Will}} \bibnamefont{and}
  \bibinfo{author}{\bibfnamefont{A.}~\bibnamefont{Wiseman}},
  \bibinfo{journal}{Phys. Rev. D} \textbf{\bibinfo{volume}{54}},
  \bibinfo{pages}{4813} (\bibinfo{year}{1996}).

\bibitem[{\citenamefont{Owen et~al.}(1998)\citenamefont{Owen, Tagoshi, and
  Ohashi}}]{OTO98}
\bibinfo{author}{\bibfnamefont{B.}~\bibnamefont{Owen}},
  \bibinfo{author}{\bibfnamefont{H.}~\bibnamefont{Tagoshi}}, \bibnamefont{and}
  \bibinfo{author}{\bibfnamefont{A.}~\bibnamefont{Ohashi}},
  \bibinfo{journal}{Phys. Rev. D} \textbf{\bibinfo{volume}{57}},
  \bibinfo{pages}{6168} (\bibinfo{year}{1998}).

\bibitem[{\citenamefont{{Porto} and {Rothstein}}(2006)}]{PR06}
\bibinfo{author}{\bibfnamefont{R.~A.} \bibnamefont{{Porto}}} \bibnamefont{and}
  \bibinfo{author}{\bibfnamefont{I.~Z.} \bibnamefont{{Rothstein}}},
  \bibinfo{journal}{Phys.\ Rev.\ Lett.} \textbf{\bibinfo{volume}{97}},
  \bibinfo{pages}{021101} (\bibinfo{year}{2006}).

\bibitem[{\citenamefont{{Porto} and {Rothstein}}(2007)}]{PR07}
\bibinfo{author}{\bibfnamefont{R.~A.} \bibnamefont{{Porto}}} \bibnamefont{and}
  \bibinfo{author}{\bibfnamefont{I.~Z.} \bibnamefont{{Rothstein}}}
  (\bibinfo{year}{2007}), \eprint{arXiv:0712.2032[gr-qc]}.

\bibitem[{\citenamefont{{Porto} and {Rothstein}}(2008{\natexlab{a}})}]{PR08b}
\bibinfo{author}{\bibfnamefont{R.~A.} \bibnamefont{{Porto}}} \bibnamefont{and}
  \bibinfo{author}{\bibfnamefont{I.~Z.} \bibnamefont{{Rothstein}}}
  (\bibinfo{year}{2008}{\natexlab{a}}), \eprint{arXiv:0802.0720 [gr-qc]}.

\bibitem[{\citenamefont{{Levi}}(2008)}]{ML08}
\bibinfo{author}{\bibfnamefont{M.}~\bibnamefont{{Levi}}}
  (\bibinfo{year}{2008}), \eprint{arXiv:0802.1508 [gr-qc]}.

\bibitem[{\citenamefont{Steinhoff
  et~al.}(2008{\natexlab{a}})\citenamefont{Steinhoff, Hergt, and
  Schaefer}}]{SHS07}
\bibinfo{author}{\bibfnamefont{J.}~\bibnamefont{Steinhoff}},
  \bibinfo{author}{\bibfnamefont{S.}~\bibnamefont{Hergt}}, \bibnamefont{and}
  \bibinfo{author}{\bibfnamefont{G.}~\bibnamefont{Schaefer}},
  \bibinfo{journal}{Phys.\ Rev.\ D} \textbf{\bibinfo{volume}{77}},
  \bibinfo{pages}{081501(R)} (\bibinfo{year}{2008}{\natexlab{a}}).

\bibitem[{\citenamefont{Steinhoff
  et~al.}(2008{\natexlab{b}})\citenamefont{Steinhoff, Schaefer, and
  Hergt}}]{SSH08}
\bibinfo{author}{\bibfnamefont{J.}~\bibnamefont{Steinhoff}},
  \bibinfo{author}{\bibfnamefont{G.}~\bibnamefont{Schaefer}}, \bibnamefont{and}
  \bibinfo{author}{\bibfnamefont{S.}~\bibnamefont{Hergt}},
  \bibinfo{journal}{Phys.\ Rev.\ D} \textbf{\bibinfo{volume}{77}},
  \bibinfo{pages}{104018} (\bibinfo{year}{2008}{\natexlab{b}}).

\bibitem[{\citenamefont{Steinhoff
  et~al.}(2008{\natexlab{c}})\citenamefont{Steinhoff, Hergt, and
  Schafer}}]{SHS08}
\bibinfo{author}{\bibfnamefont{J.}~\bibnamefont{Steinhoff}},
  \bibinfo{author}{\bibfnamefont{S.}~\bibnamefont{Hergt}}, \bibnamefont{and}
  \bibinfo{author}{\bibfnamefont{G.}~\bibnamefont{Schafer}}
  (\bibinfo{year}{2008}{\natexlab{c}}), \eprint{arXiv:0809.2200[gr-qc]}.

\bibitem[{\citenamefont{{Porto} and {Rothstein}}(2008{\natexlab{b}})}]{PR08a}
\bibinfo{author}{\bibfnamefont{R.~A.} \bibnamefont{{Porto}}} \bibnamefont{and}
  \bibinfo{author}{\bibfnamefont{I.~Z.} \bibnamefont{{Rothstein}}}
  (\bibinfo{year}{2008}{\natexlab{b}}), \eprint{arXiv:0804.0260 [astro-ph]}.

\bibitem[{\citenamefont{Sintes and Vecchio}(2000)}]{SinVecc00a}
\bibinfo{author}{\bibfnamefont{A.~M.} \bibnamefont{Sintes}} \bibnamefont{and}
  \bibinfo{author}{\bibfnamefont{A.}~\bibnamefont{Vecchio}}, in
  \emph{\bibinfo{booktitle}{Rencontres de Moriond:Gravitational waves and
  experimental gravity}}, edited by
  \bibinfo{editor}{\bibfnamefont{J.}~\bibnamefont{Dumarchez}}
  (\bibinfo{publisher}{Frontières, Paris}, \bibinfo{year}{2000}),
  \eprint{gr-qc/0005058}.

\bibitem[{\citenamefont{Van Den~Broeck}(2006)}]{Chris06}
\bibinfo{author}{\bibfnamefont{C.}~\bibnamefont{Van Den~Broeck}},
  \bibinfo{journal}{Class.~Quantum Grav.} \textbf{\bibinfo{volume}{23}},
  \bibinfo{pages}{L51} (\bibinfo{year}{2006}).

\bibitem[{\citenamefont{Van Den~Broeck and Sengupta}(2007)}]{ChrisAnand06}
\bibinfo{author}{\bibfnamefont{C.}~\bibnamefont{Van Den~Broeck}}
  \bibnamefont{and} \bibinfo{author}{\bibfnamefont{A.}~\bibnamefont{Sengupta}},
  \bibinfo{journal}{Class.~Quantum Grav.} \textbf{\bibinfo{volume}{24}},
  \bibinfo{pages}{155} (\bibinfo{year}{2007}).

\bibitem[{\citenamefont{{Van Den Broeck} and {Sengupta}}(2007)}]{ChrisAnand06b}
\bibinfo{author}{\bibfnamefont{C.}~\bibnamefont{{Van Den Broeck}}}
  \bibnamefont{and} \bibinfo{author}{\bibfnamefont{A.~S.}
  \bibnamefont{{Sengupta}}}, \bibinfo{journal}{Class.~Quantum Grav.}
  \textbf{\bibinfo{volume}{24}}, \bibinfo{pages}{1089} (\bibinfo{year}{2007}).

\bibitem[{\citenamefont{{Sintes} and {Vecchio}}(2000)}]{SinVecc00b}
\bibinfo{author}{\bibfnamefont{A.~M.} \bibnamefont{{Sintes}}} \bibnamefont{and}
  \bibinfo{author}{\bibfnamefont{A.}~\bibnamefont{{Vecchio}}}, in
  \emph{\bibinfo{booktitle}{Third Amaldi conference on Gravitational Waves}},
  edited by \bibinfo{editor}{\bibfnamefont{S.}~\bibnamefont{{Meshkov}}}
  (\bibinfo{publisher}{American Institute of Physics Conference Series},
  \bibinfo{year}{2000}), p. \bibinfo{pages}{403}, \eprint{gr-qc/0005059}.

\bibitem[{\citenamefont{Moore and Hellings}(2002)}]{MH02}
\bibinfo{author}{\bibfnamefont{T.~A.} \bibnamefont{Moore}} \bibnamefont{and}
  \bibinfo{author}{\bibfnamefont{R.~W.} \bibnamefont{Hellings}},
  \bibinfo{journal}{Phys. Rev.~D} \textbf{\bibinfo{volume}{65}},
  \bibinfo{pages}{062001} (\bibinfo{year}{2002}).

\bibitem[{\citenamefont{Hellings and Moore}(2003)}]{HM03}
\bibinfo{author}{\bibfnamefont{R.~W.} \bibnamefont{Hellings}} \bibnamefont{and}
  \bibinfo{author}{\bibfnamefont{T.~A.} \bibnamefont{Moore}},
  \bibinfo{journal}{Class.~Quant.~Grav.} \textbf{\bibinfo{volume}{20}},
  \bibinfo{pages}{S181} (\bibinfo{year}{2003}).

\bibitem[{\citenamefont{{Arun} et~al.}(2007)\citenamefont{{Arun}, {Iyer},
  {Sathyaprakash}, and {Sinha}}}]{AISS07}
\bibinfo{author}{\bibfnamefont{K.~G.} \bibnamefont{{Arun}}},
  \bibinfo{author}{\bibfnamefont{B.~R.} \bibnamefont{{Iyer}}},
  \bibinfo{author}{\bibfnamefont{B.~S.} \bibnamefont{{Sathyaprakash}}},
  \bibnamefont{and} \bibinfo{author}{\bibfnamefont{S.}~\bibnamefont{{Sinha}}},
  \bibinfo{journal}{Phys.~Rev.~D} \textbf{\bibinfo{volume}{75}},
  \bibinfo{pages}{124002} (\bibinfo{year}{2007}).

\bibitem[{\citenamefont{Arun et~al.}(2007)\citenamefont{Arun, Iyer,
  Sathyaprakash, Sinha, and Van Den~Broeck}}]{AISSV07}
\bibinfo{author}{\bibfnamefont{K.~G.} \bibnamefont{Arun}},
  \bibinfo{author}{\bibfnamefont{B.~R.} \bibnamefont{Iyer}},
  \bibinfo{author}{\bibfnamefont{B.~S.} \bibnamefont{Sathyaprakash}},
  \bibinfo{author}{\bibfnamefont{S.}~\bibnamefont{Sinha}}, \bibnamefont{and}
  \bibinfo{author}{\bibfnamefont{C.}~\bibnamefont{Van Den~Broeck}},
  \bibinfo{journal}{Phys.~Rev.~D} \textbf{\bibinfo{volume}{76}},
  \bibinfo{pages}{104016} (\bibinfo{year}{2007}).

\bibitem[{\citenamefont{Trias and Sintes}(2008)}]{TriasSintes07}
\bibinfo{author}{\bibfnamefont{M.}~\bibnamefont{Trias}} \bibnamefont{and}
  \bibinfo{author}{\bibfnamefont{A.~M.} \bibnamefont{Sintes}},
  \bibinfo{journal}{Phys. Rev. D} \textbf{\bibinfo{volume}{77}},
  \bibinfo{pages}{024030} (\bibinfo{year}{2008}).

\bibitem[{\citenamefont{{Babak} et~al.}(2008)\citenamefont{{Babak}, {Hannam},
  {Husa}, and {Schutz}}}]{BHHS}
\bibinfo{author}{\bibfnamefont{S.}~\bibnamefont{{Babak}}},
  \bibinfo{author}{\bibfnamefont{M.}~\bibnamefont{{Hannam}}},
  \bibinfo{author}{\bibfnamefont{S.}~\bibnamefont{{Husa}}}, \bibnamefont{and}
  \bibinfo{author}{\bibfnamefont{B.}~\bibnamefont{{Schutz}}}
  (\bibinfo{year}{2008}), \eprint{arXiv:0806.1591 [gr-qc]}.

\bibitem[{\citenamefont{Vecchio}(2004)}]{Vecchio04}
\bibinfo{author}{\bibfnamefont{A.}~\bibnamefont{Vecchio}},
  \bibinfo{journal}{Phys. Rev. D} \textbf{\bibinfo{volume}{70}},
  \bibinfo{pages}{042001} (\bibinfo{year}{2004}).

\bibitem[{\citenamefont{Lang and Hughes}(2006)}]{LangHughes06}
\bibinfo{author}{\bibfnamefont{R.~N.} \bibnamefont{Lang}} \bibnamefont{and}
  \bibinfo{author}{\bibfnamefont{S.~A.} \bibnamefont{Hughes}},
  \bibinfo{journal}{Phys. Rev.~D} \textbf{\bibinfo{volume}{74}},
  \bibinfo{pages}{122001} (\bibinfo{year}{2006}),
  \bibinfo{note}{erratum-ibid.~{D} {\bf 75}, 089902 (2007)}.

\bibitem[{\citenamefont{Lang and Hughes}(2007)}]{LangHughes07}
\bibinfo{author}{\bibfnamefont{R.~N.} \bibnamefont{Lang}} \bibnamefont{and}
  \bibinfo{author}{\bibfnamefont{S.~A.} \bibnamefont{Hughes}},
  \bibinfo{journal}{Astrophys.\ J.} \textbf{\bibinfo{volume}{677}},
  \bibinfo{pages}{1184} (\bibinfo{year}{2007}).

\bibitem[{\citenamefont{{Maj{\'a}r} and {Vas{\'u}th}}(2008)}]{MV08}
\bibinfo{author}{\bibfnamefont{J.}~\bibnamefont{{Maj{\'a}r}}} \bibnamefont{and}
  \bibinfo{author}{\bibfnamefont{M.}~\bibnamefont{{Vas{\'u}th}}},
  \bibinfo{journal}{Phys.\ Rev.\ D} \textbf{\bibinfo{volume}{77}},
  \bibinfo{pages}{104005} (\bibinfo{year}{2008}).

\bibitem[{\citenamefont{Buonanno
  et~al.}(2007{\natexlab{a}})\citenamefont{Buonanno, Cook, and
  Pretorius}}]{BCP07NR}
\bibinfo{author}{\bibfnamefont{A.}~\bibnamefont{Buonanno}},
  \bibinfo{author}{\bibfnamefont{G.~B.} \bibnamefont{Cook}}, \bibnamefont{and}
  \bibinfo{author}{\bibfnamefont{F.}~\bibnamefont{Pretorius}},
  \bibinfo{journal}{Phys.\ Rev.\ D} \textbf{\bibinfo{volume}{75}},
  \bibinfo{pages}{124018} (\bibinfo{year}{2007}{\natexlab{a}}).

\bibitem[{\citenamefont{{Berti} et~al.}(2007)\citenamefont{{Berti}, {Cardoso},
  {Gonzalez}, {Sperhake}, {Hannam}, {Husa}, and {Br{\"u}gmann}}}]{Bertietal07}
\bibinfo{author}{\bibfnamefont{E.}~\bibnamefont{{Berti}}},
  \bibinfo{author}{\bibfnamefont{V.}~\bibnamefont{{Cardoso}}},
  \bibinfo{author}{\bibfnamefont{J.~A.} \bibnamefont{{Gonzalez}}},
  \bibinfo{author}{\bibfnamefont{U.}~\bibnamefont{{Sperhake}}},
  \bibinfo{author}{\bibfnamefont{M.}~\bibnamefont{{Hannam}}},
  \bibinfo{author}{\bibfnamefont{S.}~\bibnamefont{{Husa}}}, \bibnamefont{and}
  \bibinfo{author}{\bibfnamefont{B.}~\bibnamefont{{Br{\"u}gmann}}},
  \bibinfo{journal}{Phys.\ Rev.\ D} \textbf{\bibinfo{volume}{76}},
  \bibinfo{pages}{064034} (\bibinfo{year}{2007}).

\bibitem[{\citenamefont{Baker et~al.}(2007)\citenamefont{Baker, van Meter,
  McWilliams, Centrella, and Kelly}}]{NRPNGoddard07}
\bibinfo{author}{\bibfnamefont{J.~G.} \bibnamefont{Baker}},
  \bibinfo{author}{\bibfnamefont{J.~R.} \bibnamefont{van Meter}},
  \bibinfo{author}{\bibfnamefont{S.~T.} \bibnamefont{McWilliams}},
  \bibinfo{author}{\bibfnamefont{J.}~\bibnamefont{Centrella}},
  \bibnamefont{and} \bibinfo{author}{\bibfnamefont{B.~J.} \bibnamefont{Kelly}},
  \bibinfo{journal}{Phys.\ Rev.\ Lett.} \textbf{\bibinfo{volume}{99}},
  \bibinfo{pages}{181101} (\bibinfo{year}{2007}).

\bibitem[{\citenamefont{Buonanno
  et~al.}(2007{\natexlab{b}})\citenamefont{Buonanno, Pan, Baker, Centrella,
  Kelly, McWilliams, and van Meter}}]{BuonEOB07}
\bibinfo{author}{\bibfnamefont{A.}~\bibnamefont{Buonanno}},
  \bibinfo{author}{\bibfnamefont{Y.}~\bibnamefont{Pan}},
  \bibinfo{author}{\bibfnamefont{J.~G.} \bibnamefont{Baker}},
  \bibinfo{author}{\bibfnamefont{J.}~\bibnamefont{Centrella}},
  \bibinfo{author}{\bibfnamefont{B.~J.} \bibnamefont{Kelly}},
  \bibinfo{author}{\bibfnamefont{S.~T.} \bibnamefont{McWilliams}},
  \bibnamefont{and} \bibinfo{author}{\bibfnamefont{J.~R.} \bibnamefont{van
  Meter}}, \bibinfo{journal}{Phys.\ Rev.\ D} \textbf{\bibinfo{volume}{76}},
  \bibinfo{pages}{104049} (\bibinfo{year}{2007}{\natexlab{b}}).

\bibitem[{\citenamefont{Hannam et~al.}(2008)\citenamefont{Hannam, Husa,
  Sperhake, Brugmann, and Gonzalez}}]{NRPNJena07}
\bibinfo{author}{\bibfnamefont{M.}~\bibnamefont{Hannam}},
  \bibinfo{author}{\bibfnamefont{S.}~\bibnamefont{Husa}},
  \bibinfo{author}{\bibfnamefont{U.}~\bibnamefont{Sperhake}},
  \bibinfo{author}{\bibfnamefont{B.}~\bibnamefont{Brugmann}}, \bibnamefont{and}
  \bibinfo{author}{\bibfnamefont{J.~A.} \bibnamefont{Gonzalez}},
  \bibinfo{journal}{Phys.\ Rev.\ D} \textbf{\bibinfo{volume}{77}},
  \bibinfo{pages}{044020} (\bibinfo{year}{2008}).

\bibitem[{\citenamefont{Boyle et~al.}(2007)}]{NRPNCaltech07}
\bibinfo{author}{\bibfnamefont{M.}~\bibnamefont{Boyle}} \bibnamefont{et~al.},
  \bibinfo{journal}{Phys.\ Rev.\ D} \textbf{\bibinfo{volume}{76}},
  \bibinfo{pages}{124038} (\bibinfo{year}{2007}).

\bibitem[{\citenamefont{{Hannam} et~al.}(2007)\citenamefont{{Hannam}, {Husa},
  {Br{\"u}gmann}, and {Gopakumar}}}]{HHBG07}
\bibinfo{author}{\bibfnamefont{M.}~\bibnamefont{{Hannam}}},
  \bibinfo{author}{\bibfnamefont{S.}~\bibnamefont{{Husa}}},
  \bibinfo{author}{\bibfnamefont{B.}~\bibnamefont{{Br{\"u}gmann}}},
  \bibnamefont{and}
  \bibinfo{author}{\bibfnamefont{A.}~\bibnamefont{{Gopakumar}}}
  (\bibinfo{year}{2007}), \eprint{arXiv:0712.3787 [gr-qc]}.

\bibitem[{\citenamefont{{Damour} and {Nagar}}(2008)}]{Damour2007a}
\bibinfo{author}{\bibfnamefont{T.}~\bibnamefont{{Damour}}} \bibnamefont{and}
  \bibinfo{author}{\bibfnamefont{A.}~\bibnamefont{{Nagar}}},
  \bibinfo{journal}{Phys.\ Rev.\ D} \textbf{\bibinfo{volume}{77}},
  \bibinfo{pages}{024043} (\bibinfo{year}{2008}).

\bibitem[{\citenamefont{Damour et~al.}(2008)\citenamefont{Damour, Nagar,
  Dorband, Pollney, and Rezzolla}}]{DN2007b}
\bibinfo{author}{\bibfnamefont{T.}~\bibnamefont{Damour}},
  \bibinfo{author}{\bibfnamefont{A.}~\bibnamefont{Nagar}},
  \bibinfo{author}{\bibfnamefont{E.~N.} \bibnamefont{Dorband}},
  \bibinfo{author}{\bibfnamefont{D.}~\bibnamefont{Pollney}}, \bibnamefont{and}
  \bibinfo{author}{\bibfnamefont{L.}~\bibnamefont{Rezzolla}},
  \bibinfo{journal}{Phys.\ Rev.\ D} \textbf{\bibinfo{volume}{77}},
  \bibinfo{pages}{084017} (\bibinfo{year}{2008}).

\bibitem[{\citenamefont{{Damour} et~al.}(2008)\citenamefont{{Damour}, {Nagar},
  {Hannam}, {Husa}, and {Br{\"u}gmann}}}]{DN2008}
\bibinfo{author}{\bibfnamefont{T.}~\bibnamefont{{Damour}}},
  \bibinfo{author}{\bibfnamefont{A.}~\bibnamefont{{Nagar}}},
  \bibinfo{author}{\bibfnamefont{M.}~\bibnamefont{{Hannam}}},
  \bibinfo{author}{\bibfnamefont{S.}~\bibnamefont{{Husa}}}, \bibnamefont{and}
  \bibinfo{author}{\bibfnamefont{B.}~\bibnamefont{{Br{\"u}gmann}}},
  \bibinfo{journal}{Phys.\ Rev.\ D} \textbf{\bibinfo{volume}{78}},
  \bibinfo{pages}{044039} (\bibinfo{year}{2008}).

\bibitem[{\citenamefont{{Boyle} et~al.}(2008)\citenamefont{{Boyle}, {Buonanno},
  {Kidder}, {Mrou{\'e}}, {Pan}, {Pfeiffer}, and {Scheel}}}]{UMD-CC08}
\bibinfo{author}{\bibfnamefont{M.}~\bibnamefont{{Boyle}}},
  \bibinfo{author}{\bibfnamefont{A.}~\bibnamefont{{Buonanno}}},
  \bibinfo{author}{\bibfnamefont{L.~E.} \bibnamefont{{Kidder}}},
  \bibinfo{author}{\bibfnamefont{A.~H.} \bibnamefont{{Mrou{\'e}}}},
  \bibinfo{author}{\bibfnamefont{Y.}~\bibnamefont{{Pan}}},
  \bibinfo{author}{\bibfnamefont{H.~P.} \bibnamefont{{Pfeiffer}}},
  \bibnamefont{and} \bibinfo{author}{\bibfnamefont{M.~A.}
  \bibnamefont{{Scheel}}} (\bibinfo{year}{2008}), \eprint{arXiv:0804.4184
  [gr-qc]}.

\bibitem[{\citenamefont{{Berti} et~al.}(2008)\citenamefont{{Berti}, {Cardoso},
  {Gonz{\'a}lez}, {Sperhake}, and {Br{\"u}gmann}}}]{BCGSB07}
\bibinfo{author}{\bibfnamefont{E.}~\bibnamefont{{Berti}}},
  \bibinfo{author}{\bibfnamefont{V.}~\bibnamefont{{Cardoso}}},
  \bibinfo{author}{\bibfnamefont{J.~A.} \bibnamefont{{Gonz{\'a}lez}}},
  \bibinfo{author}{\bibfnamefont{U.}~\bibnamefont{{Sperhake}}},
  \bibnamefont{and}
  \bibinfo{author}{\bibfnamefont{B.}~\bibnamefont{{Br{\"u}gmann}}},
  \bibinfo{journal}{Class. Quant. Grav.} \textbf{\bibinfo{volume}{25}},
  \bibinfo{pages}{114035} (\bibinfo{year}{2008}).

\bibitem[{\citenamefont{{Ajith} et~al.}(2008)\citenamefont{{Ajith}, {Babak},
  {Chen}, {Hewitson}, {Krishnan}, {Sintes}, {Whelan}, {Br{\"u}gmann}, {Diener},
  {Dorband} et~al.}}]{AjithNR07a}
\bibinfo{author}{\bibfnamefont{P.}~\bibnamefont{{Ajith}}},
  \bibinfo{author}{\bibfnamefont{S.}~\bibnamefont{{Babak}}},
  \bibinfo{author}{\bibfnamefont{Y.}~\bibnamefont{{Chen}}},
  \bibinfo{author}{\bibfnamefont{M.}~\bibnamefont{{Hewitson}}},
  \bibinfo{author}{\bibfnamefont{B.}~\bibnamefont{{Krishnan}}},
  \bibinfo{author}{\bibfnamefont{A.~M.} \bibnamefont{{Sintes}}},
  \bibinfo{author}{\bibfnamefont{J.~T.} \bibnamefont{{Whelan}}},
  \bibinfo{author}{\bibfnamefont{B.}~\bibnamefont{{Br{\"u}gmann}}},
  \bibinfo{author}{\bibfnamefont{P.}~\bibnamefont{{Diener}}},
  \bibinfo{author}{\bibfnamefont{N.}~\bibnamefont{{Dorband}}},
  \bibnamefont{et~al.}, \bibinfo{journal}{Phys.\ Rev.\ D}
  \textbf{\bibinfo{volume}{77}}, \bibinfo{pages}{104017}
  (\bibinfo{year}{2008}).

\bibitem[{\citenamefont{Pan et~al.}(2008)\citenamefont{Pan, Buonanno, Baker,
  Centrella, Kelly, McWilliams, Pretorius, and van Meter}}]{Pan07comparison}
\bibinfo{author}{\bibfnamefont{Y.}~\bibnamefont{Pan}},
  \bibinfo{author}{\bibfnamefont{A.}~\bibnamefont{Buonanno}},
  \bibinfo{author}{\bibfnamefont{J.~G.} \bibnamefont{Baker}},
  \bibinfo{author}{\bibfnamefont{J.}~\bibnamefont{Centrella}},
  \bibinfo{author}{\bibfnamefont{B.~J.} \bibnamefont{Kelly}},
  \bibinfo{author}{\bibfnamefont{S.~T.} \bibnamefont{McWilliams}},
  \bibinfo{author}{\bibfnamefont{F.}~\bibnamefont{Pretorius}},
  \bibnamefont{and} \bibinfo{author}{\bibfnamefont{J.~R.} \bibnamefont{van
  Meter}}, \bibinfo{journal}{Phys.\ Rev.\ D} \textbf{\bibinfo{volume}{77}},
  \bibinfo{pages}{024014} (\bibinfo{year}{2008}).

\bibitem[{\citenamefont{Finn and Chernoff}(1993)}]{FinnCh93}
\bibinfo{author}{\bibfnamefont{L.}~\bibnamefont{Finn}} \bibnamefont{and}
  \bibinfo{author}{\bibfnamefont{D.}~\bibnamefont{Chernoff}},
  \bibinfo{journal}{Phys. Rev. D} \textbf{\bibinfo{volume}{47}},
  \bibinfo{pages}{2198} (\bibinfo{year}{1993}).

\bibitem[{\citenamefont{{Poisson}}(1998)}]{Poisson98}
\bibinfo{author}{\bibfnamefont{E.}~\bibnamefont{{Poisson}}},
  \bibinfo{journal}{Phys.\ Rev.\ D} \textbf{\bibinfo{volume}{57}},
  \bibinfo{pages}{5287} (\bibinfo{year}{1998}).

\bibitem[{\citenamefont{Apostolatos et~al.}(1994)\citenamefont{Apostolatos,
  Cutler, Sussman, and Thorne}}]{ACST94}
\bibinfo{author}{\bibfnamefont{T.~A.} \bibnamefont{Apostolatos}},
  \bibinfo{author}{\bibfnamefont{C.}~\bibnamefont{Cutler}},
  \bibinfo{author}{\bibfnamefont{G.~J.} \bibnamefont{Sussman}},
  \bibnamefont{and} \bibinfo{author}{\bibfnamefont{K.~S.}
  \bibnamefont{Thorne}}, \bibinfo{journal}{Phys. Rev.~D}
  \textbf{\bibinfo{volume}{49}}, \bibinfo{pages}{6274} (\bibinfo{year}{1994}).

\bibitem[{\citenamefont{Goldberg et~al.}(1967)\citenamefont{Goldberg,
  Macfarlane, Newman, Rohrlich, and Sudarshan}}]{GMNRS67}
\bibinfo{author}{\bibfnamefont{J.~N.} \bibnamefont{Goldberg}},
  \bibinfo{author}{\bibfnamefont{A.~J.} \bibnamefont{Macfarlane}},
  \bibinfo{author}{\bibfnamefont{E.~T.} \bibnamefont{Newman}},
  \bibinfo{author}{\bibfnamefont{F.}~\bibnamefont{Rohrlich}}, \bibnamefont{and}
  \bibinfo{author}{\bibfnamefont{E.~C.~G.} \bibnamefont{Sudarshan}},
  \bibinfo{journal}{J. Math. Phys.} \textbf{\bibinfo{volume}{8}},
  \bibinfo{pages}{2155} (\bibinfo{year}{1967}).

\bibitem[{\citenamefont{Landau and Lifshitz}(1977)}]{LLvol3}
\bibinfo{author}{\bibfnamefont{L.}~\bibnamefont{Landau}} \bibnamefont{and}
  \bibinfo{author}{\bibfnamefont{E.}~\bibnamefont{Lifshitz}},
  \emph{\bibinfo{title}{Quantum Mechanics}} (\bibinfo{publisher}{Pergamon},
  \bibinfo{address}{Oxford}, \bibinfo{year}{1977}).

\bibitem[{\citenamefont{{Gualtieri} et~al.}(2008)\citenamefont{{Gualtieri},
  {Berti}, {Cardoso}, and {Sperhake}}}]{GBCS}
\bibinfo{author}{\bibfnamefont{L.}~\bibnamefont{{Gualtieri}}},
  \bibinfo{author}{\bibfnamefont{E.}~\bibnamefont{{Berti}}},
  \bibinfo{author}{\bibfnamefont{V.}~\bibnamefont{{Cardoso}}},
  \bibnamefont{and}
  \bibinfo{author}{\bibfnamefont{U.}~\bibnamefont{{Sperhake}}},
  \bibinfo{journal}{Class. Quantum Grav.} \textbf{\bibinfo{volume}{805}}
  (\bibinfo{year}{2008}).

\bibitem[{\citenamefont{{Campanelli} et~al.}(2008)\citenamefont{{Campanelli},
  {Lousto}, {Nakano}, and {Zlochower}}}]{CLNZ08}
\bibinfo{author}{\bibfnamefont{M.}~\bibnamefont{{Campanelli}}},
  \bibinfo{author}{\bibfnamefont{C.~O.} \bibnamefont{{Lousto}}},
  \bibinfo{author}{\bibfnamefont{H.}~\bibnamefont{{Nakano}}}, \bibnamefont{and}
  \bibinfo{author}{\bibfnamefont{Y.}~\bibnamefont{{Zlochower}}}
  (\bibinfo{year}{2008}), \eprint{arXiv:0808.0713 [gr-qc]}.

\bibitem[{\citenamefont{{Racine}}(2008)}]{ER08}
\bibinfo{author}{\bibfnamefont{E.}~\bibnamefont{{Racine}}},
  \bibinfo{journal}{Phys.\ Rev.\ D} \textbf{\bibinfo{volume}{78}},
  \bibinfo{pages}{044021} (\bibinfo{year}{2008}).

\bibitem[{\citenamefont{Jaranowski and Sch\"afer}(1999)}]{JS99}
\bibinfo{author}{\bibfnamefont{P.}~\bibnamefont{Jaranowski}} \bibnamefont{and}
  \bibinfo{author}{\bibfnamefont{G.}~\bibnamefont{Sch\"afer}},
  \bibinfo{journal}{Phys. Rev. D} \textbf{\bibinfo{volume}{60}},
  \bibinfo{pages}{124003} (\bibinfo{year}{1999}).

\bibitem[{\citenamefont{Jaranowski and Sch\"afer}(2000)}]{JaraS00}
\bibinfo{author}{\bibfnamefont{P.}~\bibnamefont{Jaranowski}} \bibnamefont{and}
  \bibinfo{author}{\bibfnamefont{G.}~\bibnamefont{Sch\"afer}},
  \bibinfo{journal}{Ann. Phys. (Berlin)} \textbf{\bibinfo{volume}{9}},
  \bibinfo{pages}{378} (\bibinfo{year}{2000}).

\bibitem[{\citenamefont{{de Andrade} et~al.}(2001)\citenamefont{{de Andrade},
  {Blanchet}, and {Faye}}}]{Andrade01}
\bibinfo{author}{\bibfnamefont{V.~C.} \bibnamefont{{de Andrade}}},
  \bibinfo{author}{\bibfnamefont{L.}~\bibnamefont{{Blanchet}}},
  \bibnamefont{and} \bibinfo{author}{\bibfnamefont{G.}~\bibnamefont{{Faye}}},
  \bibinfo{journal}{Class.\ Quantum Grav.} \textbf{\bibinfo{volume}{18}},
  \bibinfo{pages}{753} (\bibinfo{year}{2001}).

\bibitem[{\citenamefont{{Blanchet} and {Faye}}(2001)}]{Blanchet01a}
\bibinfo{author}{\bibfnamefont{L.}~\bibnamefont{{Blanchet}}} \bibnamefont{and}
  \bibinfo{author}{\bibfnamefont{G.}~\bibnamefont{{Faye}}},
  \bibinfo{journal}{Phys.\ Rev.\ D} \textbf{\bibinfo{volume}{63}},
  \bibinfo{pages}{062005} (\bibinfo{year}{2001}).

\bibitem[{\citenamefont{{Damour} et~al.}(2001)\citenamefont{{Damour},
  {Jaranowski}, and {Sch{\"a}fer}}}]{Damour01}
\bibinfo{author}{\bibfnamefont{T.}~\bibnamefont{{Damour}}},
  \bibinfo{author}{\bibfnamefont{P.}~\bibnamefont{{Jaranowski}}},
  \bibnamefont{and}
  \bibinfo{author}{\bibfnamefont{G.}~\bibnamefont{{Sch{\"a}fer}}},
  \bibinfo{journal}{Phys.\ Lett.} \textbf{\bibinfo{volume}{513B}},
  \bibinfo{pages}{147} (\bibinfo{year}{2001}).

\bibitem[{\citenamefont{Arun et~al.}(2005)\citenamefont{Arun, Iyer,
  Sathyaprakash, and Sundararajan}}]{AISS05}
\bibinfo{author}{\bibfnamefont{K.~G.} \bibnamefont{Arun}},
  \bibinfo{author}{\bibfnamefont{B.~R.} \bibnamefont{Iyer}},
  \bibinfo{author}{\bibfnamefont{B.~S.} \bibnamefont{Sathyaprakash}},
  \bibnamefont{and} \bibinfo{author}{\bibfnamefont{P.~A.}
  \bibnamefont{Sundararajan}}, \bibinfo{journal}{Phys.~Rev.~D}
  \textbf{\bibinfo{volume}{71}}, \bibinfo{pages}{084008}
  (\bibinfo{year}{2005}), \bibinfo{note}{erratum-ibid. ~{ D } {\bf 72}, 069903
  (2005)}.

\bibitem[{\citenamefont{Porter and Cornish}(2008)}]{PorterCornish08}
\bibinfo{author}{\bibfnamefont{E.~K.} \bibnamefont{Porter}} \bibnamefont{and}
  \bibinfo{author}{\bibfnamefont{N.~J.} \bibnamefont{Cornish}}
  (\bibinfo{year}{2008}), \eprint{arXiv:0804.0332 [gr-qc]}.

\bibitem[{\citenamefont{{Berti} et~al.}(2005)\citenamefont{{Berti}, {Buonanno},
  and {Will}}}]{BBW05a}
\bibinfo{author}{\bibfnamefont{E.}~\bibnamefont{{Berti}}},
  \bibinfo{author}{\bibfnamefont{A.}~\bibnamefont{{Buonanno}}},
  \bibnamefont{and} \bibinfo{author}{\bibfnamefont{C.~M.}
  \bibnamefont{{Will}}}, \bibinfo{journal}{Phys.~Rev.~D}
  \textbf{\bibinfo{volume}{71}}, \bibinfo{pages}{084025}
  (\bibinfo{year}{2005}).

\bibitem[{\citenamefont{Cutler}(1998)}]{Cutler98}
\bibinfo{author}{\bibfnamefont{C.}~\bibnamefont{Cutler}},
  \bibinfo{journal}{Phys. Rev. D} \textbf{\bibinfo{volume}{57}},
  \bibinfo{pages}{7089} (\bibinfo{year}{1998}).

\bibitem[{\citenamefont{Campanelli
  et~al.}(2006{\natexlab{a}})\citenamefont{Campanelli, Lousto, and
  Zlochower}}]{Campanelli2006c}
\bibinfo{author}{\bibfnamefont{M.}~\bibnamefont{Campanelli}},
  \bibinfo{author}{\bibfnamefont{C.~O.} \bibnamefont{Lousto}},
  \bibnamefont{and}
  \bibinfo{author}{\bibfnamefont{Y.}~\bibnamefont{Zlochower}},
  \bibinfo{journal}{Phys.\ Rev.\ D} \textbf{\bibinfo{volume}{74}},
  \bibinfo{pages}{041501} (\bibinfo{year}{2006}{\natexlab{a}}).

\bibitem[{\citenamefont{Campanelli
  et~al.}(2006{\natexlab{b}})\citenamefont{Campanelli, Lousto, and
  Zlochower}}]{Campanelli2006d}
\bibinfo{author}{\bibfnamefont{M.}~\bibnamefont{Campanelli}},
  \bibinfo{author}{\bibfnamefont{C.~O.} \bibnamefont{Lousto}},
  \bibnamefont{and}
  \bibinfo{author}{\bibfnamefont{Y.}~\bibnamefont{Zlochower}},
  \bibinfo{journal}{Phys.\ Rev.\ D} \textbf{\bibinfo{volume}{74}},
  \bibinfo{pages}{084023} (\bibinfo{year}{2006}{\natexlab{b}}).

\bibitem[{\citenamefont{Campanelli et~al.}(2007)\citenamefont{Campanelli,
  Lousto, Zlochower, Krishnan, and Merritt}}]{Campanelli2007b}
\bibinfo{author}{\bibfnamefont{M.}~\bibnamefont{Campanelli}},
  \bibinfo{author}{\bibfnamefont{C.~O.} \bibnamefont{Lousto}},
  \bibinfo{author}{\bibfnamefont{Y.}~\bibnamefont{Zlochower}},
  \bibinfo{author}{\bibfnamefont{B.}~\bibnamefont{Krishnan}}, \bibnamefont{and}
  \bibinfo{author}{\bibfnamefont{D.}~\bibnamefont{Merritt}},
  \bibinfo{journal}{Phys.\ Rev.\ D} \textbf{\bibinfo{volume}{75}},
  \bibinfo{pages}{064030} (\bibinfo{year}{2007}).

\bibitem[{\citenamefont{Herrmann
  et~al.}(2007{\natexlab{a}})\citenamefont{Herrmann, Hinder, Shoemaker, Laguna,
  and Matzner}}]{Herrmann2007}
\bibinfo{author}{\bibfnamefont{F.}~\bibnamefont{Herrmann}},
  \bibinfo{author}{\bibfnamefont{I.}~\bibnamefont{Hinder}},
  \bibinfo{author}{\bibfnamefont{D.}~\bibnamefont{Shoemaker}},
  \bibinfo{author}{\bibfnamefont{P.}~\bibnamefont{Laguna}}, \bibnamefont{and}
  \bibinfo{author}{\bibfnamefont{R.~A.} \bibnamefont{Matzner}},
  \bibinfo{journal}{Astrophys.\ J.} \textbf{\bibinfo{volume}{661}},
  \bibinfo{pages}{430} (\bibinfo{year}{2007}{\natexlab{a}}).

\bibitem[{\citenamefont{Herrmann
  et~al.}(2007{\natexlab{b}})\citenamefont{Herrmann, Hinder, Shoemaker, Laguna,
  and Matzner}}]{Herrmann2007c}
\bibinfo{author}{\bibfnamefont{F.}~\bibnamefont{Herrmann}},
  \bibinfo{author}{\bibfnamefont{I.}~\bibnamefont{Hinder}},
  \bibinfo{author}{\bibfnamefont{D.~M.} \bibnamefont{Shoemaker}},
  \bibinfo{author}{\bibfnamefont{P.}~\bibnamefont{Laguna}}, \bibnamefont{and}
  \bibinfo{author}{\bibfnamefont{R.~A.} \bibnamefont{Matzner}},
  \bibinfo{journal}{Phys.\ Rev.\ D} \textbf{\bibinfo{volume}{76}},
  \bibinfo{pages}{084032} (\bibinfo{year}{2007}{\natexlab{b}}).

\bibitem[{\citenamefont{{Alvi}}(2001)}]{KA}
\bibinfo{author}{\bibfnamefont{K.}~\bibnamefont{{Alvi}}},
  \bibinfo{journal}{\prd} \textbf{\bibinfo{volume}{64}},
  \bibinfo{pages}{104020} (\bibinfo{year}{2001}).

\end{thebibliography}
\end{document}